\documentclass[useAMS,usenatbib]{mn2e}

\usepackage{graphicx}
\usepackage{rotating}
\usepackage{hyperref}
\usepackage{multirow}

\title[Statistics of the Sunyaev-Zel'dovich Effect power spectrum]{Statistics of the Sunyaev-Zel'dovich Effect power spectrum}
\author[Michael W. Peel, Richard A. Battye and Scott T. Kay]{Michael W. Peel, Richard A.
Battye and Scott T. Kay\\
Jodrell Bank Centre for Astrophysics, School of Physics and Astronomy, University of Manchester, Manchester, M13 9PL}
\begin{document}

\date{Accepted 2009 May 21. Received 2009 May 13; in original form 2009 April 08}

\pagerange{\pageref{firstpage}--\pageref{lastpage}} \pubyear{2009}

\maketitle

\label{firstpage}

\begin{abstract}
Using large numbers of simulations of the microwave sky, incorporating the Cosmic Microwave Background (CMB) and the Sunyaev-Zel'dovich (SZ) effect due to clusters, we investigate the statistics of the power spectrum at microwave frequencies between spherical multipoles of 1000 and 10000. From these virtual sky maps, we find that the spectrum of the SZ effect has a larger standard deviation by a factor of 3 than would be expected from purely Gaussian realizations, and has a distribution that is significantly skewed towards higher values, especially when small map sizes are used. The standard deviation is also increased by around 10 percent compared to the trispectrum calculation due to the clustering of galaxy clusters. We also consider the effects of including residual point sources and uncertainties in the gas physics. This has implications for the excess power measured in the CMB power spectrum by the Cosmic Background Imager and BIMA experiments. Our results indicate that the observed excess could be explained using a lower value of $\sigma_8$ than previously suggested, however the effect is not enough to match $\sigma_8=0.825$. The uncertainties in the gas physics could also play a substantial role. We have made our maps of the SZ effect available online.
\end{abstract}


\section{Introduction}
The discovery of the Cosmic Microwave Background (CMB) by \citet{1965Penzias} helped provide the foundations of modern cosmology. The discovery of the primordial anisotropies on large scales by {\it COBE} \citep{1992Smoot}, followed by the precision measurements by the Wilkinson Microwave Anisotropy Probe \citep[{\it WMAP}; see for example][]{2009Hinshaw} and other experiments, pinned down the values of the principle cosmological parameters. With the forthcoming measurements by the {\it Planck} satellite, the primordial anisotropies at large scales will be measured as accurately as possible \citep{2005Planck}, with cosmic variance preventing any more precise measurements of the cosmological parameters from these anisotropies. Interest is now turning towards measurements of the primordial polarization and the total power on smaller scales. 

At these smaller scales, ``secondary" anisotropies start to dominate over the primordial signal. A significant component of this secondary anisotropy will be caused by the Sunyaev-Zel'dovich (SZ) effect. This is the inverse Compton scattering of photons from the CMB by free electrons (\citealp{1970Sunyaev,1972Sunyaev}; see \citealp{1999Birkinshaw} for a review). It is expected to be dominated by groups and clusters of galaxies due to the high temperatures of the intracluster medium. At microwave frequencies ($\sim 30$ GHz) the SZ effect will contaminate the power spectrum due to the primary CMB anisotropies at spherical multipoles of $l\sim1000$ and above, making it difficult to extract cosmological information from the primary anisotropies at these scales; by $l\sim2000$ it is expected that it will dominate the power spectrum.

Several experiments have measured the power spectrum in this region of multipoles, and some of these show hints of excess power over what would be expected from the primary anisotropies alone. The Cosmic Background Interferometer \citep[CBI;][]{2003Mason,2004Readhead,2009Sievers}, which observed multipoles up to $l=4000$ at 30~GHz, was the first to measure a possible excess, and this has been confirmed by subsequent analysis. This was followed by the BIMA interferometer \citep{2006Dawson} observing around $l=5000$ at 28.5~GHz, and at higher frequencies by ACBAR, which has weakly detected a possible excess at around $l=2500$ at 150~GHz \citep{2008Reichardt}. Curiously, observations at similar multipoles by the Sunyaev-Zel'dovich Array at 30~GHz \citep[SZA;][]{2009Sharp} and QUaD at 100 and 150~GHz \citep{2009Friedman} do not show any excess power.

The amount of signal from the SZ effect depends on the amplitude of the power spectrum of the density fluctuations in the Universe. This power spectrum is normalized by $\sigma_8$, which is the variance of the fluctuations on scales of $8 h^{-1}$~Mpc. To explain the excess measured by the CBI, \citet{2009Sievers} find that $\sigma_8$ must be between 0.9 and 1.0. However, observations of the primordial CMB anisotropies by {\it WMAP} in combination with other instruments and methods yield $\sigma_8 = 0.812 \pm 0.026$ \citep{2009Komatsu}.

An important issue to consider when estimating the range of possible values of $\sigma_8$ that agree with the data is the statistical properties of the SZ effect, which will be non-Gaussian. Here we investigate those properties to see whether the excess could be the result of measuring part of the sky where the power from the SZ effect is higher than average, which would bias the measurement of $\sigma_8$ to higher than the average value. To do this, we use $3 \times 3$ degree sky maps with a resolution of 18~arcseconds (600 pixels to a side), which allow us to accurately probe the multipole range $l = 1000-10000$.

To ease comparisons with the observations by CBI and BIMA of the high multipole excess, we use a canonical frequency of 30~GHz. We use a $\Lambda \mathrm{CDM}$ cosmology throughout this paper, with $\Omega_\mathrm{m} = 0.3$, $\Omega_\mathrm{\Lambda} = 0.7$ and $\Omega_\mathrm{b} = 0.05$. The Hubble constant $H_0 = 100 \phantom{.} h \phantom{.} \mathrm{km s^{-1} Mpc^{-1}}$, where $h = 0.7$. We use three values of $\sigma_8$: 0.75, 0.825 and 0.9. These span the best-fitting values from {\it WMAP} after 1, 3 and 5 years of observations \citep{2003Spergel,2007Spergel,2009Komatsu}.

For each cosmology, we perform 100 {\sc Pinocchio}  \citep{2002Monacoa,2002Monacob,2002Taffoni} simulations to simulate the growth of large-scale structure in the Universe; these simulations are much quicker to run than N-body simulations yet provide cluster distributions with an appropriate two-point correlation function on large scales. The output of these simulations is then used to create 1000 light cone cluster catalogues. Using a simple model of the gas physics, virtual sky maps of the SZ effect are created from the cluster catalogues and combined with maps of the CMB and residual foreground point sources; the latter is based on the extrapolated source number counts from a range of surveys.

We describe the models and methods used to create the virtual sky maps in Section 2. The statistics of these maps and their dependency on the cluster model parameters are analysed in Section 3. Section 4 discusses the effects of point sources, followed by the implications for the high multipole excess in Section 5. We conclude in Section 6.

\section{Virtual Sky model}
The virtual sky maps are built up from a series of components. First, maps of the CMB are generated using a Gaussian random field. We then use {\sc Pinocchio} simulations to calculate the positions and masses of galaxy clusters, and for each cluster we use a simple, spherically symmetric model for both the SZ effect and the surface matter density, the latter of which is used for distributing point sources (see Section \ref{sec:ps}). Full details of each stage are given below.

The effects of other foregrounds, including the galaxy, are not included as the observations focus on areas of the sky where galactic foregrounds are at a minimum. Additionally, these foregrounds are expected to be fairly smooth on the scales under consideration here.

Similar work creating maps of the SZ effect has been carried out by e.g. \citet{2006Geisbusch}, \citet{2006Schafer} and \citet{2007Sehgal}; in particular, \citet{2007Holder} also use {\sc Pinocchio} simulations to create their cluster catalogues.

\subsection{Cosmic Microwave Background anisotropies}
\begin{figure}
\centering
\includegraphics[scale=0.68]{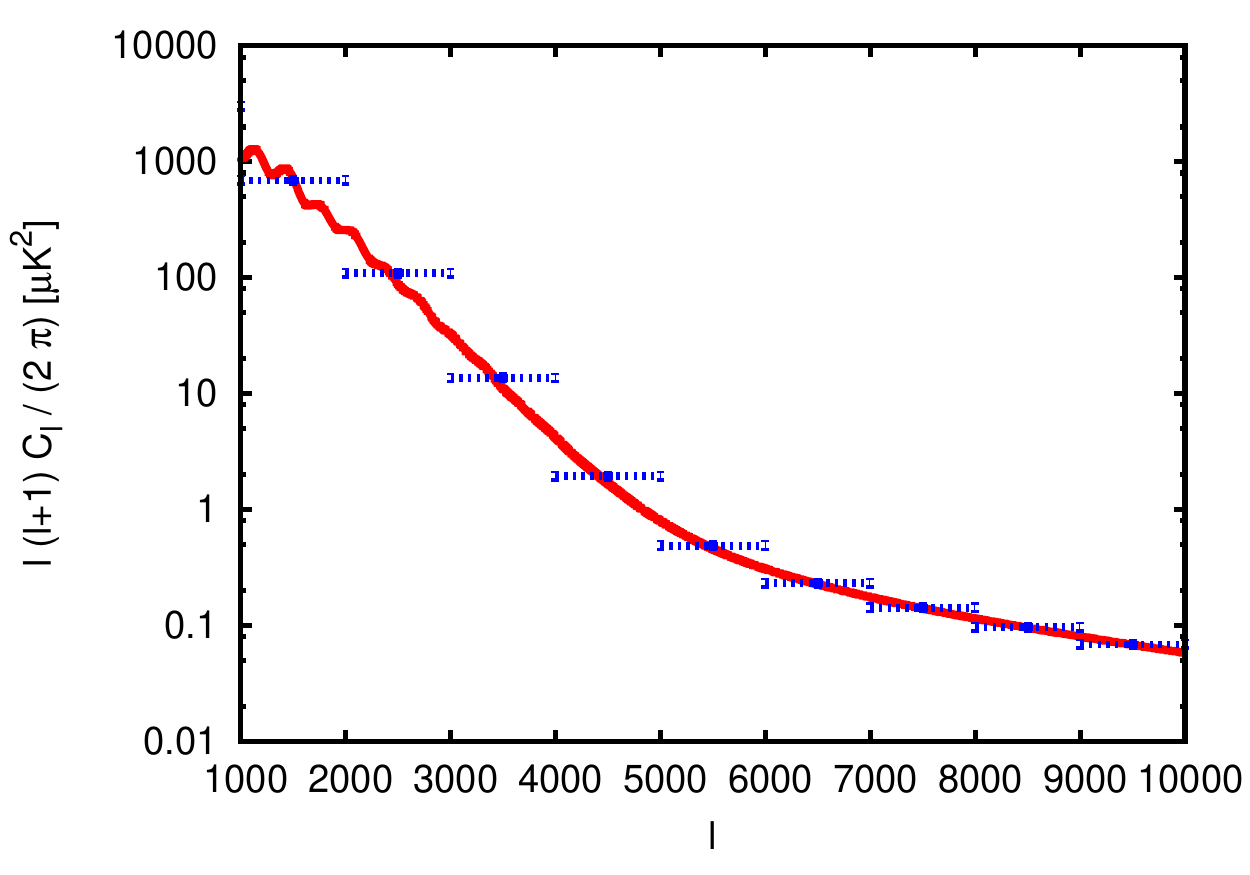}
\caption{The theoretical (red line) and binned (blue points) power spectrum of the lensed CMB between $l=$1000 and 10000, averaged over 1000 maps and binned with $\Delta l = 1000$. The binned power spectrum agrees well with the input power spectrum.}
\label{fig:cmb_powerspectrum}
\end{figure}

\begin{figure}
\centering
\includegraphics[scale=0.68]{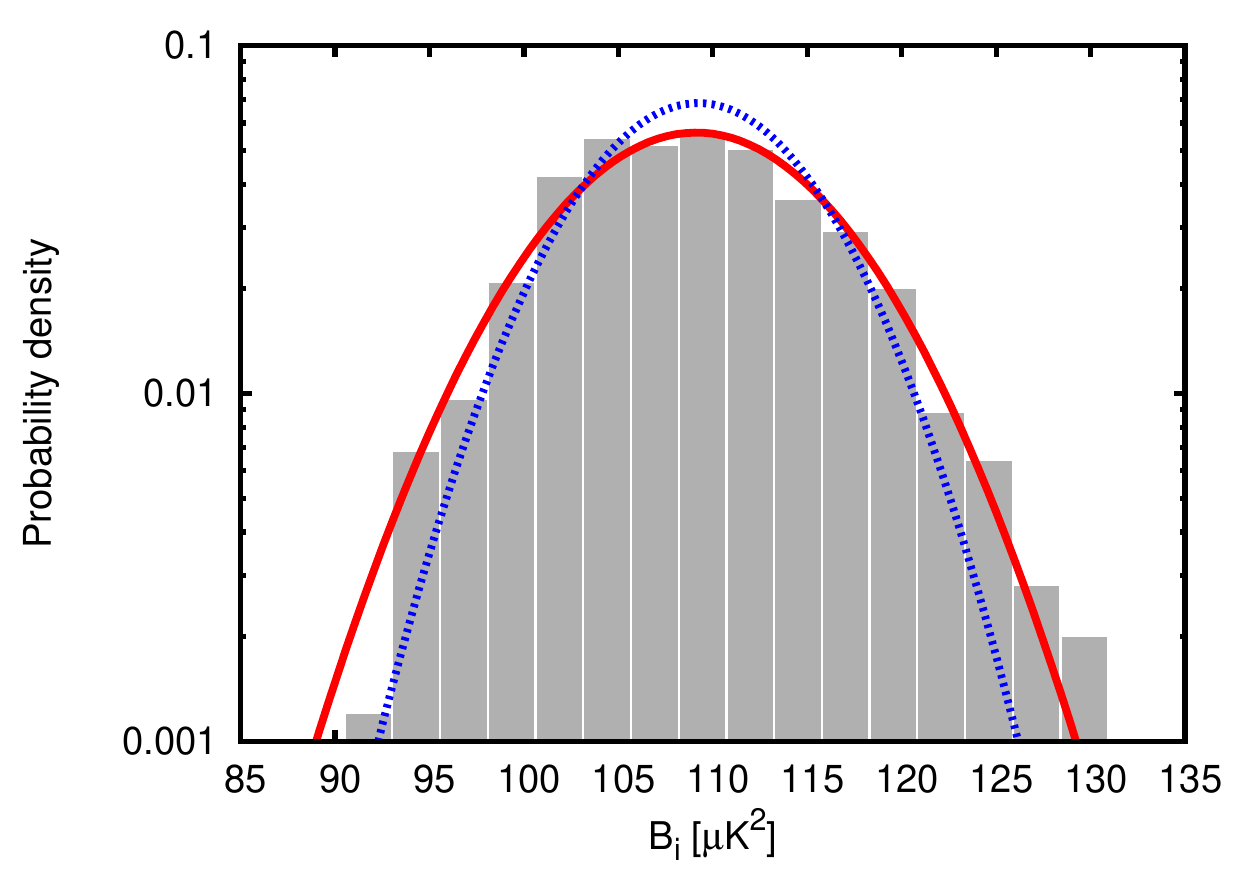}
\caption{The statistics of the CMB realizations in the multipole bin 2000-3000. The solid red line is the best fit Gaussian distribution, whilst the dotted blue line is the distribution for this bin based on the analytic formula for the variance (equation \ref{eq:cv}). The distribution is well fit by a Gaussian distribution with a slightly higher standard deviation than the analytical fomula.}
\label{fig:cmb_statistics}
\end{figure}

Maps of the small-scale CMB anisotropies can be computed using the standard assumption that the CMB is a Gaussian random field and hence is completely described by its angular power spectrum, $C_l$. We generate the expected power spectrum of the gravitationally lensed CMB from {\sc CAMB}\footnote{November 2008 version; available from \url{http://www.camb.info}} \citep{2000Lewis}. It is known that the gravitational lensing component is non-Gaussian \citep[see for example][]{2000Zaldarriaga}, however as this component is sub-dominant we assume that it is Gaussianly distributed.

When calculating the binned power spectrum, we assume that $l(l+1)C_l / (2 \pi)$ is flat over a region of size $\Delta l$ such that the power within a multipole bin centred on $\bar{l}_i$ can be calculated by
\begin{equation} \label{eq:binning}
B_i = \frac{1}{\Delta l} \sum_{l \in \mathrm{bin}} \frac{l(l+1) {C_l}}{2 \pi},
\end{equation}
where the summation is over $\bar{l}_i - \frac{1}{2} \Delta l \leq l < \bar{l}_i + \frac{1}{2} \Delta l$. Values of $C_l$ that have not been sampled due to the finite grid are assumed to have the value of the closest $C_l$ that has been sampled. This is obviously imperfect where the $C_l$'s have a large gradient or are ill- or irregularly-sampled, but is likely to be sufficient for the purposes at hand.

The expected mean in each multipole bin can be calculated from the input power spectrum, and the expected variance is calculated by
\begin{equation} \label{eq:cv}
\delta B_{i} = \frac{1}{\Delta l}\sqrt{\sum_{l \in \mathrm{bin}} \left(\frac{l(l+1)C_l}{2 \pi} \right)^2 \frac{2 }{(2 l + 1) f_\mathrm{sky}}},
\end{equation}
where $f_\mathrm{sky}$ is the fraction of the sky, $\Delta l$ is the width of the multipole bin (here, $\Delta l = 1000$) and the sum is over all multipoles within the bin. The quantity $\delta B_i \sim \bar{l}_i^{-1/2} f_\mathrm{sky}^{-1/2}$ is sometimes called the cosmic variance since it represents the limit on how well one can possibly measure $B_i$ as imposed by the Gaussian statistics. The aim of this paper is to understand this issue in the case of the non-Gaussian anisotropies created by the SZ effect.

Fig. \ref{fig:cmb_powerspectrum} shows the CMB power spectrum using bins of 1000 multipoles, and Fig. \ref{fig:cmb_statistics} shows the statistics within the multipole bin 2000-3000. The statistics are consistent with a Gaussian distribution with a slightly larger standard deviation than expected from cosmic variance due to inefficiencies in the sampling of the spectrum at different multipoles. The distribution is also slightly skewed, as the finite sampling leads to a $\chi^2$ distribution rather than a Gaussian one.

\subsection{Cluster catalogues} \label{sec:cluster_cats}
\begin{figure}
\centering
\includegraphics[scale=0.68]{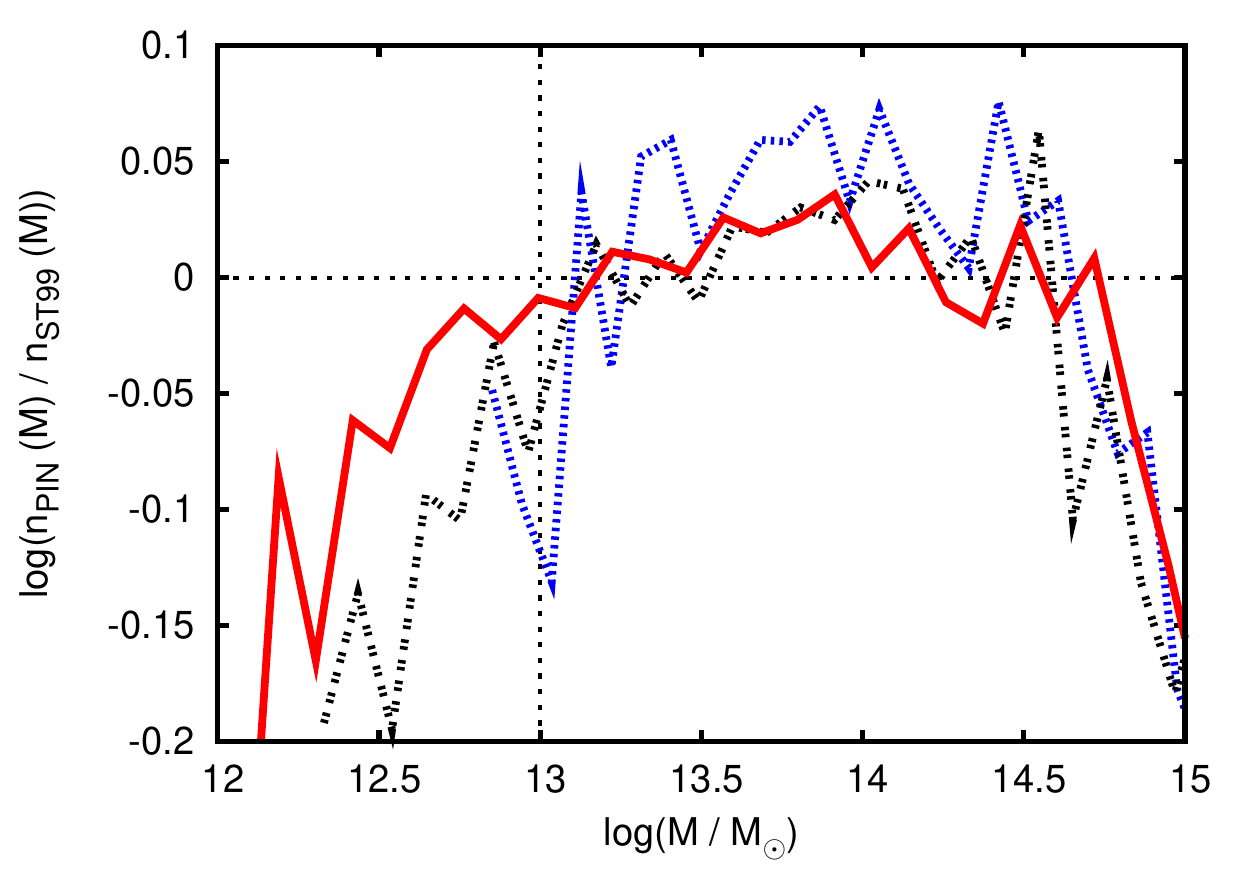}
\caption{The ratio of the {\sc Pinocchio} mass function to ST99 for different resolutions of the same {\sc Pinocchio} simulation. The resolutions are 2.5~Mpc (blue short dashed line), 1.66~Mpc (black dotted line) and 1.25~Mpc (red solid line; our adopted resolution). The latter two are consistent with each other down to $M_\mathrm{vir} \sim 10^{13} M_\odot$. At $M_\mathrm{vir}=10^{13} M_\odot$ the mass function from the highest resolution simulation is in good agreement with that from ST99 for a resolution of 1.25~Mpc. At the highest masses the scatter is caused by small number statistics, but a systematic trend downwards is still present.}
\label{fig:pinocchio_convergence}
\end{figure}

In order to create a map of the SZ effect, the positions and properties of the clusters need to be determined. We use the PINpointing Orbit-Crossing Collapsed HIerarchical Objects ({\sc Pinocchio}) algorithm\footnote{Version 2.1.2 beta, available from \\ \url{http://adlibitum.oats.inaf.it/monaco/Homepage/Pinocchio/}} \citep{2002Monacoa,2002Monacob,2002Taffoni} to generate mock cluster catalogues.

{\sc Pinocchio} uses Lagrangian perturbation theory to predict the collapse of matter into haloes. Galaxy cluster positions and masses generated by {\sc Pinocchio} will be associated with the large-scale cosmological structure, and hence will have a non-zero two-point correlation function (they will be clustered), whereas distributions created from analytical mass functions (\citealp[for example][henceforth ST99]{1999Sheth}) with random cluster positions will be Poisson distributed with zero two-point correlations. {\sc Pinocchio} does not simulate the structure within haloes, thus offering a significantly faster method of creating cluster catalogues compared with N-body techniques. In contrast to N-body simulations, requiring many thousands of CPU-hours to run, a reasonable resolution {\sc Pinocchio} run takes less than a day on a single CPU.

We have run 100 {\sc Pinocchio} simulations for each value of $\sigma_8$, each simulating comoving cubes with sides of 500~Mpc with a cell size of 1.25~Mpc (thus 400~cells to a side). Halo catalogues were created at 50~redshifts between $z=5$ and $z=0$, the centres of which are spaced 160~Mpc apart in comoving distance. The initial power spectrum used to generate the distribution of matter in the simulation is from \citet{1992Efstathiou}.

A set of 1000 lightcone catalogues for each value of $\sigma_8$ were generated from the simulation output. For each of the 50 redshifts, a halo catalogue is randomly selected from the 100~simulations. This halo catalogue is then translated by random amounts, independently in all three dimensions using a periodic box, as well as being rotated by a randomly chosen integer multiple of 90~degrees independently along each of the three axes. A 160~Mpc deep slice of the simulation is then taken, and clusters within a 3 by 3 degree block centred on the middle of the slice are included in the lightcone. The limit of 3~degrees to a side is imposed by the size of the simulation at the highest redshift; the box size subtends an angle of 3~degrees at $z=5$.

\begin{figure}
\centering
\includegraphics[scale=0.68]{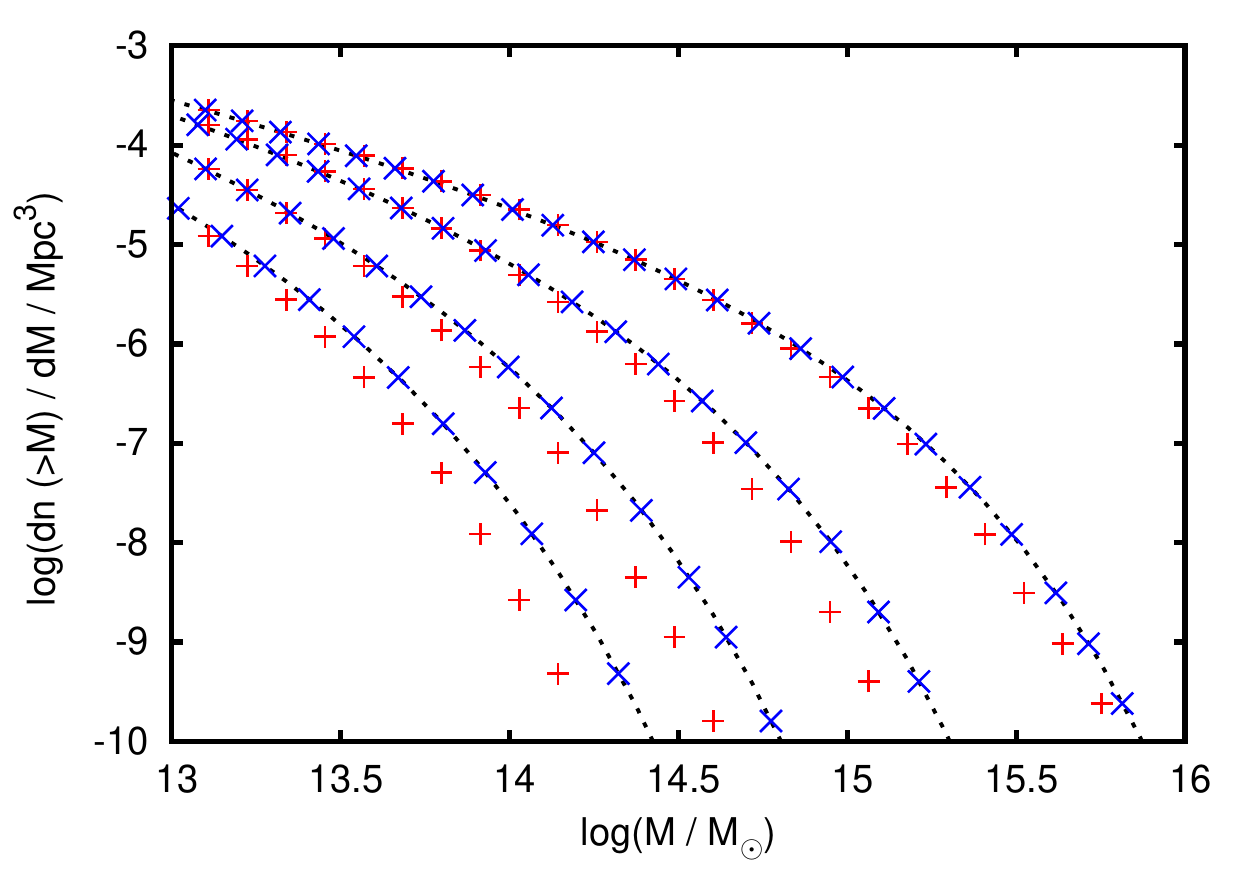}
\caption{The comoving differential number count of clusters greater than a given mass analytically from ST99 (dashed lines), from the mean mass function of all {\sc Pinocchio} catalogues with $\sigma_8=0.9$ (red +) and from the corrected {\sc Pinocchio} catalogues (blue crosses). The number counts for four redshifts are given; from the top down these are z=0, 1, 2 and 3. The correction brings the two mass functions into agreement}
\label{fig:pinocchio_discrepancy}
\end{figure}

The choice of simulation resolution means that the simulations have a high degree of completeness down to $M_\mathrm{vir} = 10^{13} M_\odot$; that is, the comoving number density of clusters $n(M) = dn/dM$ agrees between the Pinocchio simulations and that expected from ST99, as shown in Fig. \ref{fig:pinocchio_convergence}. However, {\sc Pinocchio} systematically underestimates the number of the largest mass clusters when compared to the analytical prediction of ST99; this becomes more pronounced at higher redshifts (see Fig. \ref{fig:pinocchio_discrepancy}). This is unexpected since {\sc Pinocchio} is designed to reproduce ST99, and causes a significant discrepancy in the amplitude of the power spectrum of the SZ effect compared with maps created from random cluster catalogues generated from the ST99 number count predictions.

We assume that the ST99 mass function is the more accurate of the two, although \citet{2006Warren} show that this may itself be systematically low at the highest masses compared to N-body simulations. We calculate a mass correction table for each cosmology such that the number of clusters greater than a given mass at each redshift (calculated from the average of all of the simulations with the same cosmology) matches the predicted value from ST99. The appropriate correction is then applied to all clusters in the lightcone prior to the creation of the maps.

\begin{figure}
\centering
\includegraphics[scale=0.68]{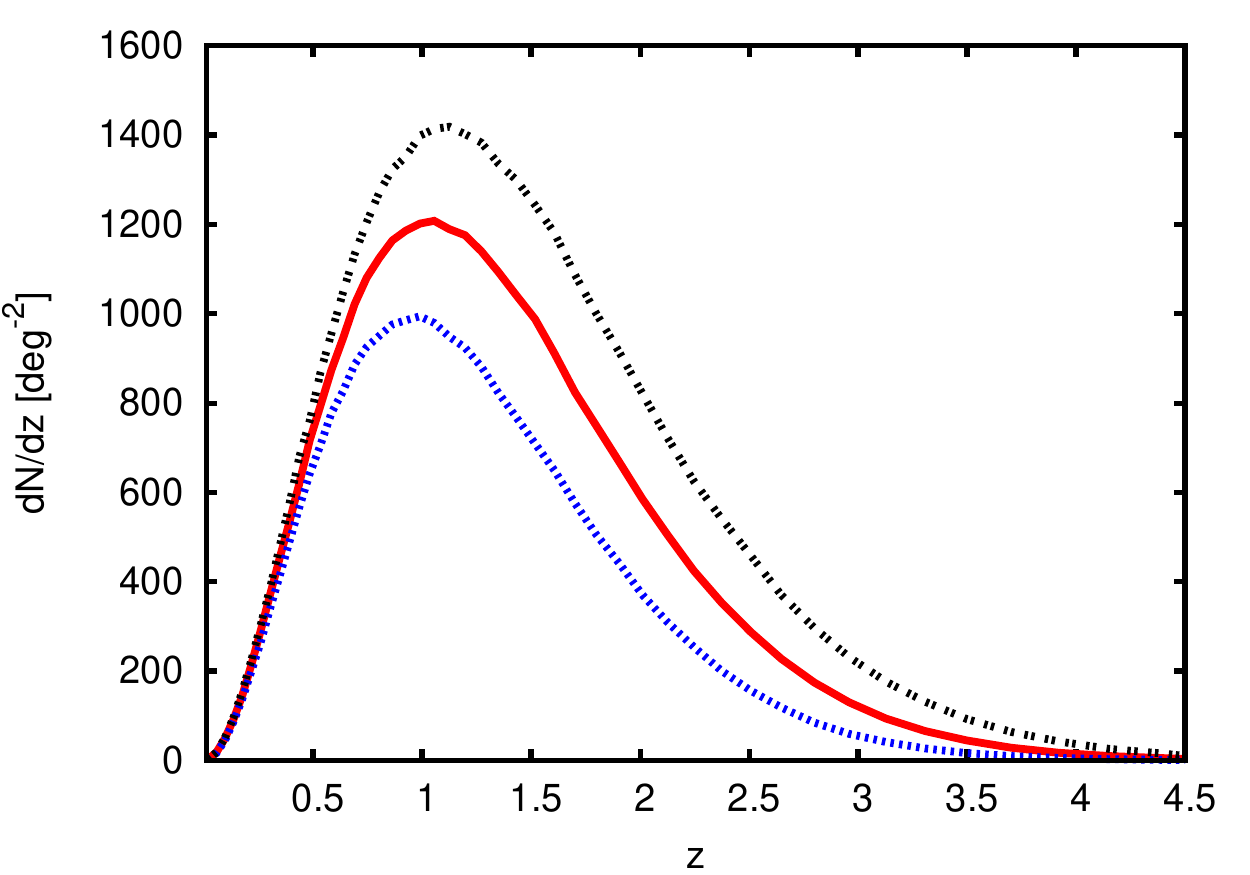}\\
\includegraphics[scale=0.68]{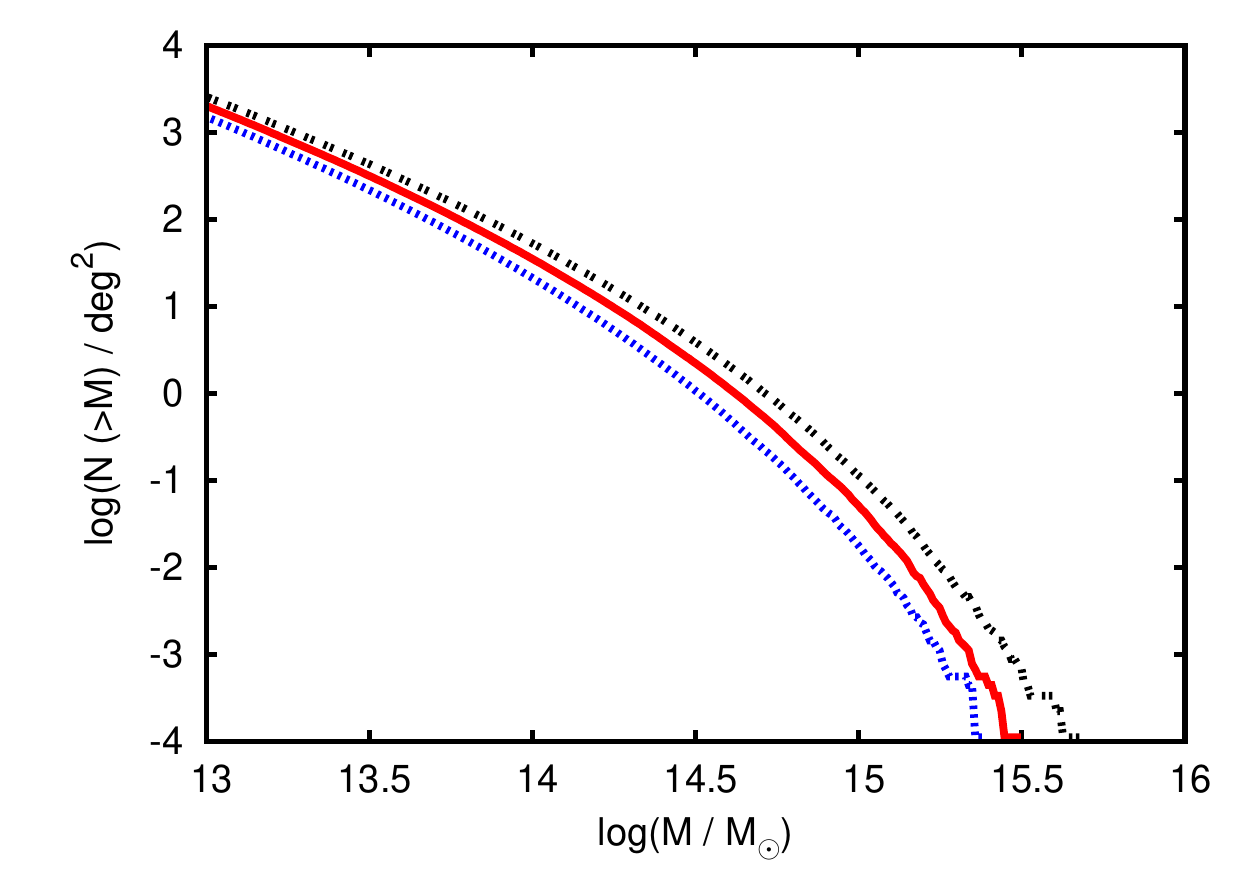}
\caption{The distribution of all clusters as a function of redshift (top) and mass (bottom) within the modified lightcone catalogues above $10^{13} M_\odot$, for each of the three values of $\sigma_8$ (from the top, 0.9 with black dots, 0.825 with the red solid line and 0.7 with blue large dots). Values from ST99 would overlay the points plotted; they are essentially the same (by construction).}
\label{fig:pinocchio_redshift_mass}
\end{figure}

The effect of this correction is shown in Fig. \ref{fig:pinocchio_discrepancy}, which gives the differential number count of clusters greater than a given mass before and after the correction has been applied. The correction brings the {\sc Pinocchio} spectrum into agreement with that generated from a set of 1000 cluster catalogues with mass and redshift values drawn from the ST99 mass function and randomly determined positions. The average mass and redshift distributions of the corrected catalogues for all three cosmologies are given in Fig. \ref{fig:pinocchio_redshift_mass}.

\subsection{Cluster model}
Although clusters are complicated in the optical regime, for the purpose of modeling the SZ effect they can be represented very simply by assuming that the clusters are relaxed, virialised objects. Although clusters undergo major mergers, these are relatively infrequent, occurring in around 10 to 20 percent of the population depending on the redshift \citep{2007Kay}.

The cluster model we utilize is spherically symmetric and consists of two components: the dark matter and the gas. The former is used to calculate the surface mass density of the cluster, used for the inclusion of point sources, and the latter gives the thermal SZ effect. A kinetic SZ effect also exists, which is created by the relative motion of the clusters to the CMB, but this is likely to be negligible compared to the thermal SZ effect at the frequencies discussed in this paper and hence is not considered here.

The virial radius, $r_{\mathrm{vir}}$, is used to define the size of a cluster. This can be calculated by using the spherical infall model \citep{1972Gunn}, which gives
\begin{equation} \label{eq:rvir}
r_{\mathrm{vir}} = \left( \frac{3 M_{\mathrm{vir}}}{4 \pi \Delta_{\mathrm{c}}(z) \rho_{\mathrm{crit}}(z)} \right) ^{1/3},
\end{equation}
where $M_{\mathrm{vir}}$ is the virial mass of the cluster, $\rho_{\mathrm{crit}}(z) = 3H^2(z) / \left(8 \pi G \right)$ is the critical density for the Universe to be flat and $H(z)$ is the Hubble parameter, given by the Friedmann equation
\begin{equation} \label{eq:hubble}
	E^{2}(z) = \frac{H^2(z)}{H_0^2} = \Omega_{\mathrm{m}} (1 + z)^3 + \Omega_{\mathrm{\Lambda}}
\end{equation}
for a flat universe (which we will assume throughout). The quantity $\Delta_{\mathrm{c}}$ is the mean matter density within the virial radius of the cluster  in units of the critical density,
\begin{equation}
\Delta_{\mathrm{c}}(z) = \frac{\rho_{\mathrm{cluster}} (z)}{\rho_{\mathrm{crit}} (z)}.
\end{equation}
The value for $\Delta_{\mathrm{c}}$ depends on the cosmological parameters; in an Einstein-de Sitter universe it is exactly $18 \pi^2 \approx 178$. For our adopted cosmology, we use the fit given by \citet{1998Bryan},
\begin{equation}
\Delta_\mathrm{c} = 18 \pi^2 + 82 (\Omega_m(z) - 1) - 39 (\Omega_m(z) - 1)^2,
\end{equation}
where $\Omega_m(z) = \Omega_m (1 + z)^3 / E^2(z)$ is the matter density at redshift $z$.

Finally, the cluster radius needs to be converted to an angular size so that the cluster can be projected on to a virtual sky map. This is done by dividing the virial radius by the angular diameter distance, $\theta_{\mathrm{vir}} = r_{\mathrm{vir}} / d_{\mathrm{A}}(z)$ in which
\begin{equation}
d_{\mathrm{A}}(z) = \frac{c \int_{0}^{z} H^{-1}(z^\prime) dz^\prime}{(1 + z)}.
\end{equation}

\subsubsection{Dark matter}
We calculate the surface mass density of each galaxy cluster for the purpose of distributing point sources on to the map (see Section \ref{sec:ps}). For the dark matter density profile, we use the NFW profile,
\begin{equation}
\rho_\mathrm{DM} \left( \frac{r}{r_\mathrm{s}} \right) = \frac{\rho_{\mathrm{crit}} \Delta_{\mathrm{c}}}{ \frac{r}{r_\mathrm{s}} \left( 1+ \frac{r}{r_\mathrm{s}} \right)^2}.
\end{equation}
This was found by fitting profiles to dark matter haloes from N-body/hydrodynamical simulations run by \citet{1995Navarro, 1996Navarro,1997Navarro}, and is now the standard profile for modeling dark matter haloes. The scale radius for the NFW profile is $r_{\mathrm{s}} = r_{\mathrm{vir}} / c_\mathrm{DM}$, where $c_\mathrm{DM}$ is the concentration parameter. Values for the concentration parameter depend weakly on the cluster mass with a certain amount of scatter \citep[see for example][]{2007Neto,2008Duffy}. However, as we are only using the surface mass density maps indirectly these effects are unimportant here, so we assume a constant value, $c_\mathrm{DM}=5$, which is appropriate for clusters.

The surface mass density at distance $\varphi_s = \theta / \theta_{\mathrm{s}}$ from the centre of the cluster, where $\theta_{\mathrm{s}} = r_{\mathrm{s}} / d_{\mathrm{A}}$ and $\theta$ is the angular distance from any given position in the sky to the centre of the cluster, is given by
\begin{equation}
\Sigma (\varphi_s) = \Sigma_{0} \zeta_{\mathrm{DM}} (\varphi_s).
\end{equation}
In this, the central surface mass density of a cluster can be calculated using (see for example \citealp{2001Lokas})
\begin{equation}
\Sigma_{0} = \frac{M_{\mathrm{vir}}}{2 \pi r_{s}^{2} (\ln(1 + c) - (c / (1 + c)))},
\end{equation}
and the NFW profile can be projected on to a 2D sky to give \citep{1996Bartelmann}
\begin{equation}
\zeta_{\mathrm{DM}} = \frac{2}{{\varphi_s^2 - 1}}\left\{ {\begin{array}{*{20}c}
  {1 - \frac{2}{{\sqrt {\varphi_s^2 - 1} }}\arctan \left( {\sqrt {\frac{{\varphi_s - 1}}{{\varphi_s + 1}}} } \right)} & {\varphi_s > 1}, \\
  {1 - \frac{2}{{\sqrt {1 - \varphi_s^2 } }}\arctan \left( {\sqrt {\frac{{1 - \varphi_s}}{{1 + \varphi_s}}} } \right)} & {\varphi_s < 1}, \\
  1 & {\varphi_s = 1}. \\
\end{array}} \right.
\end{equation}

\subsubsection{SZ effect}
Our gas model consists of two parts: a profile, and a normalization for the level of the SZ effect, which is dependent on the mass and redshift of the cluster. The method is analogous to that used for creating maps of the dark matter distribution.

We use the isothermal $\beta$-model \citep{1976Cavaliere} for the cluster profile,
\begin{equation} \label{eq:betamodel}
\xi_\mathrm{gas} \left(\frac{r}{r_\mathrm{c}} \right) = \left( 1+ \frac{r^2}{r_\mathrm{c}^2} \right)^{-\frac{3\beta}{2}},
\end{equation}
in which $\beta$ is a dimensionless parameter that measures the outer slope of the profile, $r_{\mathrm{c}} = r_{\mathrm{vir}} / c_\mathrm{gas}$ is the core radius that describes the turn-over point between the core and the power-law slope. The concentration of the SZ effect around the cluster centre is controlled by $c_\mathrm{gas}$. Fiducial values of $c_\mathrm{gas}=10$ and $\beta = 2/3$ are used, following \citet{2003Battye}. We truncate the profile at the virial radius, $r_{\mathrm{vir}}$, which prevents potential divergence in the gas mass \citep[see for example][]{1999Birkinshaw}. This may however underestimate the total SZ effect from an individual cluster, where infalling gas is shocked outside of the virial radius and prevented from falling inwards (see for example \citealp{1990Evrard,2005Kocsis}).

The change in the temperature of the CMB from the SZ effect can be calculated using
\begin{equation} \label{eq:conv_y_t}
\Delta T(\varphi_c) = y(\varphi_c) g(x) T_{\mathrm{CMB}}
\end{equation}
where $g(x) = (x / \tanh(x / 2)) - 4$, the dimensionless frequency $x = h_\mathrm{P} \nu / k_{\mathrm{B}}T_{\mathrm{CMB}}$, $k_{\mathrm{B}}$ is Boltzmann's constant, $h_\mathrm{P}$ is Planck's constant and $\nu$ is the frequency of interest. The present-day temperature of the CMB as measured by the FIRAS instrument on {\it COBE} is $T_{\mathrm{CMB}} = 2.728$K \citep{1996Fixsen}. The SZ effect along the line of sight from a single cluster depends on the integrated SZ effect from the cluster, $Y(M,z)$, and the cluster profile $\zeta(\varphi_c)$ as
\begin{equation} \label{eq:y_theta}
y(\varphi_c) = \frac{Y \zeta(\varphi_c) }{2 \pi \int_0^{\varphi_\mathrm{vir}} \zeta(\varphi_c^\prime ) \varphi_c^\prime d\varphi_c^\prime }
\end{equation}
where $\varphi_c = \theta / \theta_\mathrm{c}$ is the distance from the centre of the cluster in units of $\theta_{\mathrm{c}} = r_{\mathrm{c}} / d_{\mathrm{A}}$.

\begin{figure*}
\centering
\includegraphics[scale=0.26,viewport=135 145 750 750,clip]{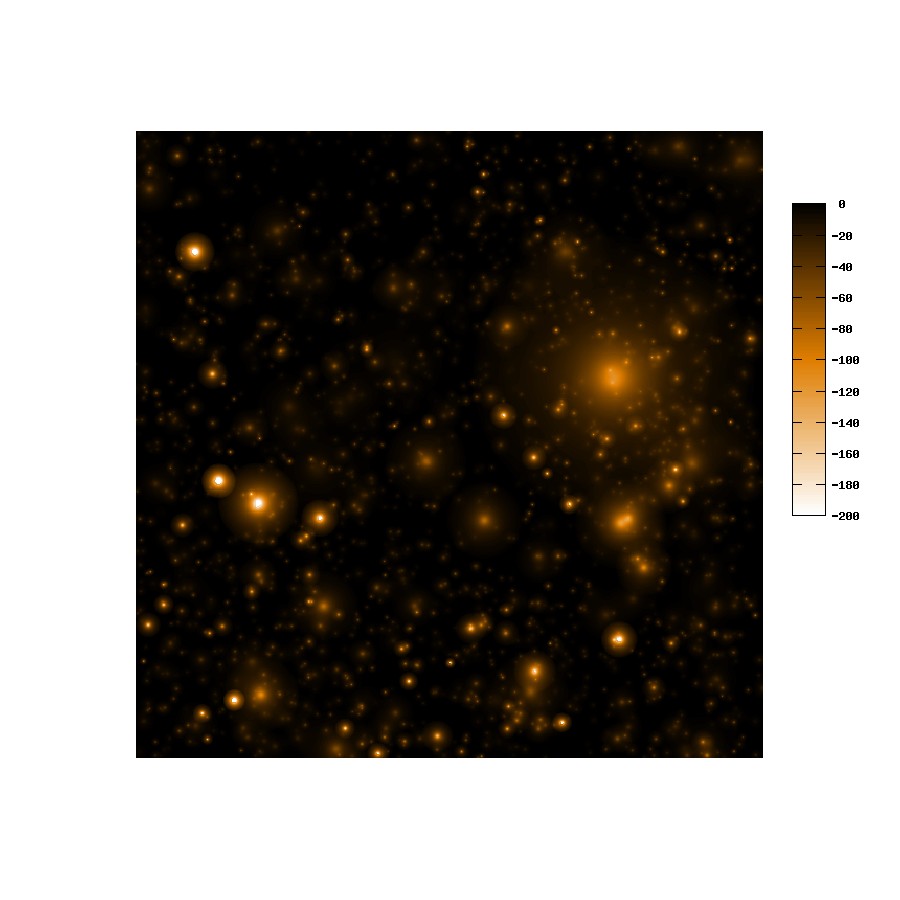}
\includegraphics[scale=0.26,viewport=135 145 750 750,clip]{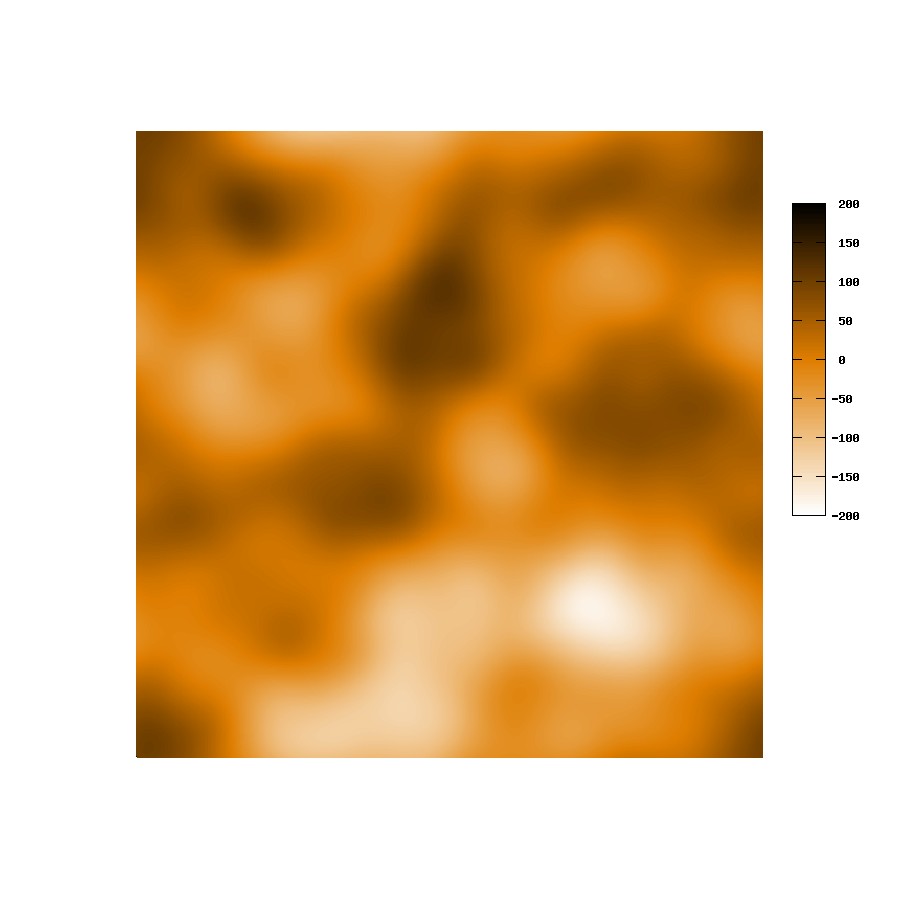}
\includegraphics[scale=0.26,viewport=135 145 750 750,clip]{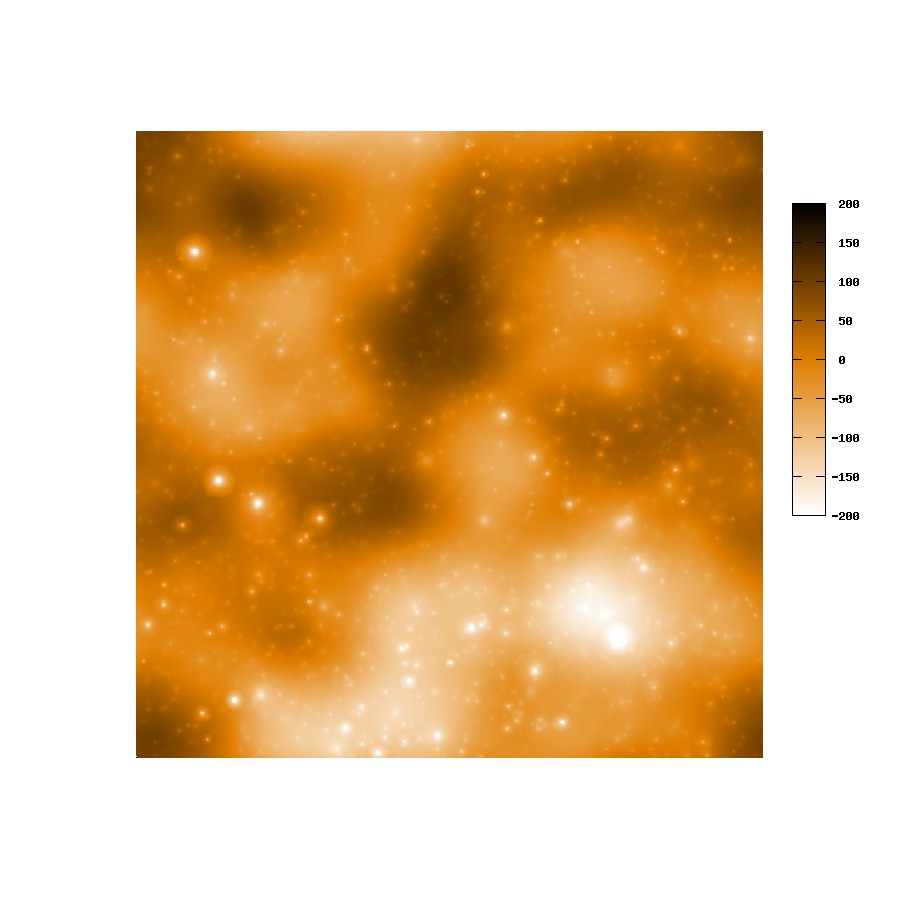}
\caption{A realization from the $\sigma_8 = 0.825$ cosmology using a clustered {\sc Pinocchio} lightcone with mass corrections. The maps are $1 \times 1$ degrees, with a resolution of 6 arcseconds. From left to right the maps are of the temperature decrement from the SZ effect, the CMB and the two combined. White represents colder areas, with black representing hotter areas. The colours range from 0 to -200 $\mu$K for the first image and 200 to -200 $\mu$K for the other two.}
\label{fig:map_set}
\end{figure*}

The integrated Y parameter is a measure of the total power of the SZ effect and is used as the normalization of our cluster profile. It is directly related to the total thermal energy of the gas. We assume a power-law relation between $Y$ and the cluster mass,
\begin{equation}
Y = \frac{Y_{*} h^{-1}}{d_A^2(z)} \left( \frac{M_\mathrm{vir}}{10^{14} h^{-1} M_\odot} \right)^\gamma \left( \frac{\Delta_c(z)}{\Delta_c(0)} E^2(z) \right)^{1/3} \left(1 + z \right)^\alpha,
\label{eq:clustermodel}
\end{equation}
where $Y_{*}, \gamma$ and $\alpha$ fix the normalisation, slope and redshift evolution respectively. For our fiducial values we adopt $Y_{*} = 2 \times 10^{-6}$ Mpc$^{2}$; the isothermal value of $\gamma = 5/3$ and $\alpha = 0$ such that the clusters are approximately self-similar. These choices are motivated by the results from recent cosmological simulations, which demonstrate that the relationship between $Y$ and $M$ is a power-law with small intrinsic scatter, being close to that predicted by the self-similar model at all redshifts, and relatively insensitive to the effects of cooling and heating of the intracluster medium (see, for example, \citealp{2004Silva}; \citealp{2005Motl}; \citealp{2006Nagai}).

Using the recent {\it Millennium Gas} simulations, Kay et al. (in prep.) obtain best-fitting values of $Y_*=2.3\times 10^{-6}$ Mpc$^2$ and $\gamma=1.64$ for a non-radiative simulation at $z=0$, and $Y_*=1.9\times 10^{-6}$ Mpc$^2$ and $\gamma=1.76$ for a simulation where the gas was preheated and allowed to cool radiatively. Both fits were applied to clusters with $M_{\rm vir}>10^{14}h^{-1}{\rm M}_{\odot}$, the objects that dominate the power spectrum over the range of multipoles of interest in this paper. These values agree well with the fiducial values above.

Finally, the $\beta$ profile can be projected from three dimensions to a 2D sky to become \citep{2003Battye}
\begin{equation}
\zeta_\mathrm{SZ}(\varphi_c) = \left( 1 + \varphi_c^2 \right)^{\frac{1}{2} - \frac{3\beta}{2}} \frac{J \left[\left(\frac{c^2 -\varphi_c^{2}}{1 + \varphi_c^{2}} \right)^{1/2}, \beta \right]}{J [c, \beta]},
\end{equation}
where
\begin{equation}
J[a, b] = \int_0^a \left( 1 + x^2 \right)^{-\frac{3 b}{2}} dx.
\end{equation}

\subsection{Realization properties} \label{sec:realization}
Fig. \ref{fig:map_set} shows $1 \times 1$ degree maps of the SZ effect and the CMB separately and combined for a single realization from the $\sigma_8 = 0.825$ cosmology. In the SZ map, an $M_\mathrm{vir} = 5.4 \times 10^{14} M_\odot$ galaxy cluster at a redshift of $z=0.13$ lies in the top right, with a virial radius of 14 arcminutes. The SZ map contains 1,933 galaxy groups and clusters in total, with a maximum decrement in a single pixel of $450 \mu$K and an average of $8.5 \mu$K per pixel across the map.

For the $3 \times 3$ realizations with $\sigma_8 = 0.75$ and a resolution of 18 arcseconds, there are $13470\pm300$ galaxy groups and clusters per realization, where the standard deviation is from the scatter between the realizations. These objects cause an average decrement per pixel of $4.3\pm0.5 \mu$K in each map (again with a standard deviation from the scatter). For $\sigma_8 = 0.825$ this increases to $18040\pm350$ galaxy groups and clusters and an average of  $6.5\pm1.3 \mu$K, and for $\sigma_8 = 0.9$ this further increases to 23000$\pm$400 galaxy groups and clusters and an average of {$9.3\pm 1.3 \mu$K. The increase in the number of objects, and hence the total amount of power from the SZ effect, is due to the increase in the clustering of matter as $\sigma_8$ increases.

\begin{figure}
\centering
\includegraphics[scale=0.68]{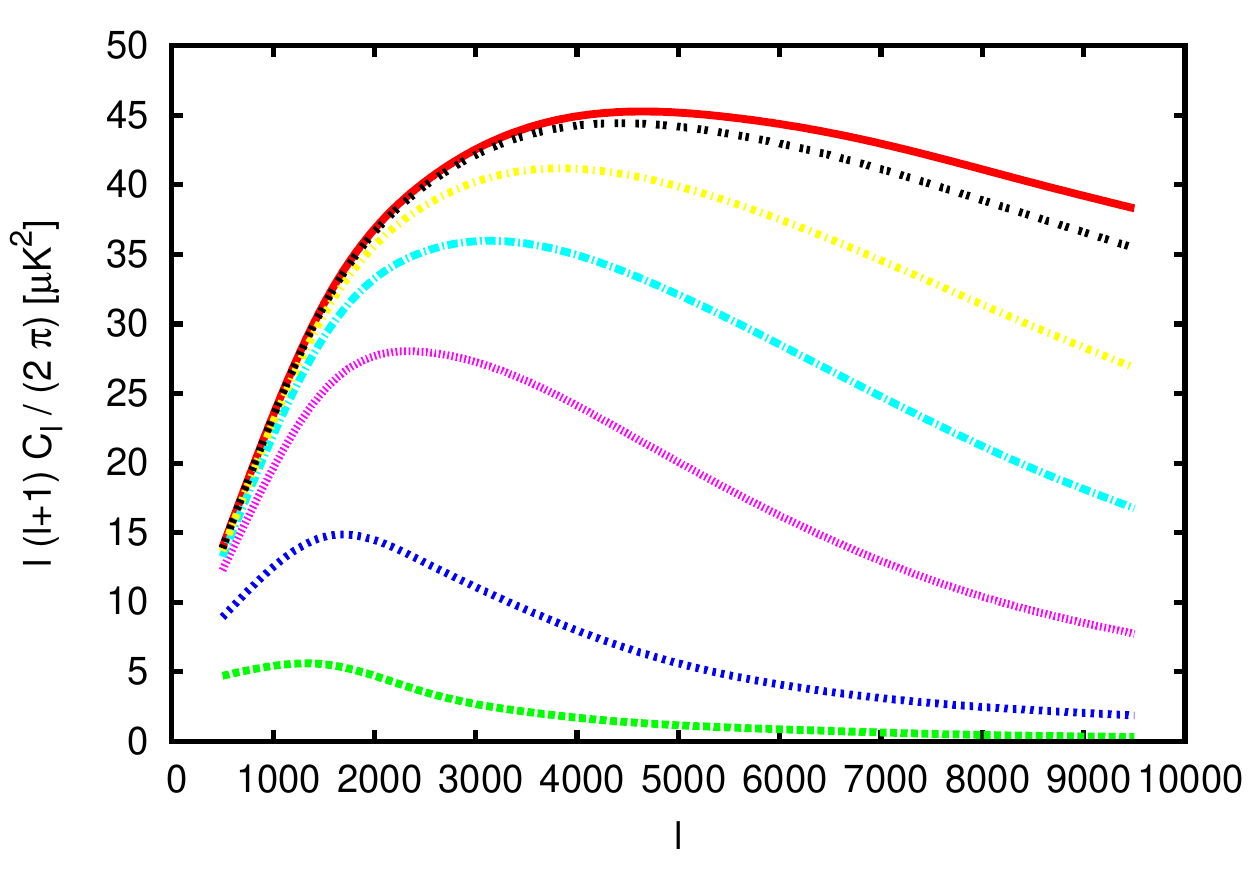}\\
\includegraphics[scale=0.68]{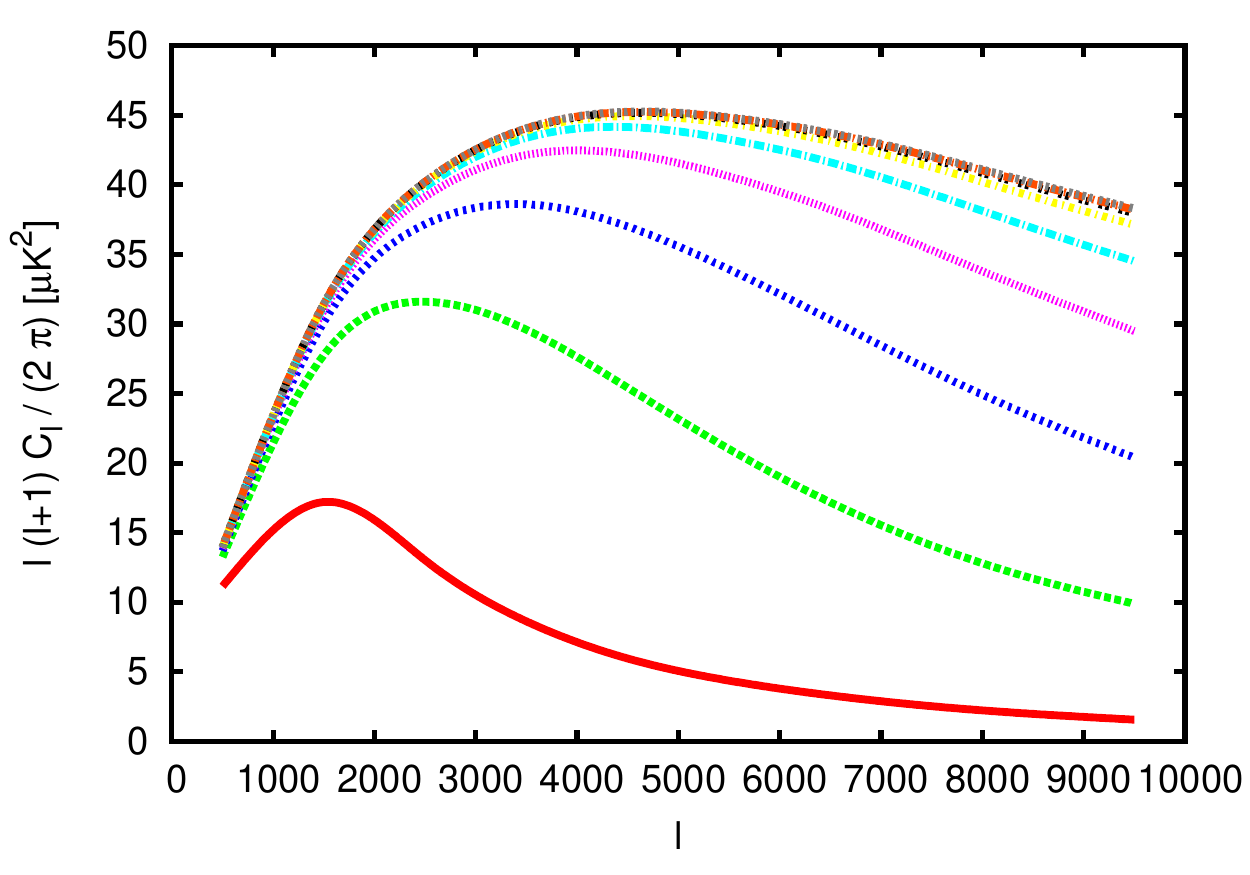}
\caption{Top panel: The mean power spectra from the SZ effect in the $\sigma_8=0.825$ cosmology imposing various minimum mass limits: from the bottom up, $1 \times 10^{15}$, $5 \times 10^{14}$, $2 \times 10^{14}$, $1 \times 10^{14}$, $5 \times 10^{13}$, $2 \times 10^{13}$ and $1 \times 10^{13} M_\odot$. Bottom panel: The same, but imposing a maximum redshift cutoff of (from the bottom up) $z=$ 0.5, 1.0, 1.5, 2.0, 2.5, 3.0, 3.5, 4.0 and 4.5. These results show that the spectrum is approximately converged for $M>10^{13} M_\odot$ and $z<3.5$.}
\label{fig:convergence_testing}
\end{figure}

To confirm that the power spectra calculated from these realizations are converged such that no significant amount of power comes from clusters with lower masses or higher redshifts, the mean spectra from realizations with various minimum mass and maximum redshift cutoffs imposed are shown in Fig. \ref{fig:convergence_testing} for $\sigma_8 = 0.825$. The spectra are converged to within a few $\mu$K$^2$ over the multipole range of interest when clusters down to $10^{13} M_\odot$ and out to a redshift $z=3.5$ are included (the effects of reducing the {\it maximum} mass cutoff will be discussed in Section \ref{sec:upper_mass}). We use $10^{13} M_\odot$ and $z=4.5$ as the fiducial limits for the realizations shown in the rest of this paper. Although the average SZ effect across the map is converged by the same redshift, it is not converged at a minimum mass of $10^{13} M_\odot$; lower mass objects than considered here will significantly contribute towards the average SZ effect.

As an additional check, we compare the ratio of the mean power spectra between realizations with different values of $\sigma_8$. \citet{2002Komatsu} found that $C_l$ scales as $\sigma_8^\alpha$, where $\alpha \approx 7$. Fitting for $\alpha$ for our realizations and averaging over the multipole bins, we find that between the $\sigma_8 = 0.75$ and $0.825$ realizations, $\alpha \approx 7.1$. Between $\sigma_8 = 0.90$ and $0.825$ the value is $\alpha \approx 6.9$. Thus there is fairly good agreement, although the relationship appears not to be a perfect power law over this range of $\sigma_8$.

\begin{figure}
\centering
\includegraphics[scale=0.68]{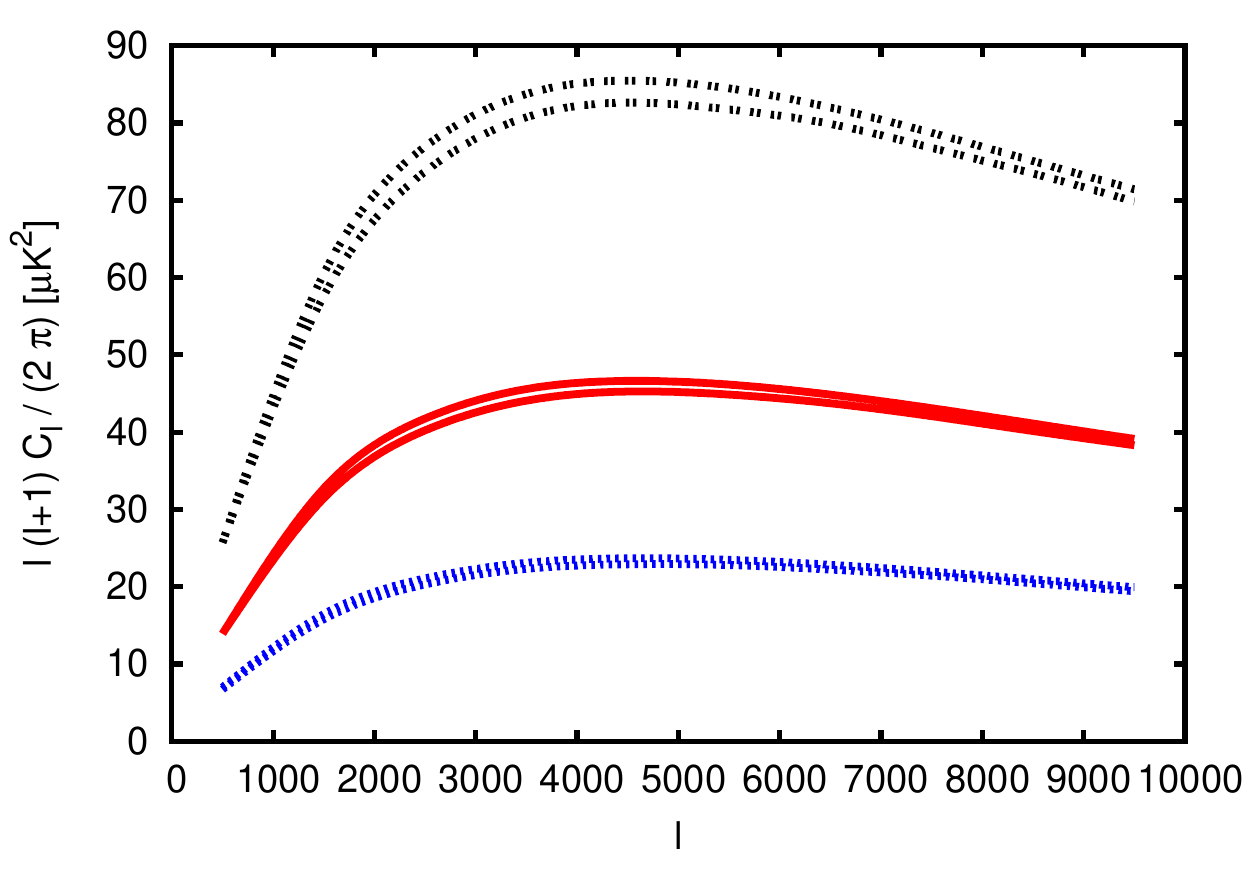}
\caption{The power spectra expected from the SZ effect as calculated using the analytical formula and the modified {\sc Pinocchio} realizations for each cosmology. The blue dashed line shows $\sigma_8 = 0.75$, the red solid line $0.825$ and the black dotted line $0.9$. The higher values of the pairs are from the theoretical spectrum, the lower from the realizations; there is consistency between the shapes of the curves and the analytical formula is $2-4$ per cent higher than the realizations.}
\label{fig:comp_theory_mean}
\end{figure}

The mean power spectrum can also be evaluated directly from the number density and cluster profile using the halo formalism \citep[see for example][]{2002Komatsu}. The details of this calculation are outlined in the Appendix, and Fig. \ref{fig:comp_theory_mean} compares the values computed using this method with those from the realizations for the three different cosmologies. The analytical method predicts $\sim2-4$ per cent more power than the realizations, which is likely due to inefficiencies in the creation of the maps, but it appears that there is broad consistency here.

We have made our maps of the SZ effect with $\sigma_8=0.825$ available online in FITS format.\footnote{\url{http://www.jodrellbank.manchester.ac.uk/~mpeel/szeffect/}}

\section{Power spectrum statistics}
Using the virtual sky maps described in Section 2, the binned power spectra from the SZ effect are calculated using bins of $\Delta l = 1000$. We analyse the histograms of these power spectra and investigate the normalized skew as well as the mean and standard deviation. Higher orders than the skewness, such as the kurtosis, tend to be significantly affected by outlying data points and hence are not considered here.

Similar work investigating the statistics of SZ effect realizations has been carried out by \citet{2002Zhang} and \citet{2009Shaw}.

\subsection{Fiducial results}
\begin{figure*}
\centering
  \includegraphics[scale=0.43,viewport=50 0 330 245,clip]{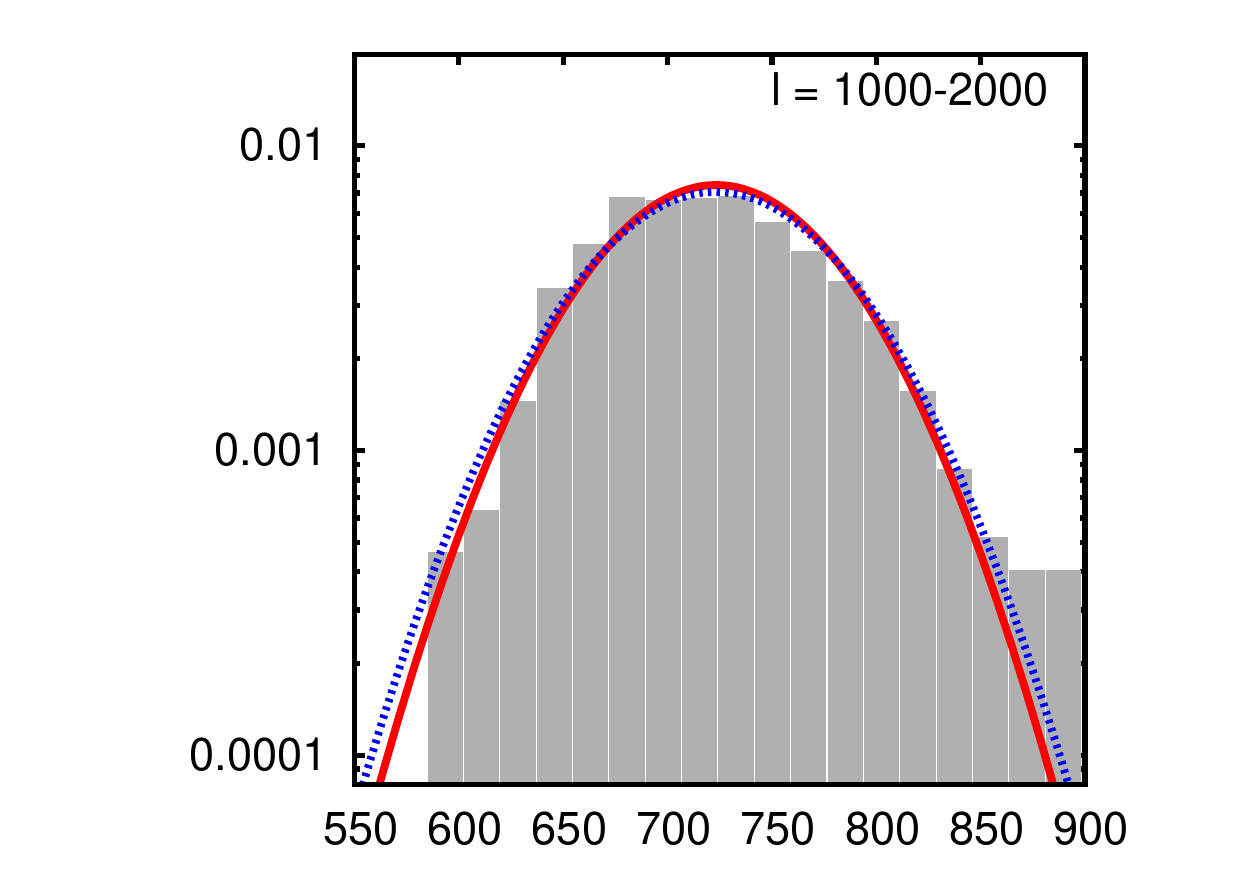}
  \includegraphics[scale=0.43,viewport=50 0 330 245,clip]{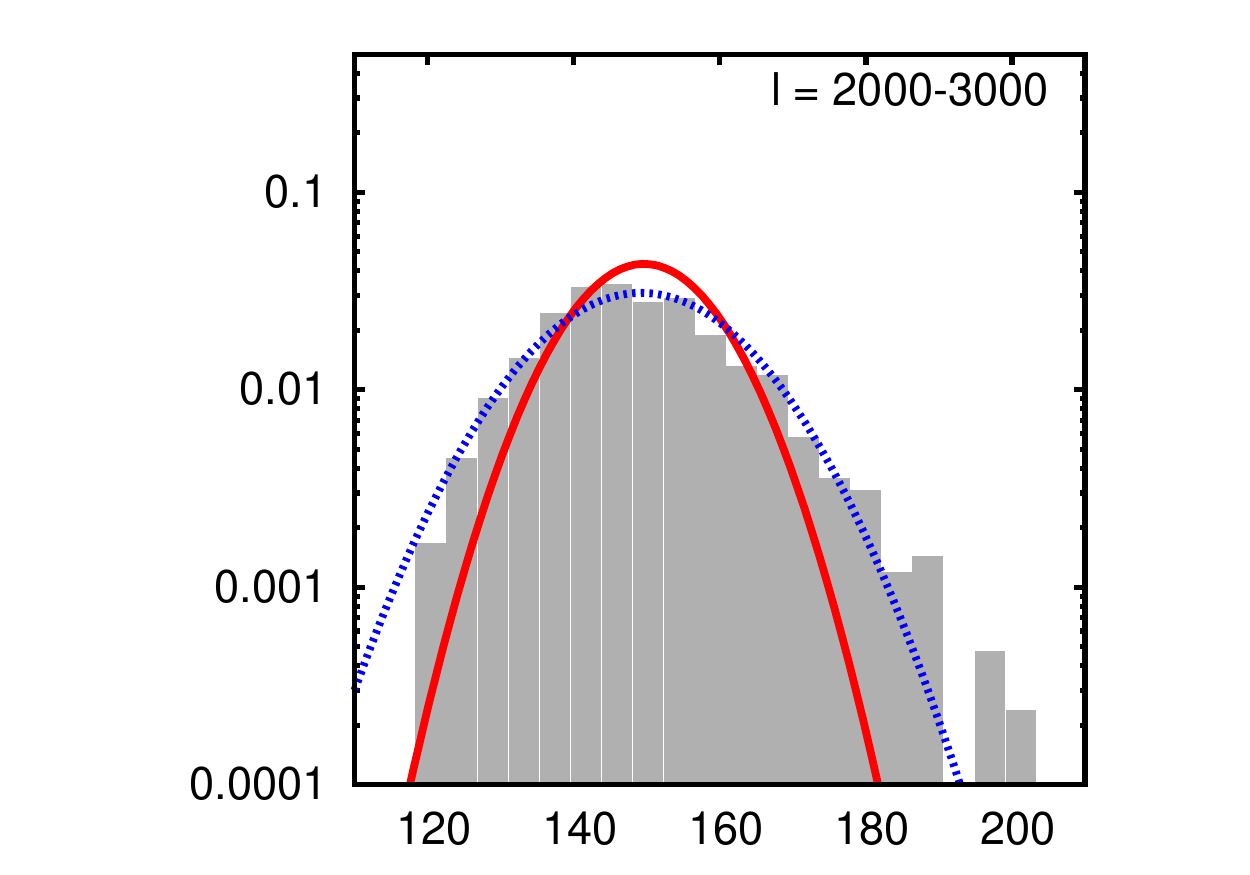}
  \includegraphics[scale=0.43,viewport=50 0 330 245,clip]{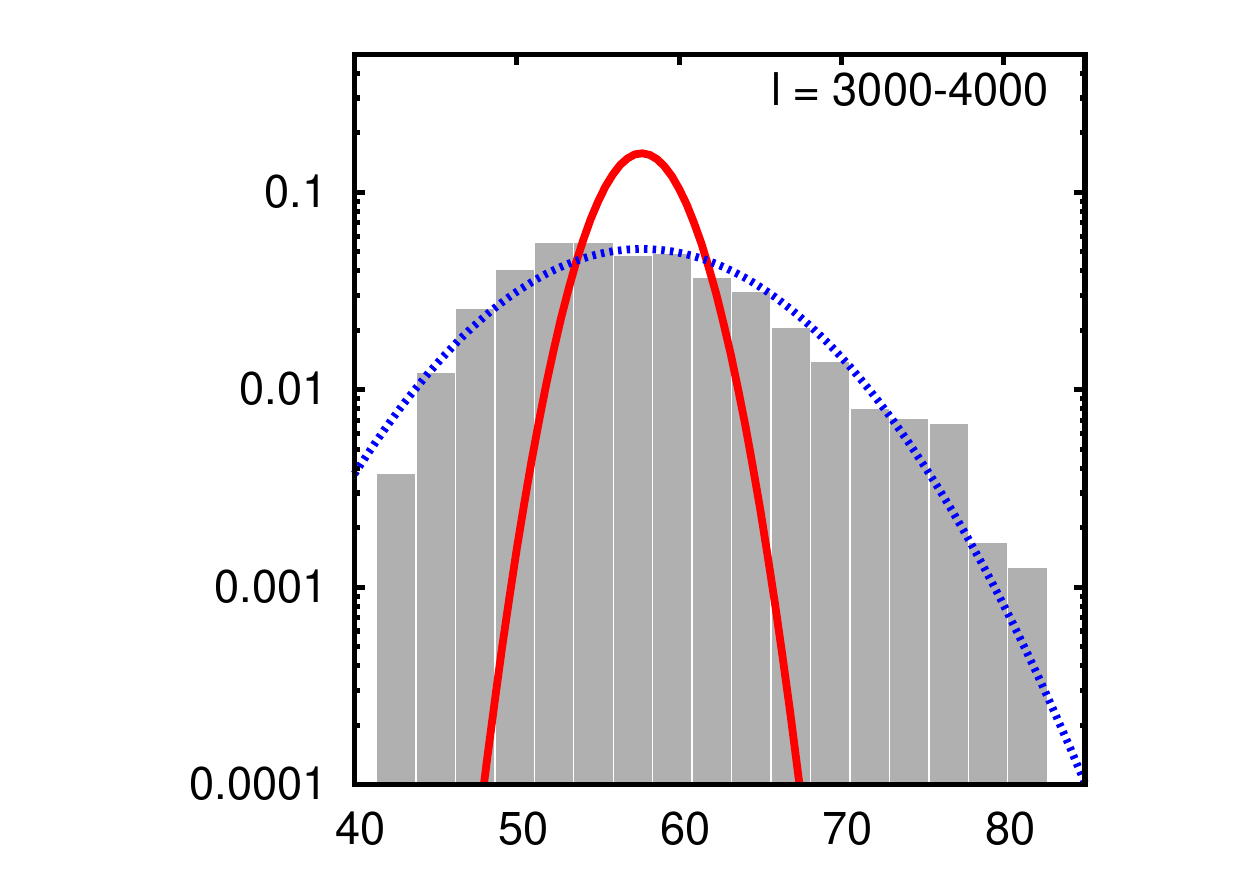}
  \includegraphics[scale=0.43,viewport=50 0 330 245,clip]{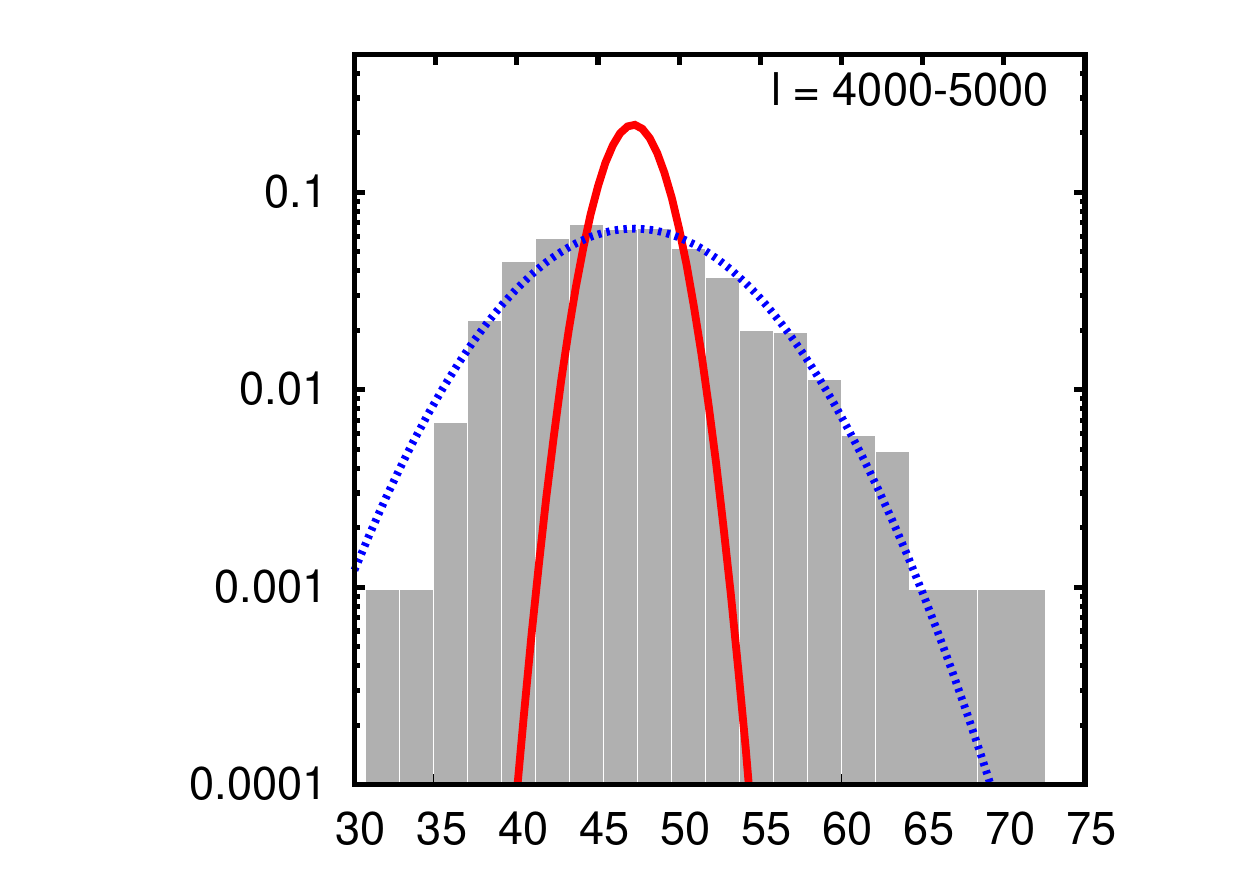}
  \\
  \includegraphics[scale=0.43,viewport=50 0 330 245,clip]{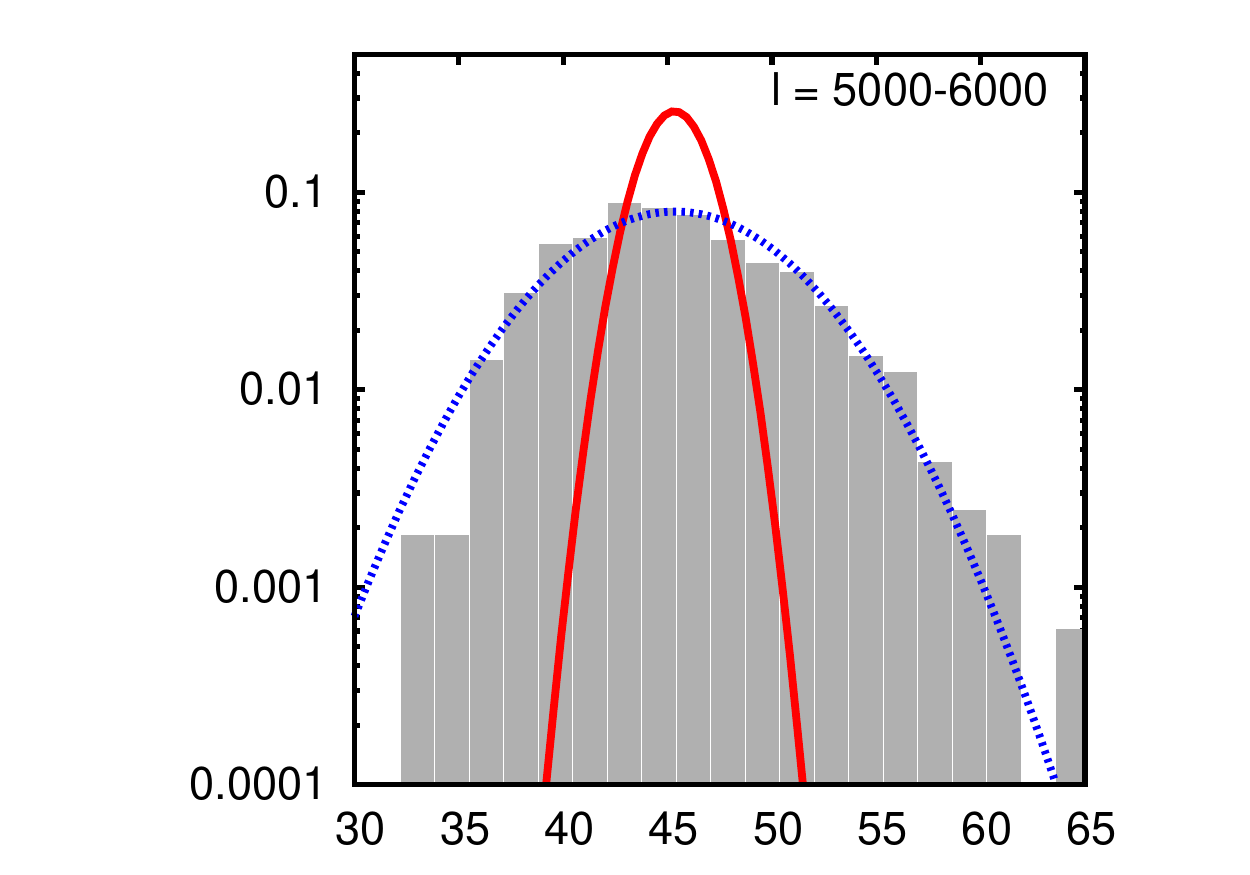}
  \includegraphics[scale=0.43,viewport=50 0 330 245,clip]{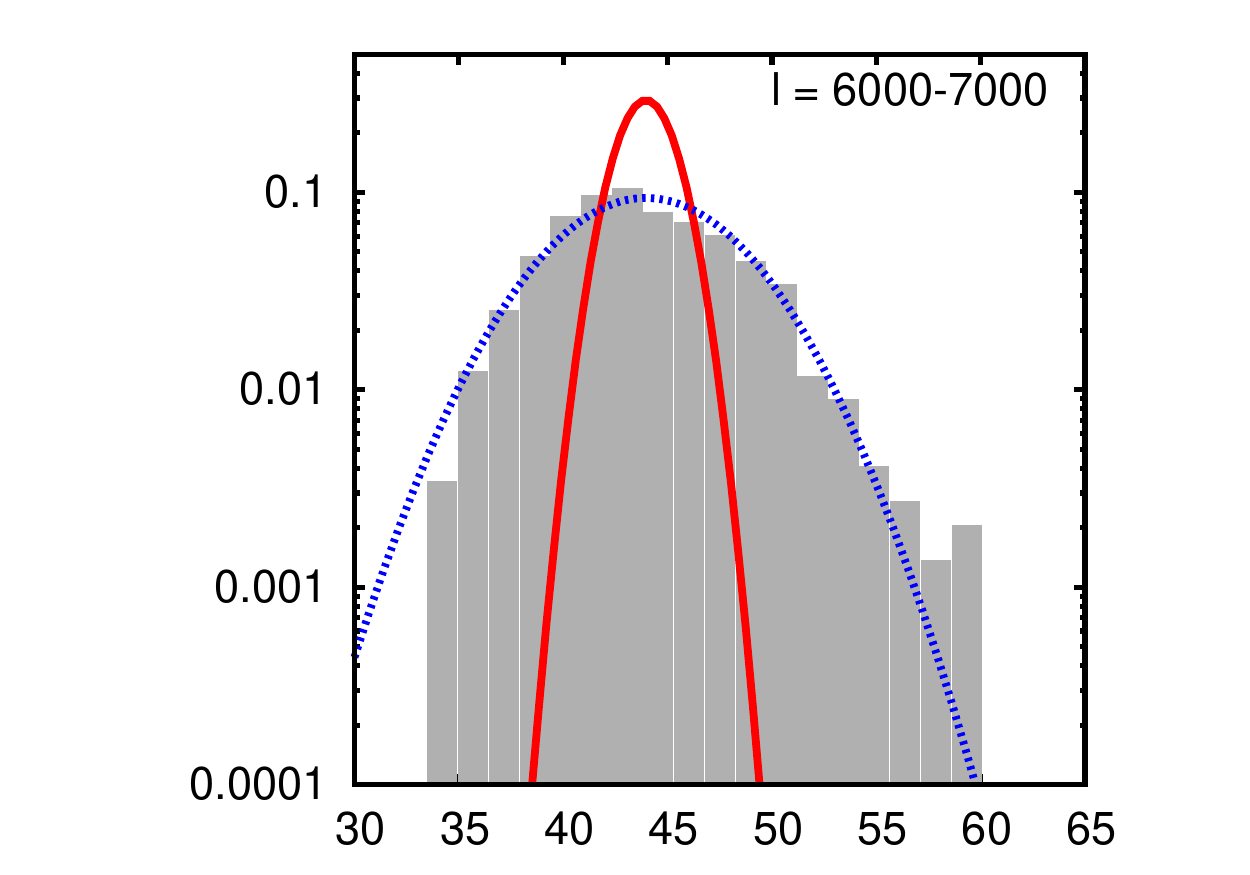}
  \includegraphics[scale=0.43,viewport=50 0 330 245,clip]{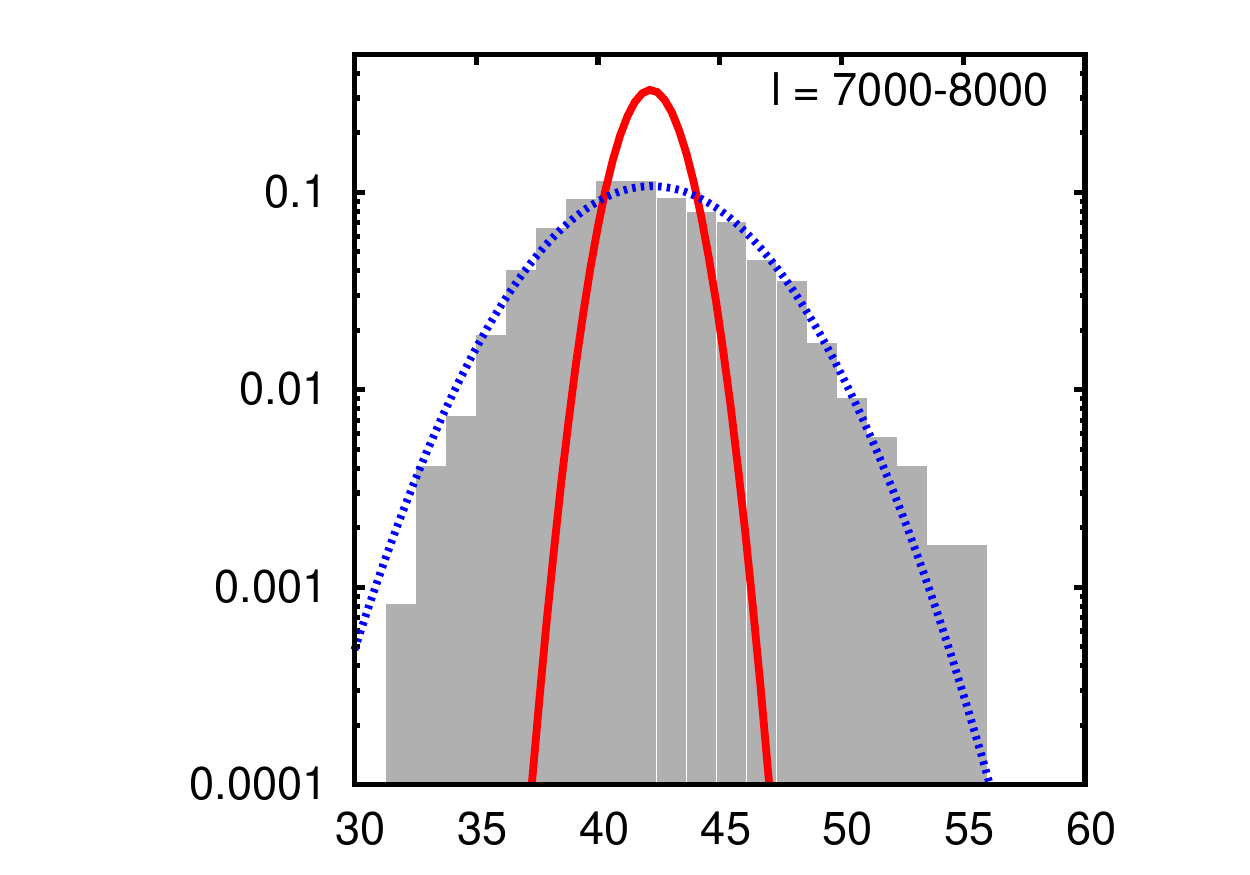}
  \includegraphics[scale=0.43,viewport=50 0 330 245,clip]{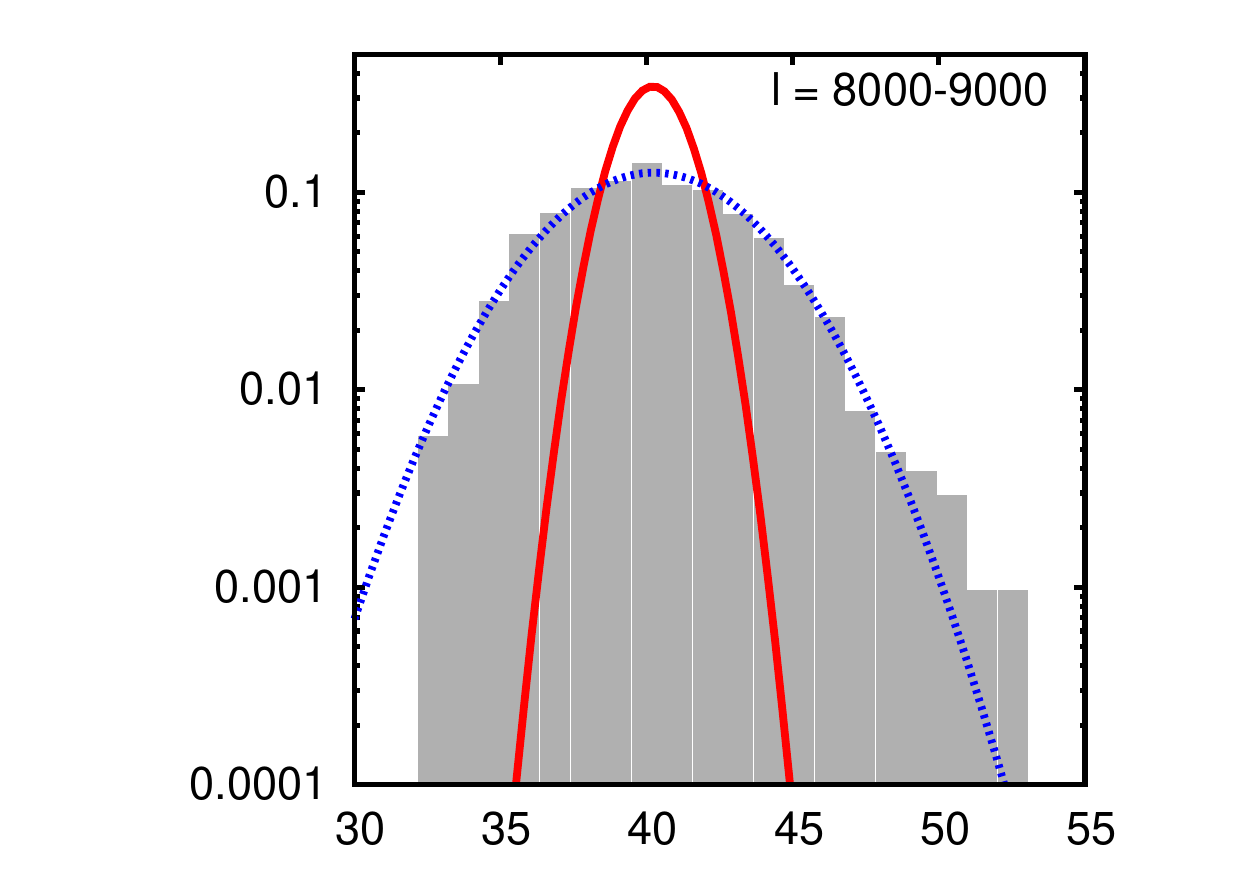}
	\caption{The statistics of the combined CMB and SZ effect power spectra for the $\sigma_8 = 0.825$ cosmology for a thousand $3\times3$ degree maps. In each histogram, the x-axis is the binned power $B_i$ in the maps in $\mu$K$^2$, and the y-axis is the probability density of that power in a map. The histograms on the top row show bins of 1000 multipoles between $l=$1000 and 4000; the second row 5000-9000. A Gaussian curve generated from the mean and standard deviation of the distribution is shown by the dotted blue line. The red solid line shows the equivalent for a set of 1000 realizations generated using the mean power spectrum from the combination of the CMB and SZ effect but assuming that all the power is Gaussianly distributed. This agrees at the lowest multipoles, but quickly becomes too narrow, indicative of the increased variance due to the SZ effect.}
	\label{fig:cmb_sz_plots_3deg}
\end{figure*}

\begin{table*}
\begin{center}\begin{tabular}{c|ccccc|ccccc|cccc}
& \multicolumn{4}{c}{$\sigma_8 = 0.75$} & & \multicolumn{4}{c}{$\sigma_8 = 0.825$} & & \multicolumn{4}{c}{$\sigma_8 = 0.9$}\\
\hline
Multipoles & $\bar{B}_i$ & $\delta B_i$ & $s_i$ & $\delta B_i / \delta B_{i,\mathrm{G}}$ & & $\bar{B}_i$ & $\delta B_i$ & $s_i$ & $\delta B_i / \delta B_{i,\mathrm{G}}$ & & $\bar{B}_i$ & $\delta B_i$ & $s_i$ & $\delta B_i / \delta B_{i,\mathrm{G}}$\\
\hline 
1000-2000 & 16 & 7.4 & 2.6 & 7.3 & & 31 & 12 & 1.8 & 6.0 & & 58 & 23 & 2.4 & 6.3\\
2000-3000 & 20 & 5.5 & 1.4 & 5.1 & & 40 & 9.3 & 0.94 & 4.4 & & 74 & 17 & 1.0 & 4.3\\
3000-4000 & 22 & 4.3 & 0.86 & 4.4 & & 44 & 7.5 & 0.72 & 4.0 & & 81 & 14 & 0.72 & 3.8\\
4000-5000 & 23 & 3.4 & 0.66 & 3.9 & & 45 & 6.0 & 0.65 & 3.5 & & 83 & 11 & 0.56 & 3.5\\
5000-6000 & 23 & 2.8 & 0.39 & 3.5 & & 45 & 5.0 & 0.49 & 3.3 & & 82 & 9.1 & 0.52 & 3.3\\
6000-7000 & 22 & 2.3 & 0.45 & 3.3 & & 44 & 4.3 & 0.50 & 3.0 & & 80 & 7.7 & 0.41 & 3.0\\
7000-8000 & 21 & 2.0 & 0.45 & 3.1 & & 42 & 3.7 & 0.40 & 3.0 & & 77 & 6.6 & 0.46 & 2.8\\
8000-9000 & 20 & 1.7 & 0.41 & 3.0 & & 40 & 3.2 & 0.40 & 2.6 & & 73 & 5.7 & 0.36 & 2.7\\
9000-10000 & 19 & 1.5 & 0.38 & 2.9 & & 38 & 2.8 & 0.41 & 2.7 & & 70 & 4.9 & 0.33 & 2.6\\
\hline 
1000-2000 & 710 & 54 & 0.37 & 1.0 & & 720 & 56 & 0.40 & 1.0 & & 750 & 62 & 0.26 & 1.1\\
2000-3000 & 130 & 9.3 & 0.24 & 1.1 & & 150 & 13 & 0.59 & 1.4 & & 180 & 19 & 0.60 & 1.9\\
3000-4000 & 36 & 4.5 & 0.77 & 2.7 & & 58 & 7.7 & 0.70 & 3.1 & & 94 & 14 & 0.72 & 3.3\\
4000-5000 & 25 & 3.4 & 0.64 & 3.6 & & 47 & 6.1 & 0.68 & 3.4 & & 85 & 11 & 0.55 & 3.4\\
5000-6000 & 23 & 2.8 & 0.39 & 4.1 & & 45 & 5.0 & 0.50 & 3.2 & & 82 & 9.1 & 0.51 & 3.2\\
6000-7000 & 22 & 2.4 & 0.45 & 3.3 & & 44 & 4.3 & 0.49 & 3.1 & & 80 & 7.7 & 0.40 & 3.1\\
7000-8000 & 22 & 2.0 & 0.44 & 3.3 & & 42 & 3.7 & 0.40 & 3.1 & & 77 & 6.6 & 0.46 & 2.9\\
8000-9000 & 21 & 1.7 & 0.40 & 2.9 & & 40 & 3.2 & 0.39 & 2.7 & & 74 & 5.7 & 0.36 & 2.6\\
9000-10000 & 19 & 1.5 & 0.37 & 2.9 & & 38 & 2.8 & 0.41 & 2.8 & & 70 & 4.9 & 0.33 & 2.7\\
\end{tabular}
\caption{The statistics of the SZ effect (top section) and combined CMB and SZ effect (bottom section), generated using the $3 \times 3$ degree clustered galaxy cluster catalogues from the modified {\sc Pinocchio} simulations. The mean ($\bar{B}_i$) and the standard deviation ($\delta B_i$) within bin $i$ are in units of $\mu$K$^2$; the skew ($s$) is dimensionless. The ratio of the standard deviation to that from the Gaussianly distributed simulations with the same power spectrum is given by $\delta B_i / \delta B_{i,\mathrm{G}}$. There is a considerable increase in the standard deviation from the SZ effect compared to Gaussian distributions, and the distribution is also skewed.}
\label{tab:sz_statistics}
\end{center}
\end{table*}

The resulting histograms for the combined $3 \times 3$ degree CMB and SZ effect maps with $\sigma_8 = 0.825$ between $l=1000$ and $9000$ are shown in Fig. \ref{fig:cmb_sz_plots_3deg}. To quantify the distributions, we compute the mean $\bar{B}_i$ for each bin $i$, as well as the standard deviation $\delta B_i$ and normalized skewness $s_i$. The latter is defined by
\begin{equation}
\mathrm{s_i} = \frac{1}{N}\sum_n \left(\frac{B_i^n - \bar{B}_i}{\sigma} \right)^3,
\end{equation}
where $B_i^n$ is the value of the binned $l(l+1) C_l / (2 \pi)$ from the nth realization, $\bar{B}_i$ is the mean of those values, N is the number of realizations and the sum is over all realizations. The computed values for all three values of $\sigma_8$, for both the power spectrum of the SZ effect on its own and that from the combined CMB and SZ effect, are given in Table \ref{tab:sz_statistics}.

\begin{figure}
\centering
\includegraphics[scale=0.68]{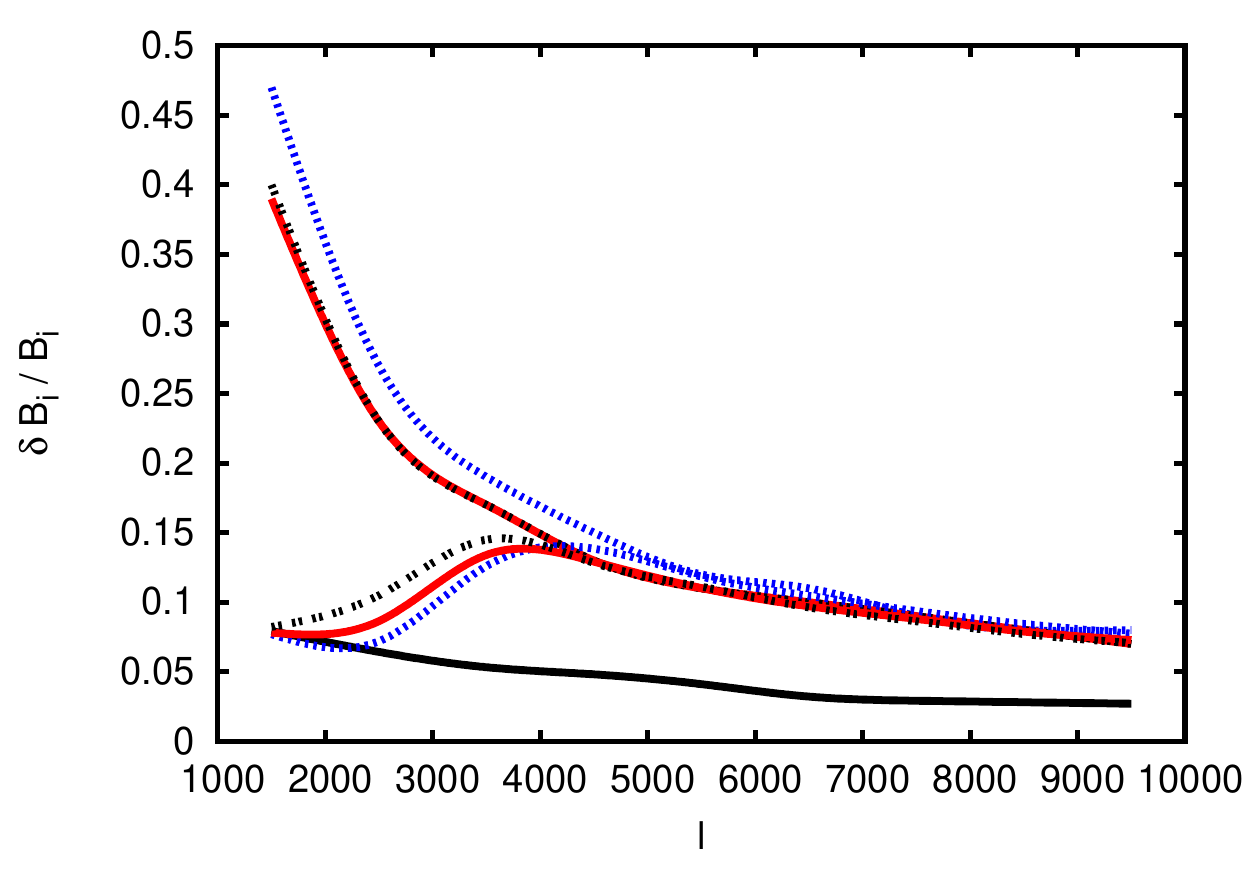}
\caption{The ratio of the standard deviation to the mean for the CMB only (solid black line), the SZ effect only (top three lines) and the combined CMB and SZ effect (central lines) for the three different values of $\sigma_8$ (0.75: blue dotted line, 0.825: red solid line, 0.9: black double-dotted line) from the $3 \times 3$ degree maps. The values for the CMB scale as $l^{-1/2}$, as per equation \ref{eq:cv}, whereas those from the SZ effect scale as $l^{-1}$ but with a higher amplitude.}
\label{fig:cmb_sz_stats}
\end{figure}

The SZ effect significantly increases the standard deviation of the realization statistics for all but the lowest multipoles. This is shown graphically in Fig. \ref{fig:cmb_sz_stats}, where the ratio of the standard deviation to the mean is given for the CMB and SZ effects separately, and then combined. The SZ effect has a much higher ratio than for the CMB on its own. When combined with the CMB, the realizations have a low standard deviation at the lowest multipoles, where the CMB is dominant, but quickly return to higher values at the higher multipoles where the CMB has decreased in power due to Silk damping and the SZ effect is closer to its peak power. The point at which the transition occurs is slightly different for the three values of $\sigma_8$; this is due to the different power levels from the SZ effect, with the $\sigma_8 = 0.9$ cosmology having more power and hence transitioning at lower multipoles.

\begin{figure*}
\centering
  \includegraphics[scale=0.43,viewport=50 0 330 245,clip]{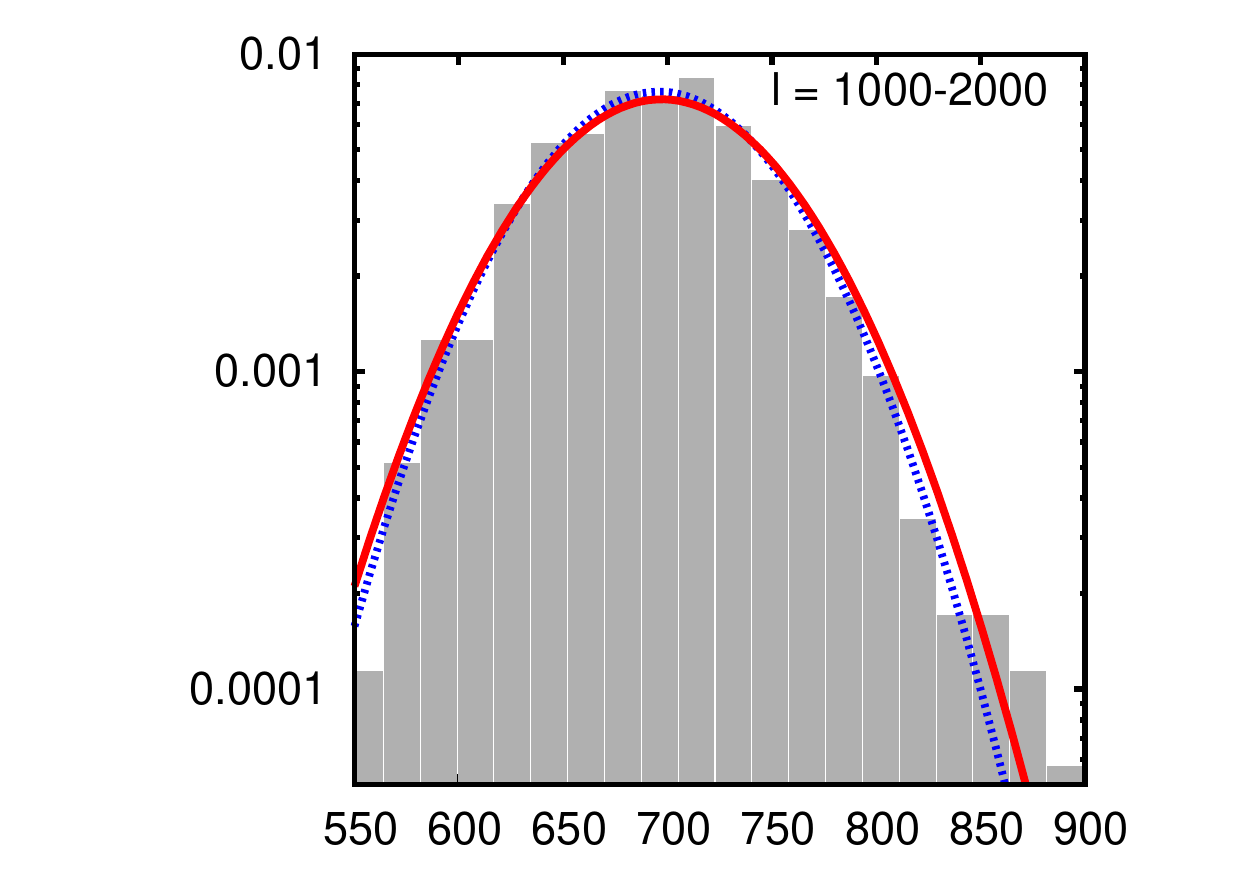}
  \includegraphics[scale=0.43,viewport=50 0 330 245,clip]{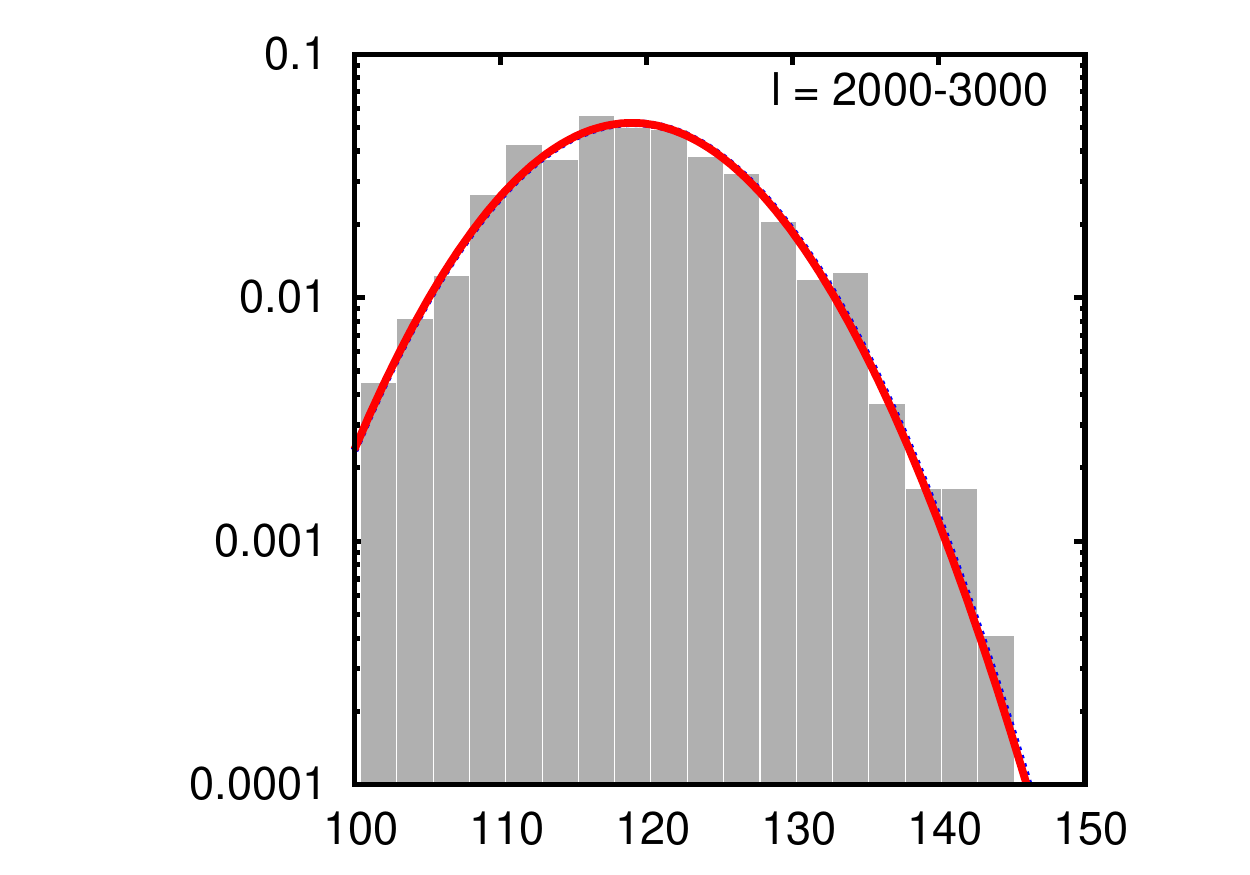}
  \includegraphics[scale=0.43,viewport=50 0 330 245,clip]{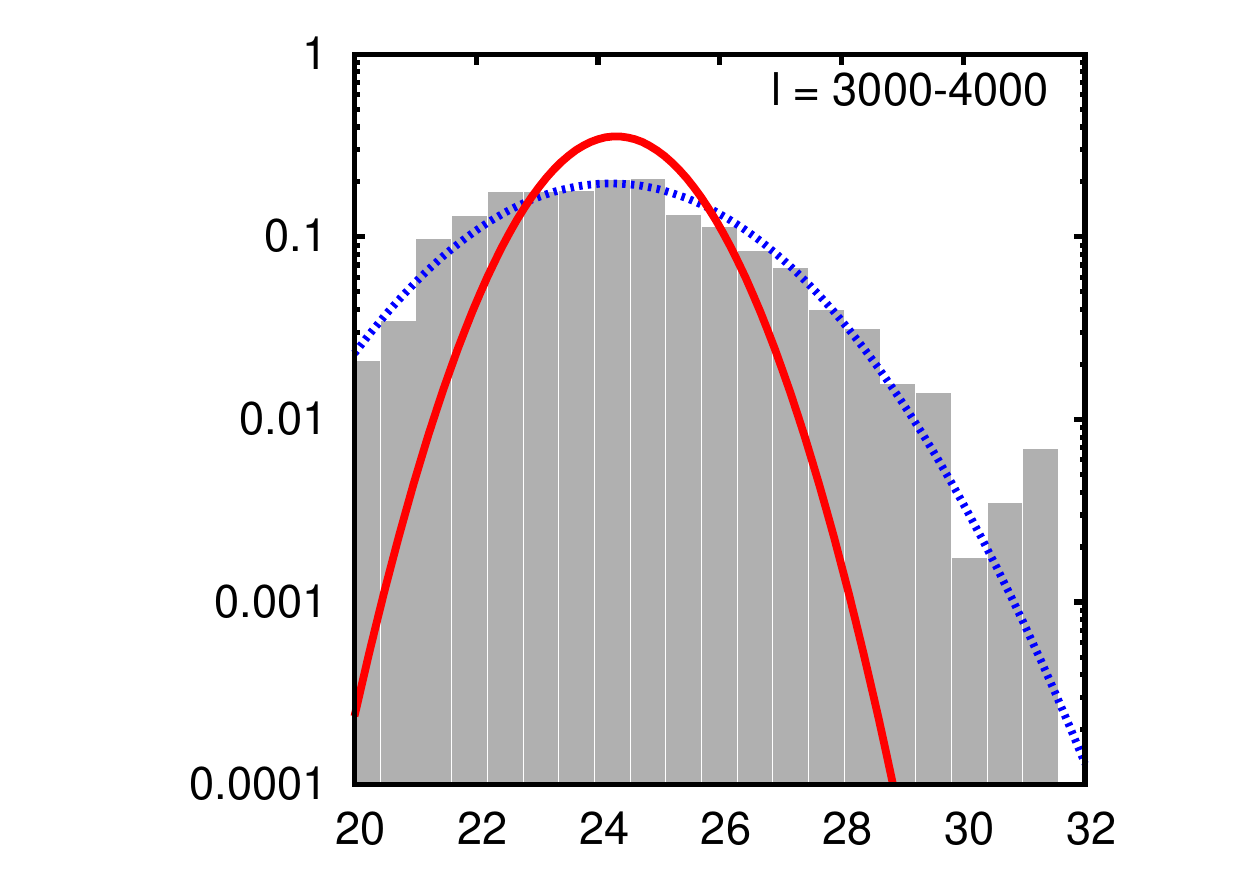}
  \includegraphics[scale=0.43,viewport=50 0 330 245,clip]{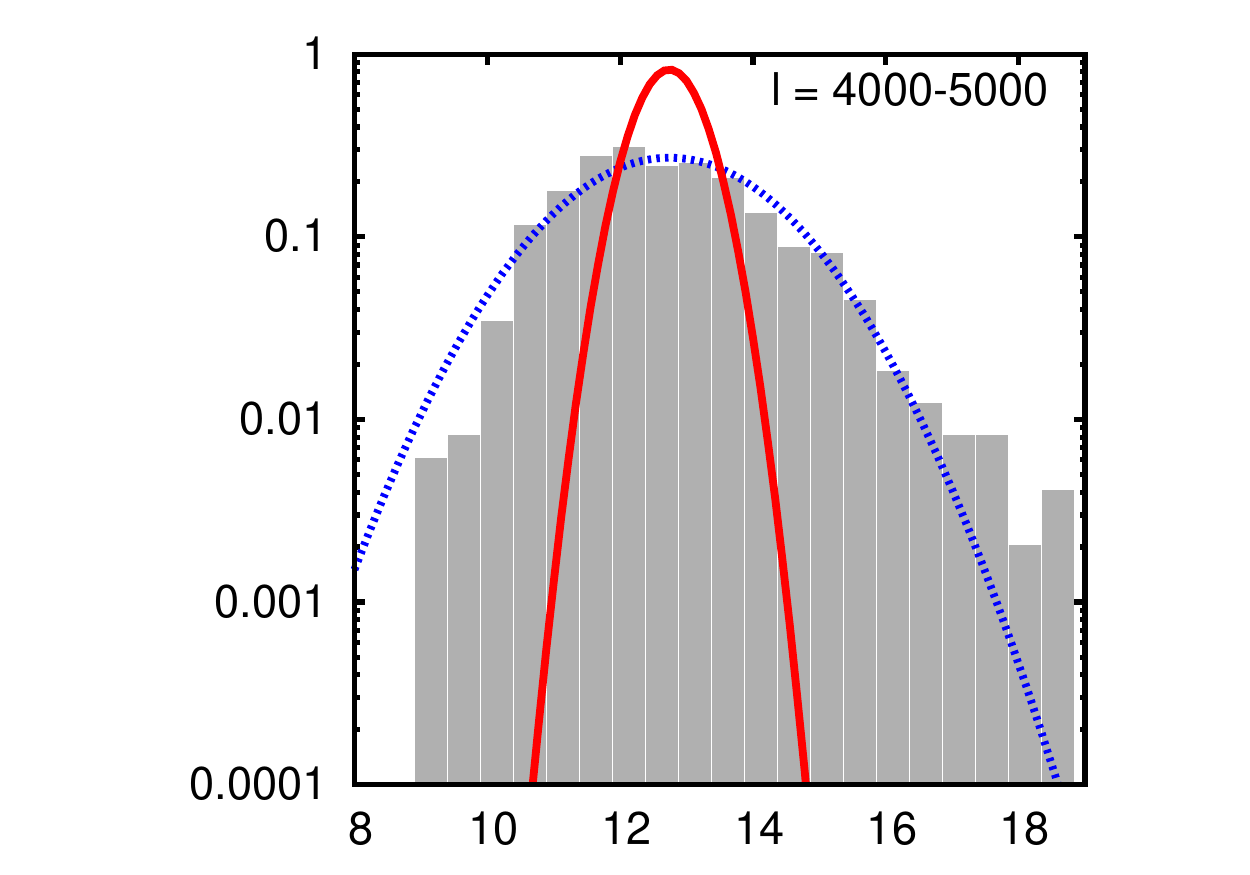}
  \\
  \includegraphics[scale=0.43,viewport=50 0 330 245,clip]{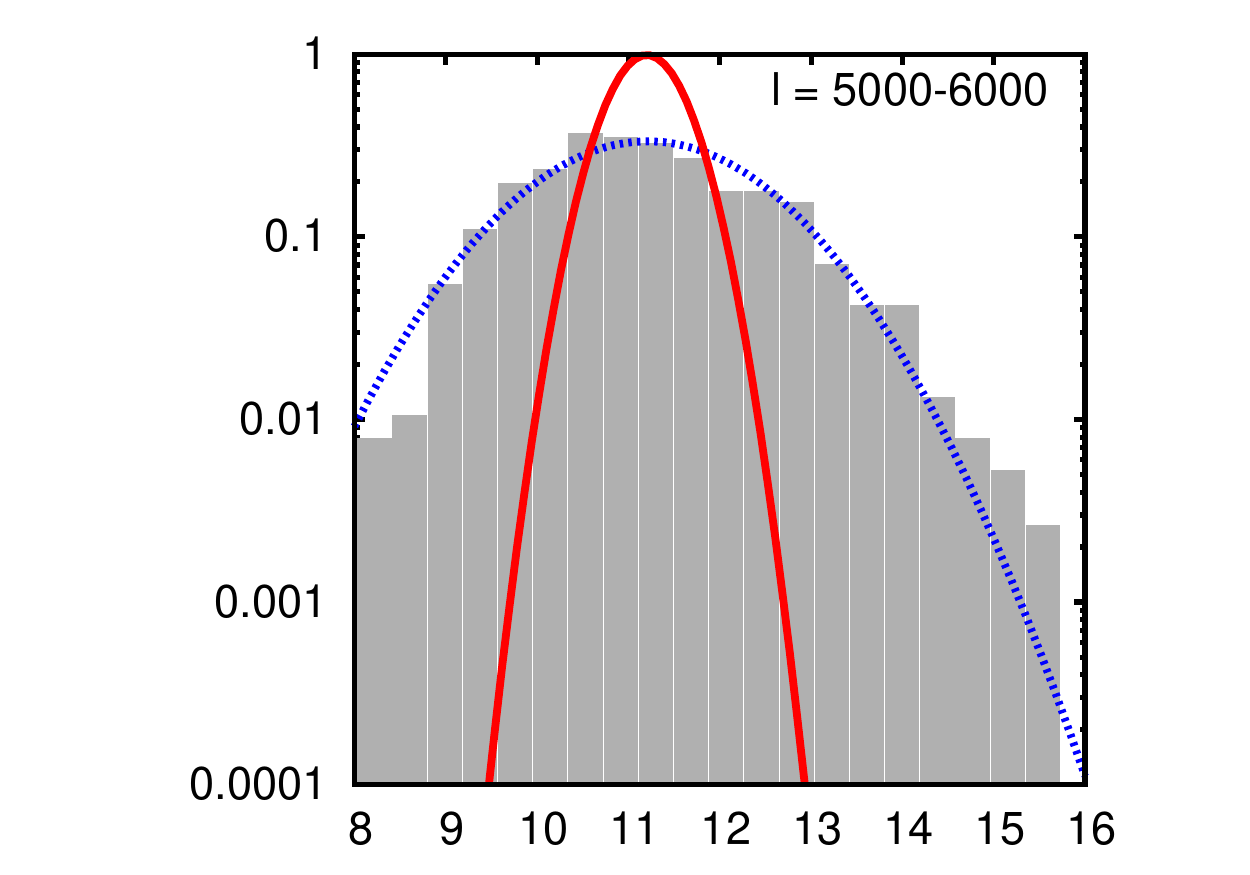}
  \includegraphics[scale=0.43,viewport=50 0 330 245,clip]{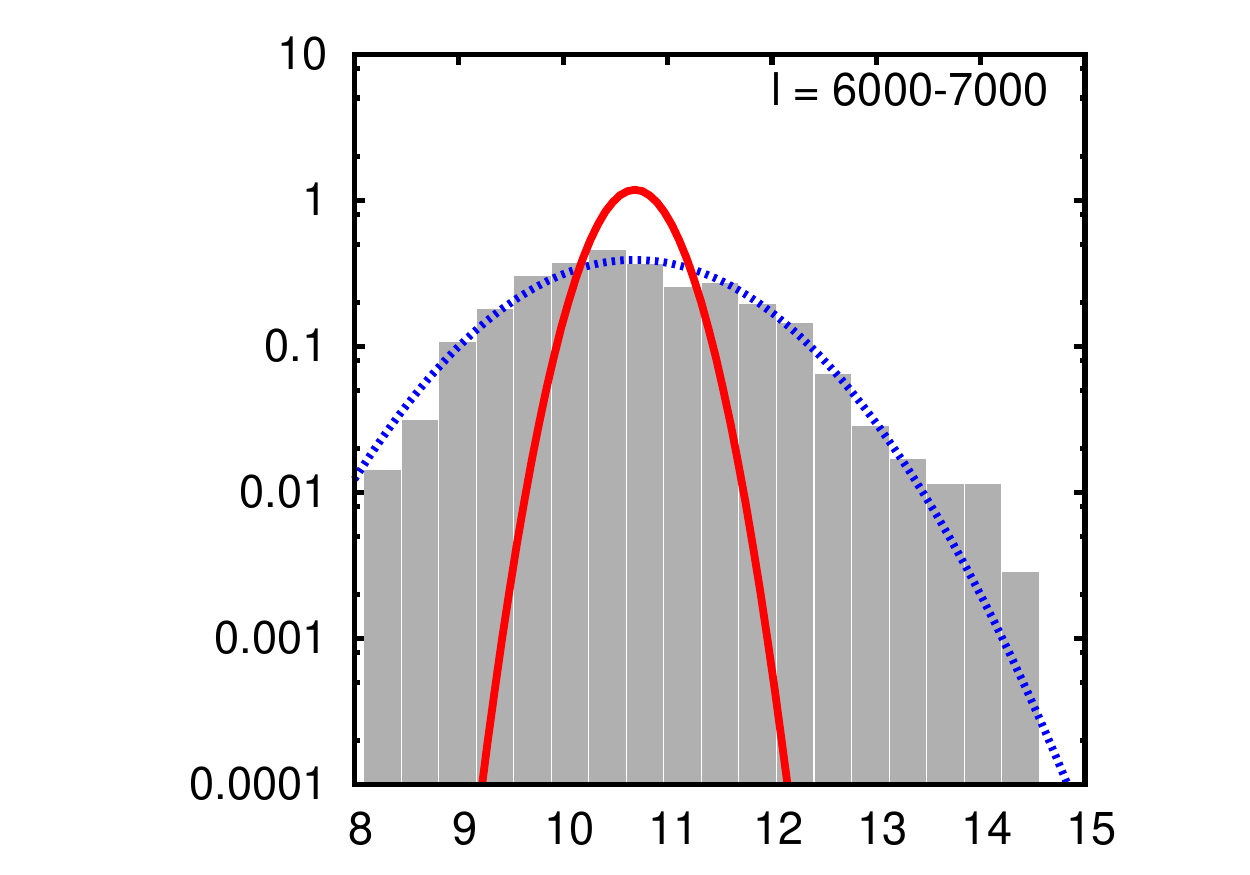}
  \includegraphics[scale=0.43,viewport=50 0 330 245,clip]{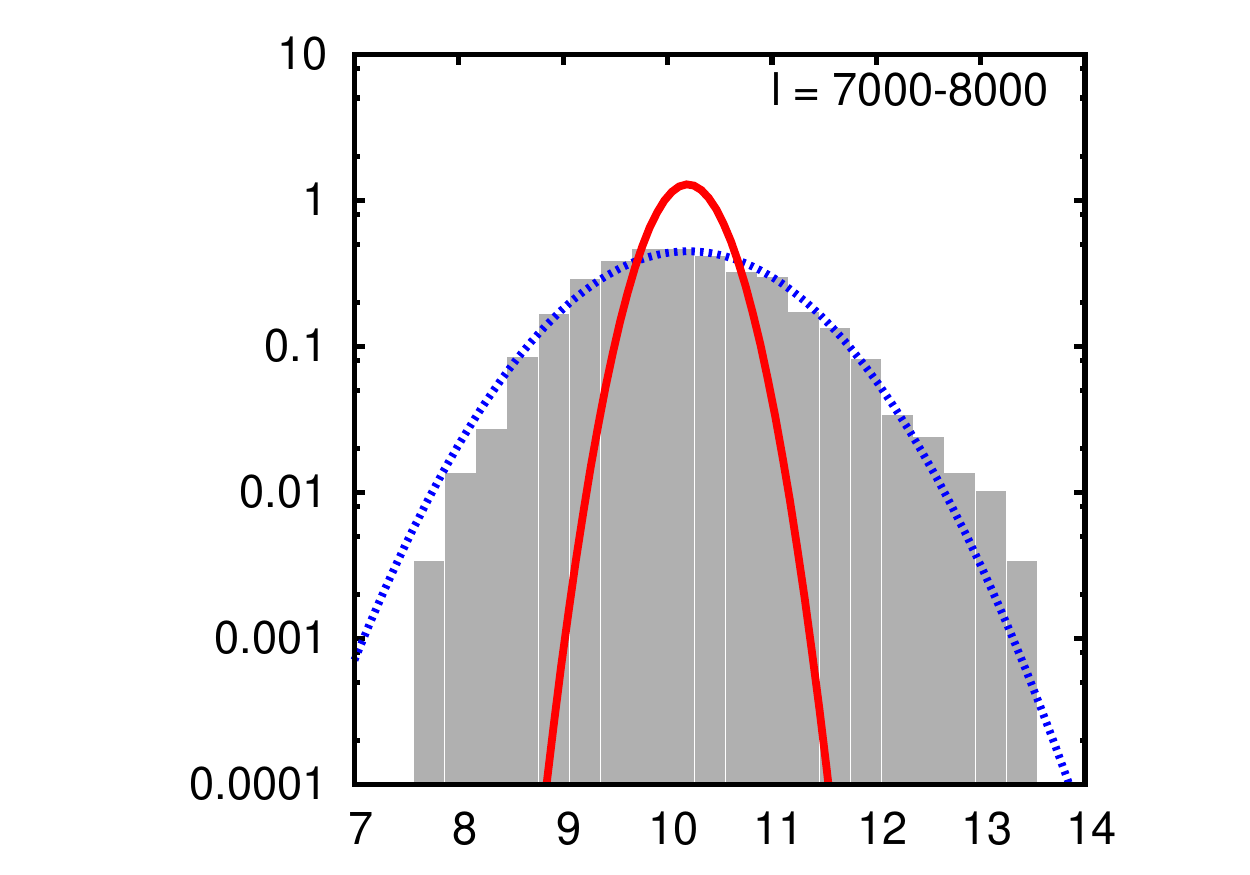}
  \includegraphics[scale=0.43,viewport=50 0 330 245,clip]{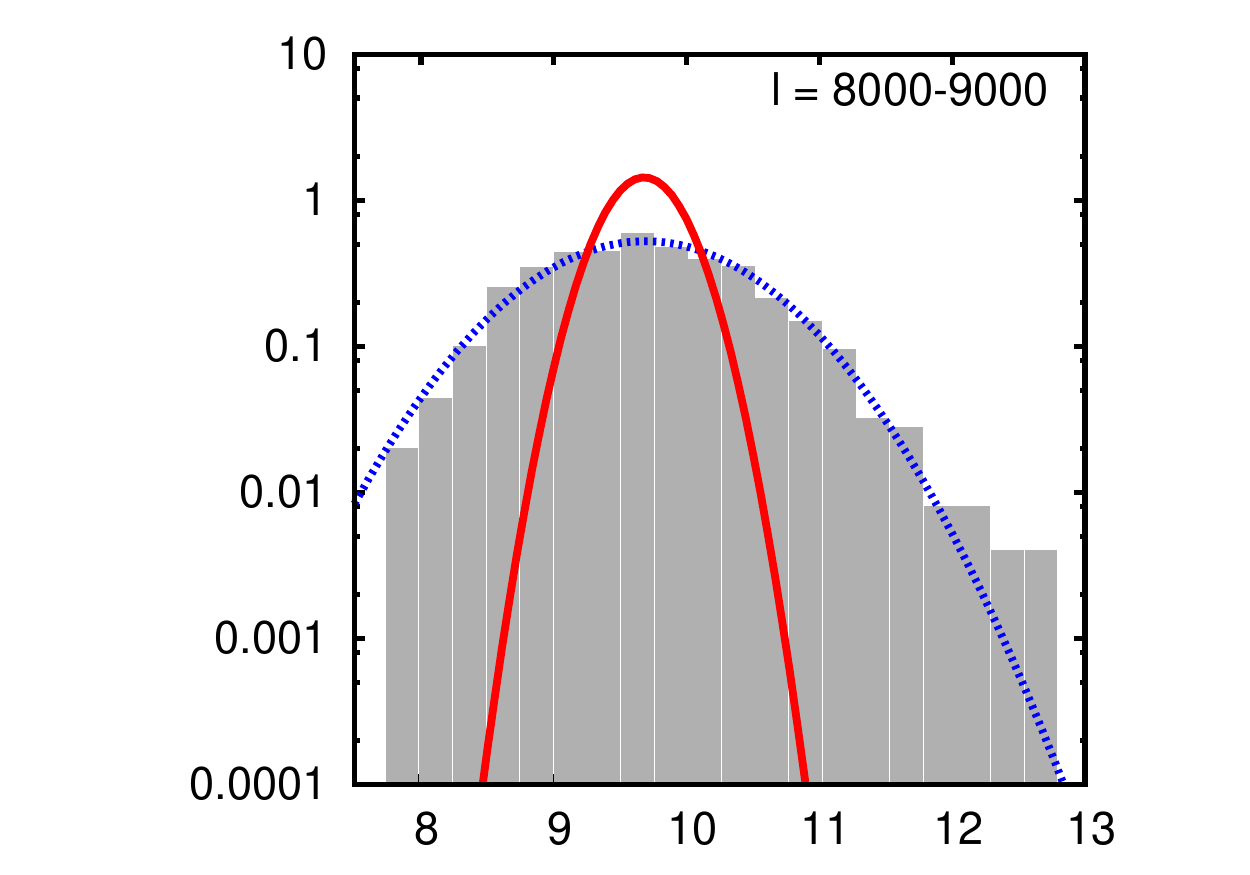}
	\caption{As Fig. \ref{fig:cmb_sz_plots_3deg}, but for 150~GHz. The CMB dominates to higher multipoles than at 30~GHz., such that the broadening of the distributions due to the SZ effect is less important at intermediate multipoles.}
	\label{fig:cmb_sz_plots_3deg_150ghz}
\end{figure*}

\begin{figure}
\centering
\includegraphics[scale=0.68]{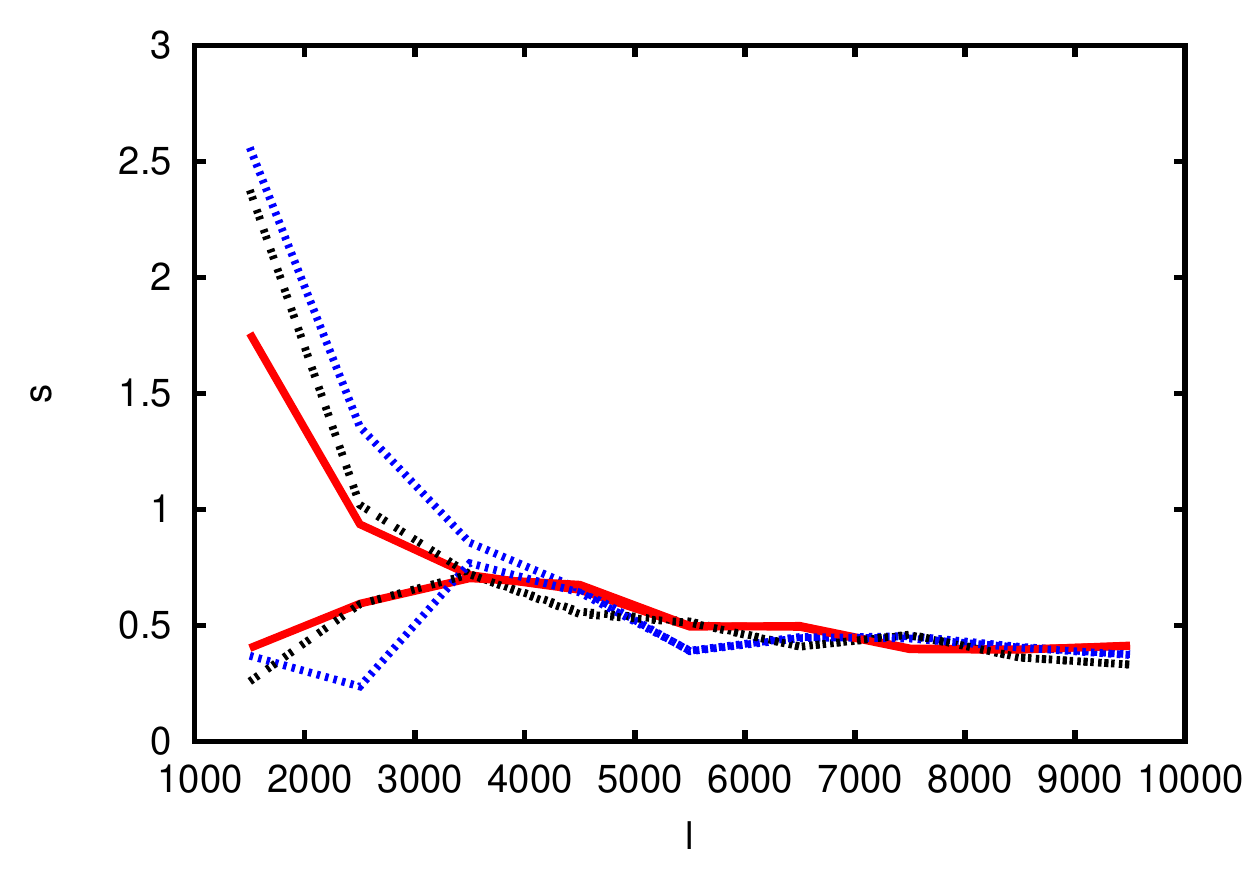}
\caption{The skewness $s$ for the $3 \times 3$ degree maps for the three values of $\sigma_8$: $0.75$ is shown by the blue dotted line, $0.825$ by the red solid line and $0.9$ by the black double-dotted line. The top set of lines are for the SZ only; the bottom set are when combined with the CMB. There is little difference between the two above $l\sim4000$. At the lowest multipoles, the SZ effect significantly skews the distribution, and this skewness remains important at intermediate to high multipoles.}
\label{fig:skewness}
\end{figure}

Fig. \ref{fig:cmb_sz_plots_3deg_150ghz} shows the equivalent histograms to Fig. \ref{fig:cmb_sz_plots_3deg} but at 150~GHz. The SZ effect at this frequency is weaker by a factor of 2 than at 30~GHz, reducing the power spectrum by a factor of 4. As a result, the CMB dominates the power spectrum to higher multipoles and the increase in standard deviation due to the SZ effect does not become important until $l \sim 4000$. The highest multipoles considered here are not greatly affected by this change in frequency as these remain dominated by the SZ effect.

To quantify more precisely this increase in standard deviation, we create realizations with the same mean power spectrum but Gaussianly distributing all of the power in the map, as if all of the power is due to the CMB. The Gaussian fit to these distributions is shown by the red solid lines in Fig. \ref{fig:cmb_sz_plots_3deg} and \ref{fig:cmb_sz_plots_3deg_150ghz}. At the lowest multipoles, where the CMB is dominant, the statistics from this method are in good agreement with those from the realizations. At higher multipoles, as expected, the standard deviation is much lower for Gaussian statistics than for the SZ effect. The ratios of the standard deviation from the two methods are given in Table \ref{tab:sz_statistics}; the standard deviation is much smaller, typically by a factor of 3 for $l>3000$, when compared with the non-Gaussian realizations.

In Section \ref{sec:realization}, the mean power spectra were found to scale approximately as $C_l \propto \sigma_8^\alpha$, where $\alpha \sim 7$, as per \citet{2002Komatsu}. Assuming that the standard deviation varies with $\sigma_8$ in the same fashion, we find that $\alpha_\mathrm{SD} \approx 6.0$ between $\sigma_8 = 0.75$ and 0.825, and $\approx 6.8$ between 0.9 and 0.825. Thus, there is a strong dependence on $\sigma_8$ for the standard deviation, although not quite as strong as for the mean power spectrum. The curve for $\sigma_8 = 0.75$ is slightly higher than for the other two values of $\sigma_8$ in Fig. \ref{fig:cmb_sz_stats}; this is an illustration of the non-universality of $\alpha_\mathrm{SD}$.

\begin{figure*}	
\centering
\includegraphics[scale=0.70,viewport=80 30 300 215,clip]{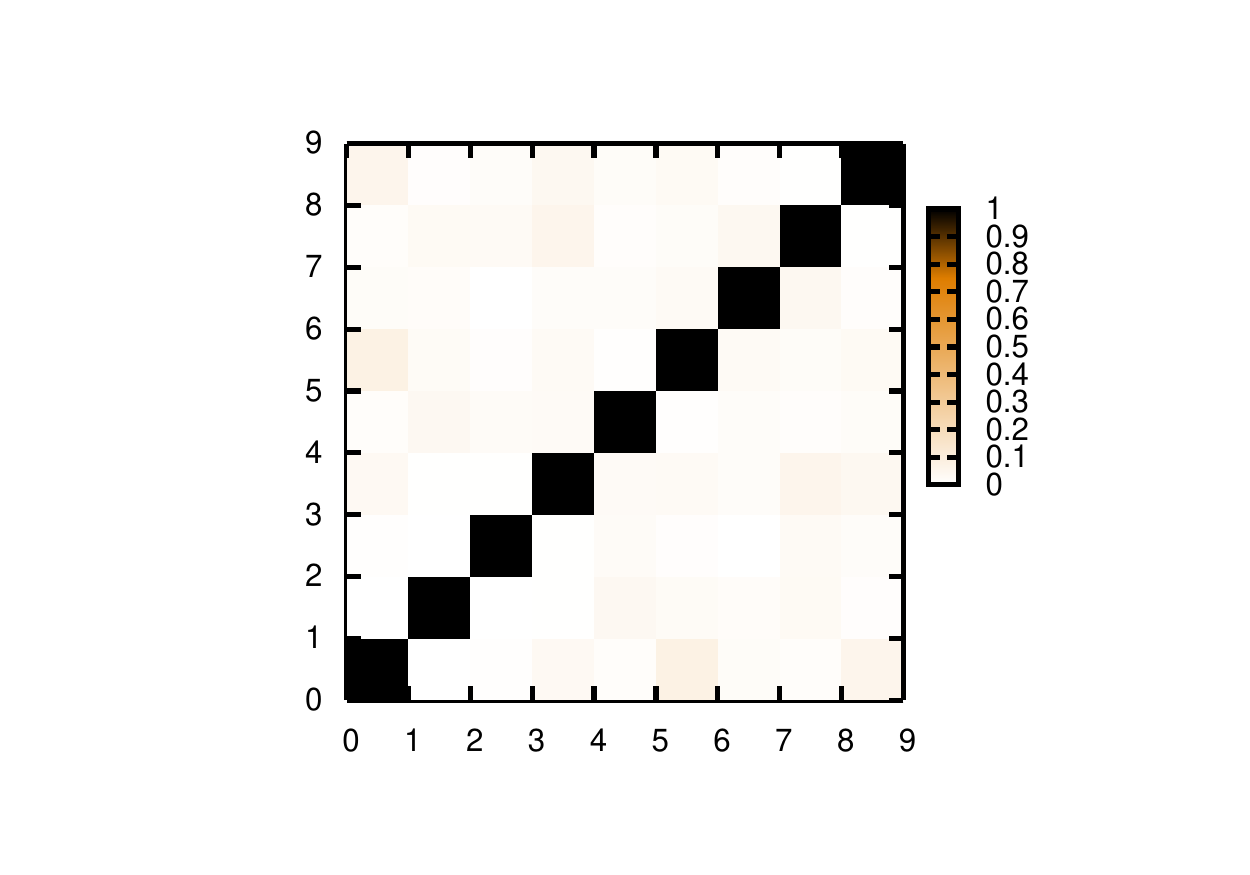}
\includegraphics[scale=0.70,viewport=80 30 300 215,clip]{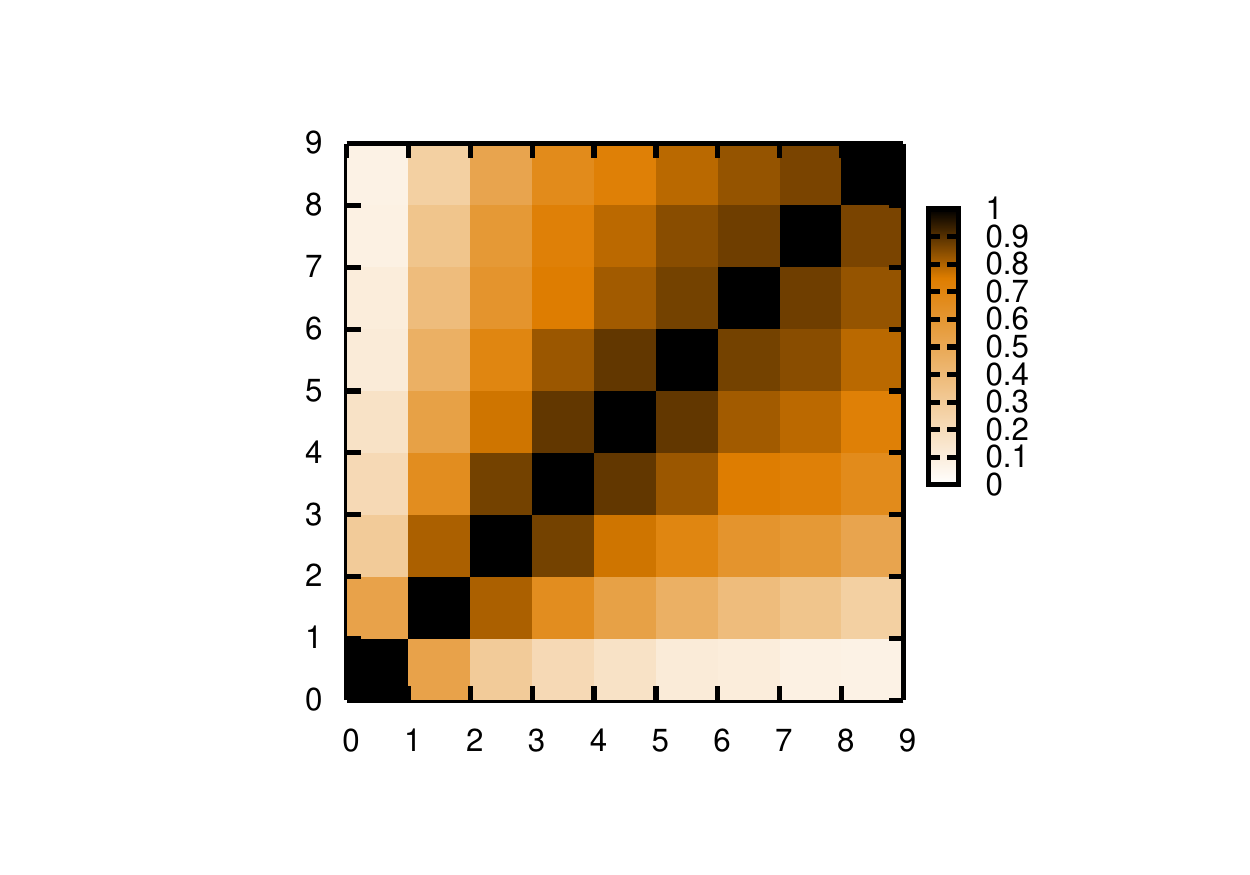}
\includegraphics[scale=0.70,viewport=80 30 300 215,clip]{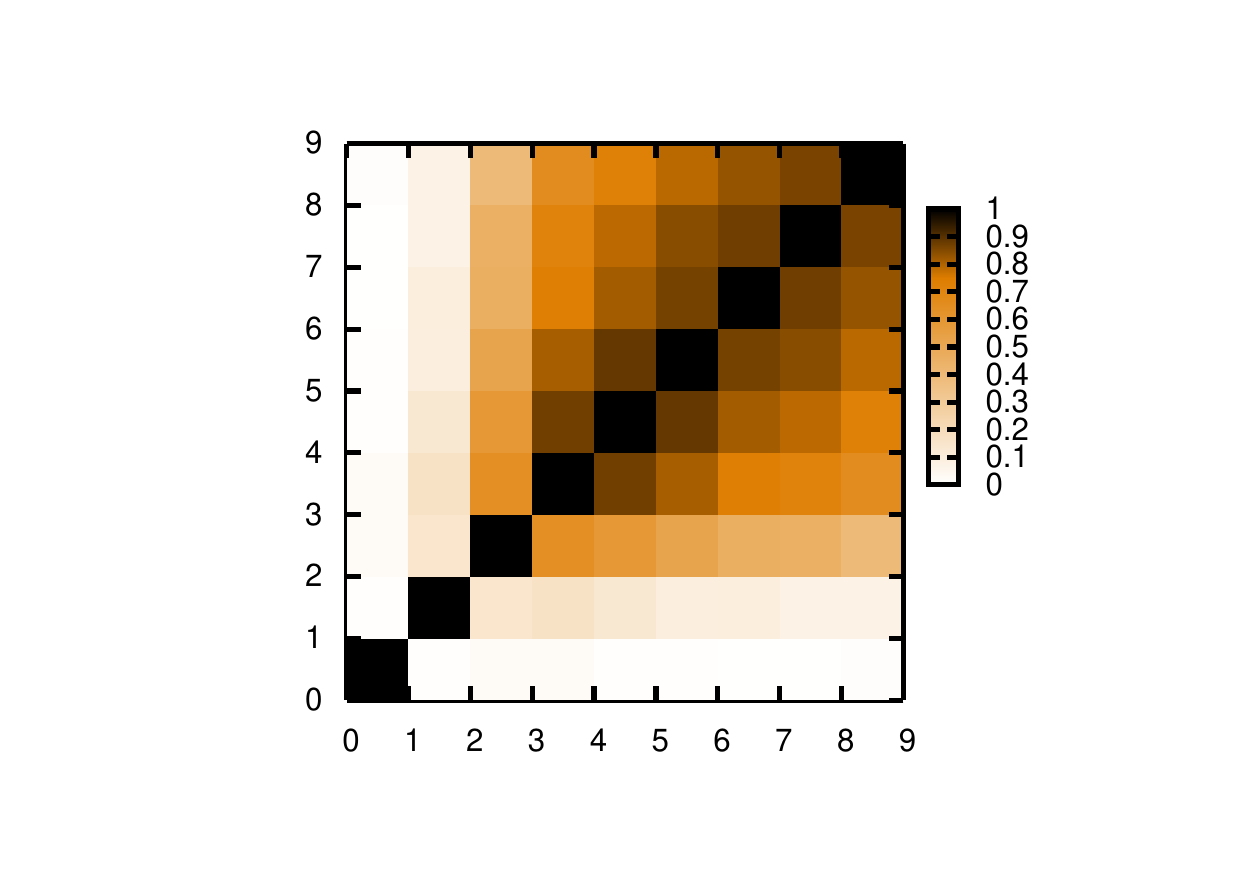}
\caption{The correlation matrix for 1000 realizations of the CMB (left), SZ effect (middle) and the combined CMB and SZ effect (right), using $3 \times 3$ degree maps with bins of $\Delta l = 1000$. The bin numbers are given in the $x$ and $y$ axis (for example, 0 is the multipole range $0-1000$). The CMB has negligible off-diagonal terms, however these terms are no longer negligible when the SZ effect is accounted for.}
\label{fig:sz_covariance}
\end{figure*}

In Section \ref{sec:realization}, the mean theoretical spectrum calculated by an analytical formula was compared to the mean from the realizations and was found to agree well. The standard deviation can also be computed; see the Appendix for the method. The standard deviation calculated solely from the Gaussian components differs from that calculated from the realizations by a factor between 3 and 7 depending on the multipole; when the angular trispectrum component is added the two agree to within 20 per cent when compared with the results from the {\sc Pinocchio} realizations presented in Fig. \ref{fig:cmb_sz_stats}. The origin of the remaining discrepancy will be discussed in section \ref{sec:clustering}.

We find that the SZ effect results in a positive skew in the statistics, resulting from an overabundance of high-power realizations when compared with the expectations from Gaussian statistics due to the presence of rare, massive clusters and associated smaller clusters. This is shown in Fig. \ref{fig:skewness}; the lowest multipoles for the SZ effect on its own are especially skewed. With the addition of the CMB, the low multipole bins become Gaussian as the CMB dominates, but some skewness remains at the higher multipoles where the SZ effect is dominant. There is a certain amount of noise in the measurements of the skew here, however; an increased number of realizations would measure this more accurately. \citet{2007Zhang} have also found that the SZ effect yields a positively skewed distribution using an analytical approach for calculating the probability distribution for a single multipole.

\subsection{Correlation matrix}
For purely Gaussian maps, for example the CMB, the power on different scales is uncorrelated such that the off-diagonal terms of the covariance matrix are zero. However, this may not be the case for the SZ effect. The covariance matrix can be calculated by
\begin{equation}
C_{ij} = \frac{1}{N} \sum_{n} (B_i^n - \bar{B}_i) (B_j^n - \bar{B}_j).
\end{equation}
This is then normalized to give the correlation matrix by
\begin{equation}
\hat{C}_{ij} = \frac{C_{ij}}{\sqrt{C_{ii} C_{jj}}}.
\end{equation}

The correlation matrix is depicted graphically for the CMB and SZ effect individually and combined in Fig. \ref{fig:sz_covariance}. The matrix is diagonal for the CMB, however the off-diagonal terms are significant for the SZ effect, in agreement with the predictions of \citet{2001Cooray}. This is due to the power from individual clusters spanning many thousands of multipoles, correlating the multipole bins considered here. This effect is reduced for widely separated multipole bins as individual clusters are not dominant at all multipoles; for example, as shown by Fig. \ref{fig:convergence_testing} the largest clusters only provide the bulk of the SZ effect at the lowest multipoles, and smaller (mostly uncorrelated) clusters provide the power at the highest multipoles.

\subsection{Dependency on map size}
\begin{figure*}
\centering
  \includegraphics[scale=0.43,viewport=50 0 330 245,clip]{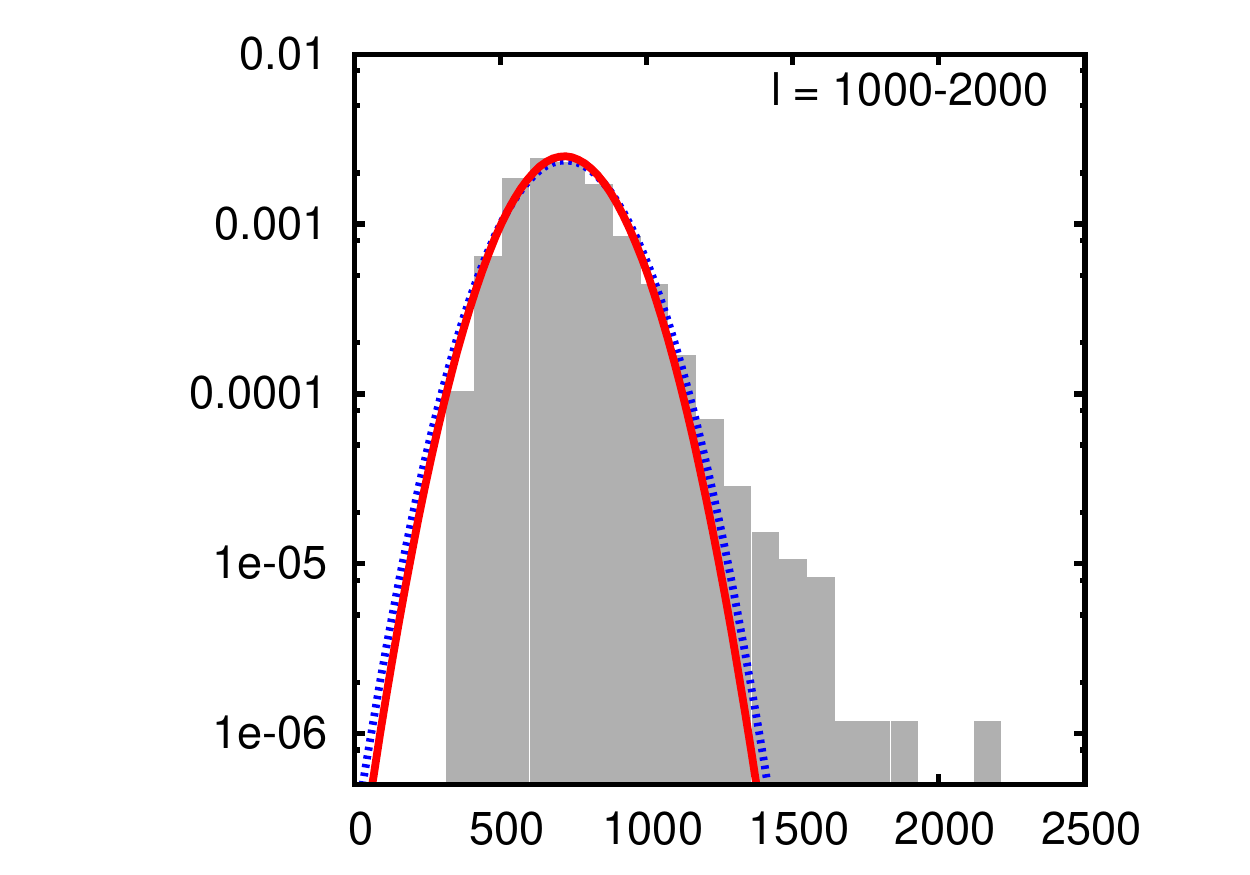}
  \includegraphics[scale=0.43,viewport=50 0 330 245,clip]{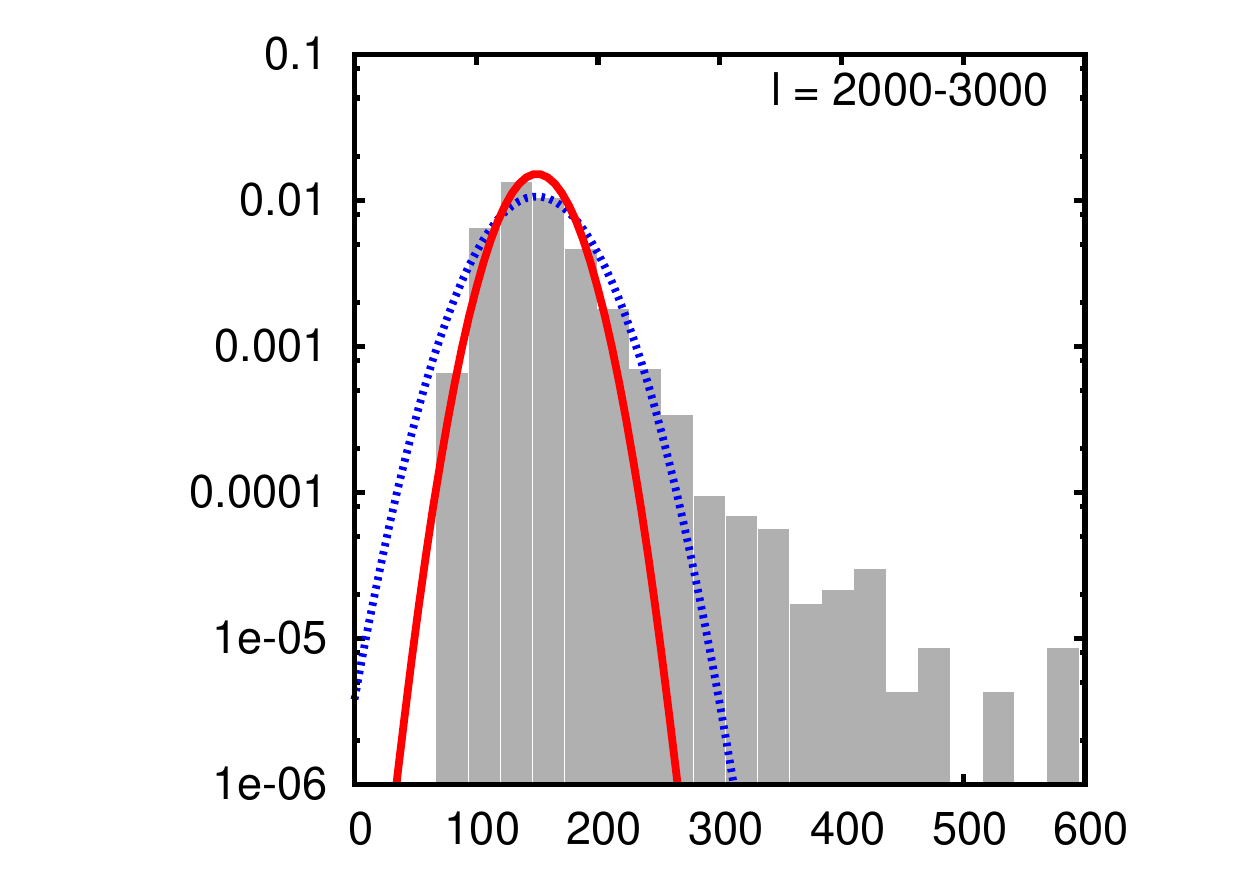}
  \includegraphics[scale=0.43,viewport=50 0 330 245,clip]{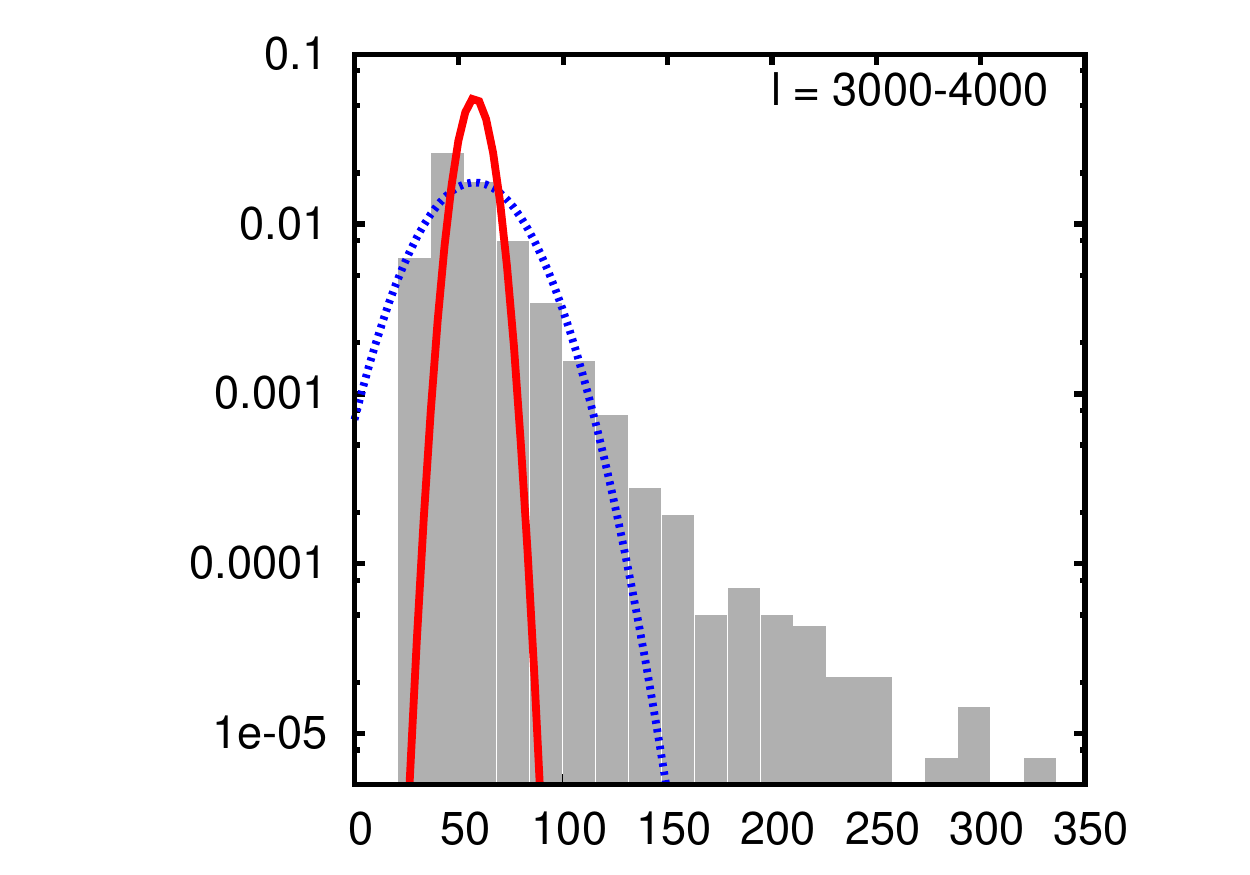}
  \includegraphics[scale=0.43,viewport=50 0 330 245,clip]{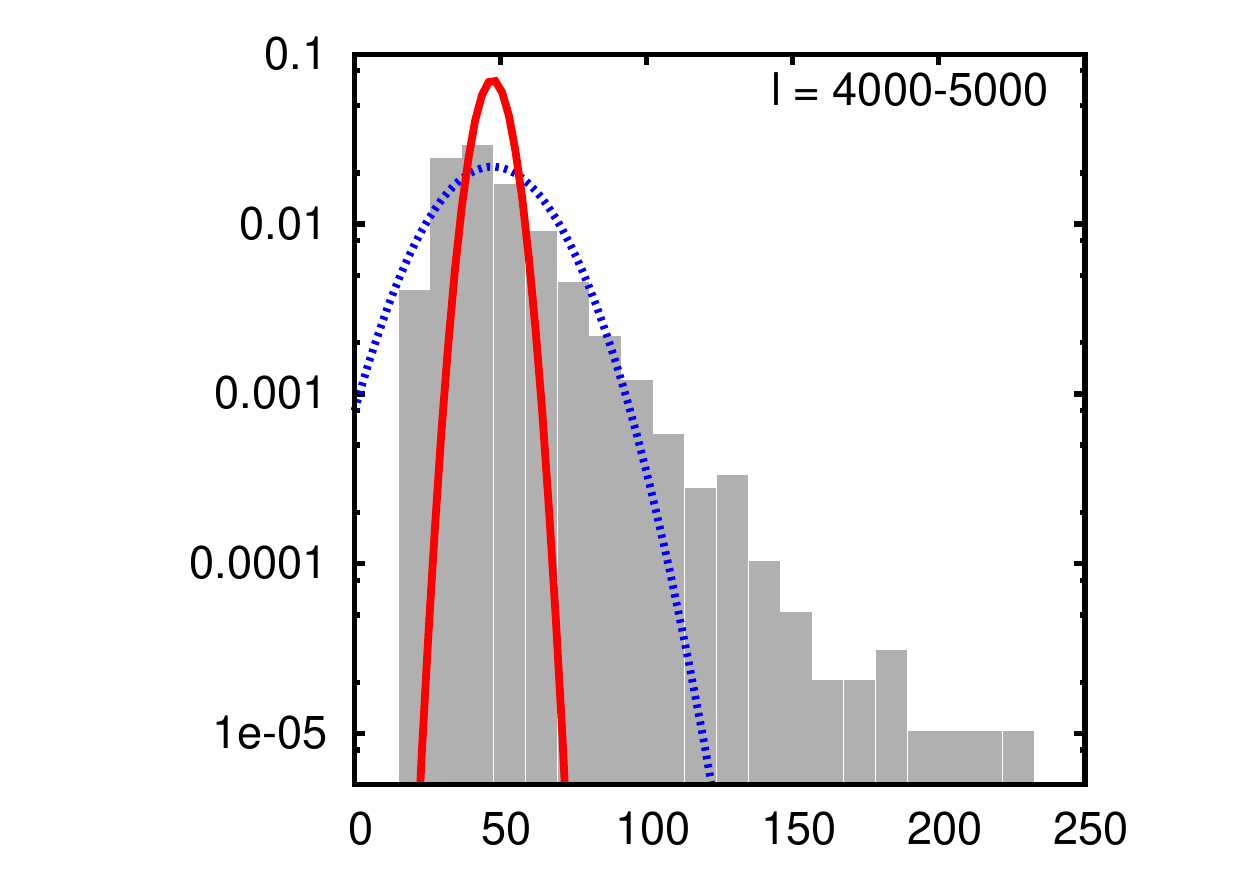}
  \\
  \includegraphics[scale=0.43,viewport=50 0 330 245,clip]{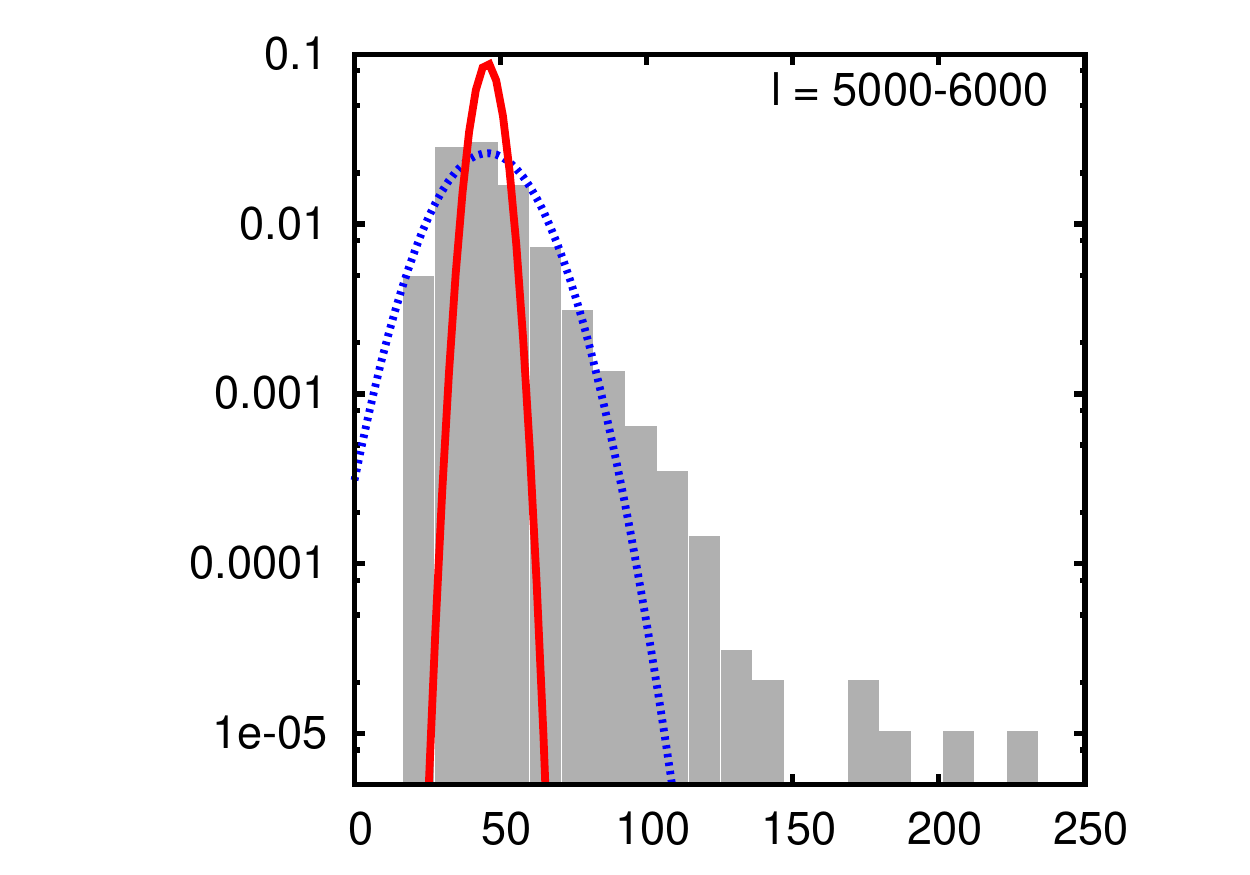}
  \includegraphics[scale=0.43,viewport=50 0 330 245,clip]{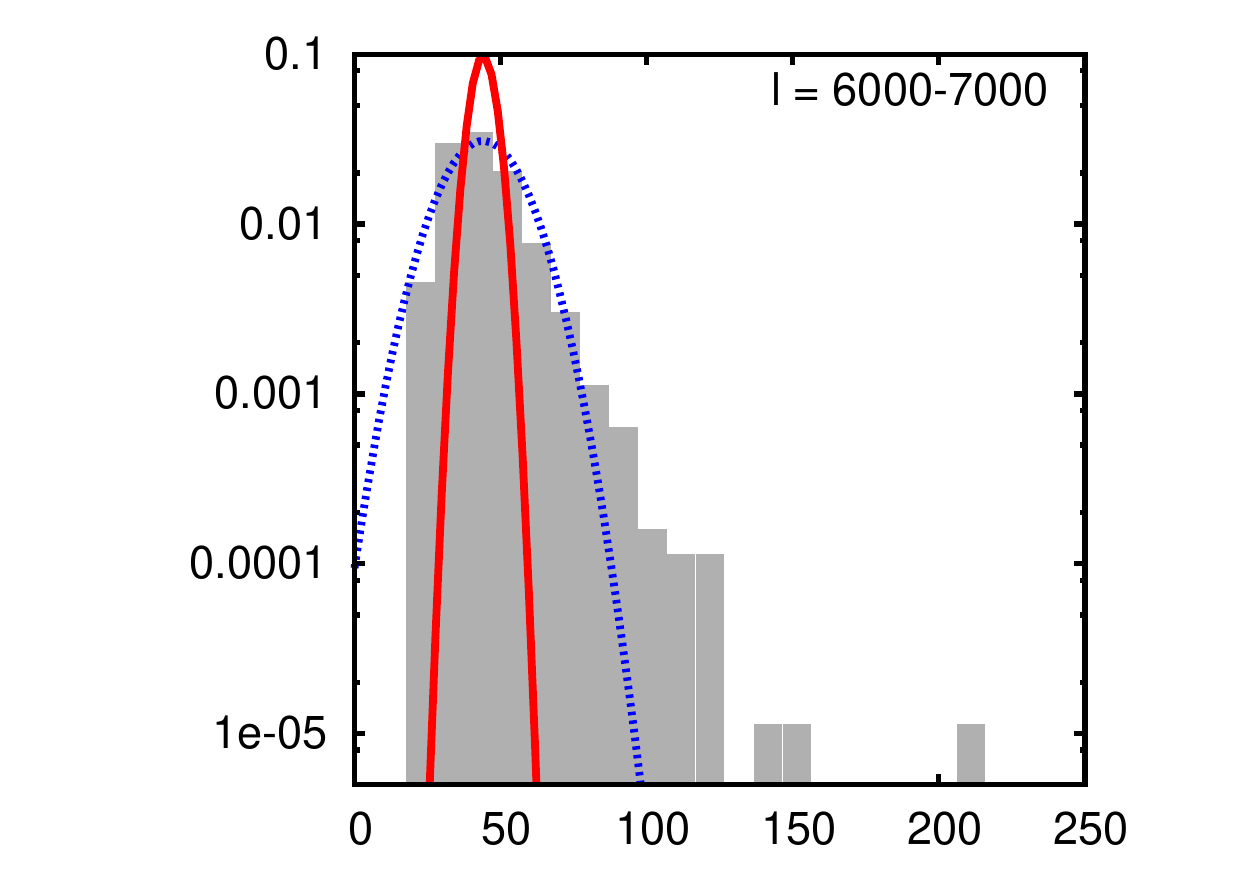}
  \includegraphics[scale=0.43,viewport=50 0 330 245,clip]{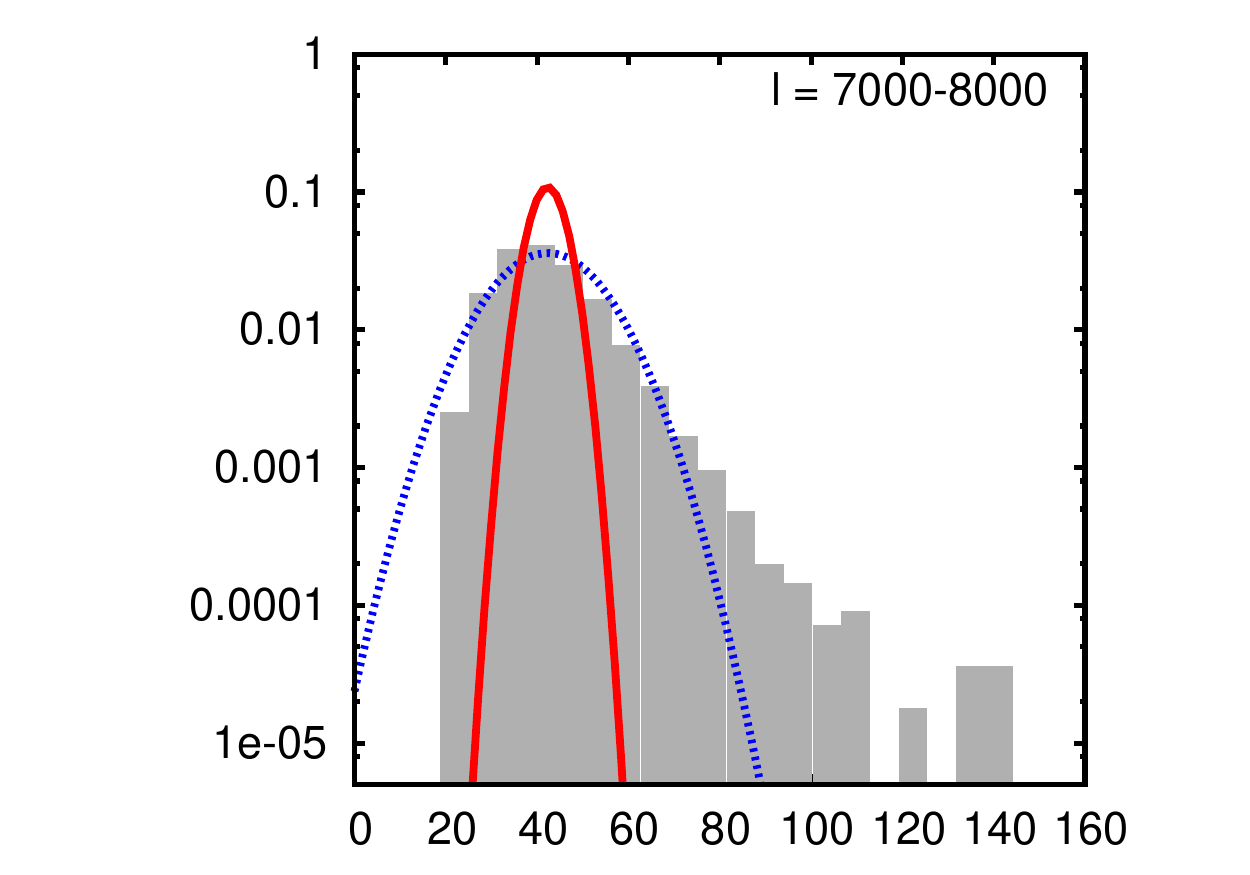}
  \includegraphics[scale=0.43,viewport=50 0 330 245,clip]{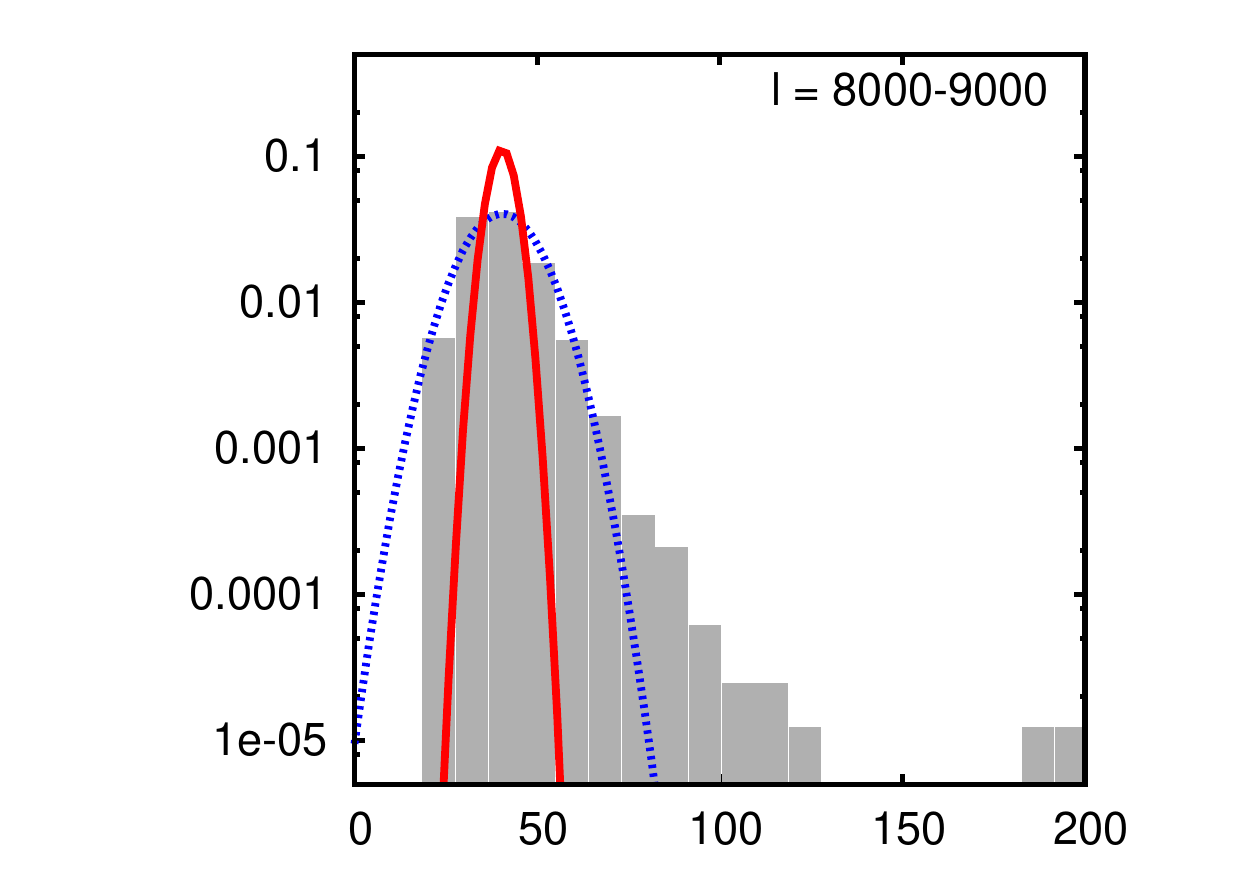}
	\caption{As Fig. \ref{fig:cmb_sz_plots_3deg}, but for 9000 $1 \times 1$ degree maps. The first few multipole bins are dominated by the CMB, hence have a distribution that is close to Gaussian; the later bins become increasingly more skewed to higher values. Extremal values are more likely than in the $3 \times 3$ degree maps.}
	\label{fig:cmb_sz_plots_1deg}
\end{figure*}

The statistics of the SZ effect power spectrum will depend on the size of the map being considered. We investigate this dependency by considering maps of $1 \times 1$ and $2 \times 2$ degrees, in addition to the $3 \times 3$ degree maps that have been investigated thus far. These smaller maps are created by selecting and mapping only the clusters from the catalogues that lie within these areas. For the $1 \times 1$ degree maps, this provides 9000 independent realizations, whereas for the $2 \times 2$ degree maps we continue using 1000 realizations. The mean power spectra for these three different map sizes remain the same.

The histograms for the $1 \times 1$ degree realizations containing both the CMB and the SZ effect are shown in Fig. \ref{fig:cmb_sz_plots_1deg}. As before, for the lowest multipoles the CMB dominates, so the histogram for the lowest bin is approximately Gaussian. For the higher multipoles, however, the histograms are significantly skewed towards positive values, much more so than for the $3 \times 3$ degree maps. This qualitatively agrees with \citet{2007Zhang}, who found that larger map sizes decrease the skewness of the probability distribution due to averaging over a larger number of clusters.

\begin{figure*}
\centering
\includegraphics[scale=0.68]{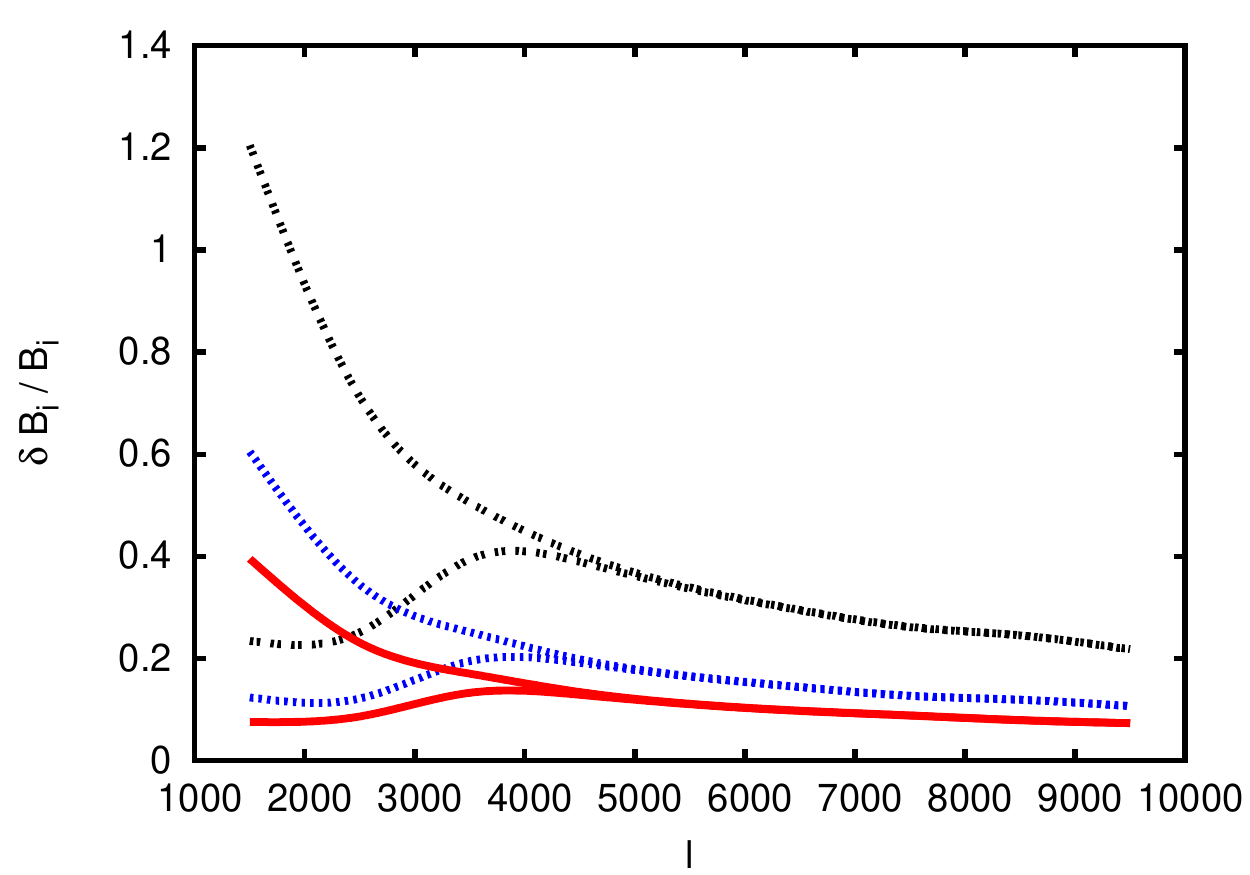}
\includegraphics[scale=0.68]{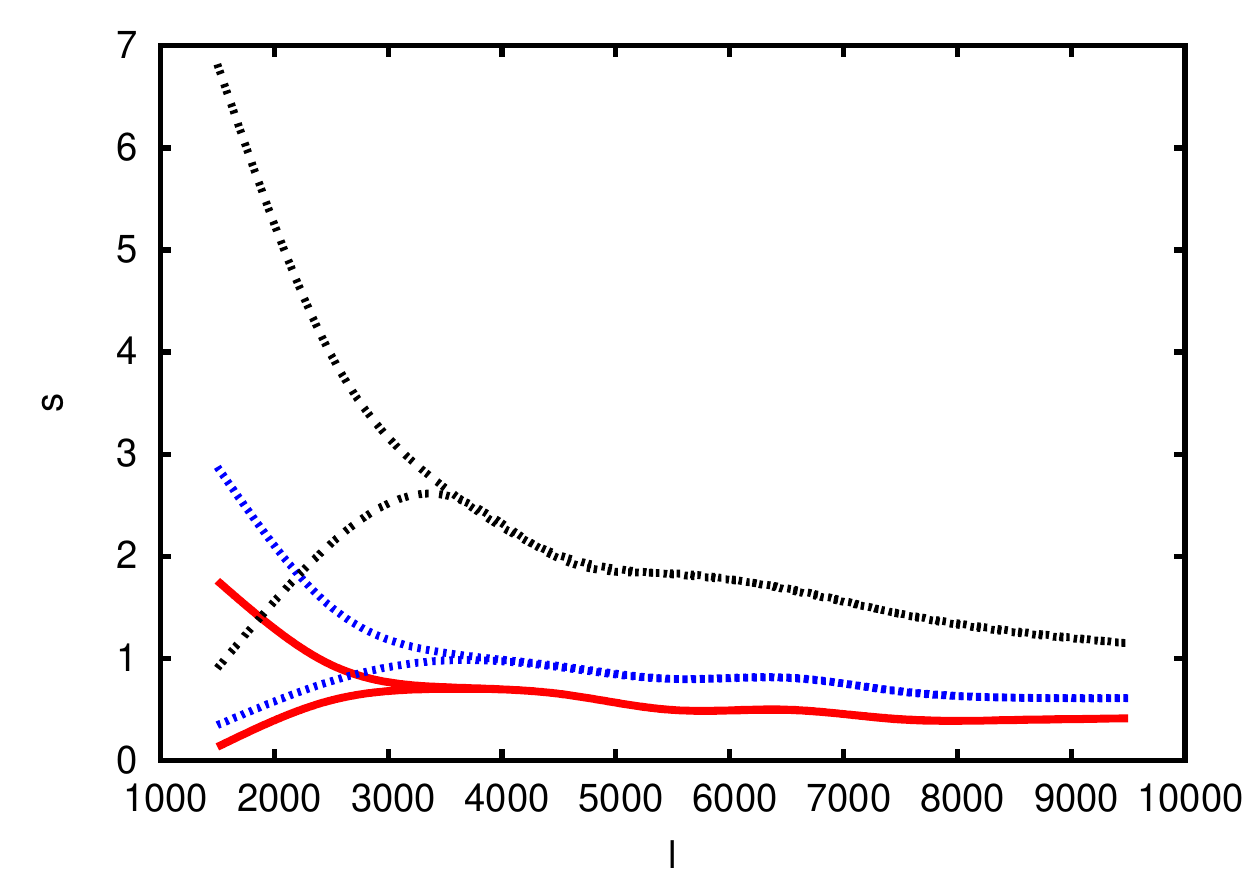}
\caption{Left: $\delta B_i / \bar{B}_i$ from the SZ effect on its own (top lines) and for the CMB and SZ effect combined (bottom lines) with $\sigma_8 = 0.825$ for three different map sizes: 3x3 degree (red solid line), 2x2 degree (blue dotted line) and 1x1 degree (black double-dotted line). The values scale as $1/f_\mathrm{sky}^{1/2}$. Right: As the left plot, but plotting the skewness for the three map sizes; this also scales roughly as $1/f_\mathrm{sky}^{1/2}$.}
\label{fig:mapsize}
\end{figure*}

For the CMB, the standard deviation from the realizations scales linearly with $1/f_\mathrm{sky}^{1/2}$. The ratio of the standard deviation to the mean for the three maps sizes is shown in Fig. \ref{fig:mapsize}; this scales as $1/f_\mathrm{sky}^{1/2}$ for both the SZ effect on its own and the combined CMB and SZ effect. This is as predicted by the analytical formula. The transition between the CMB- and SZ-dominated regimes is clear for all three map sizes. The skew for the SZ effect also scales close to $1/f_\mathrm{sky}^{1/2}$. If the dependence is parameterized as $1/f_\mathrm{sky}^{\alpha/2}$ and $\alpha$ is calculated independently for each multipole bin, then $\alpha \approx 1.2$ at the lowest multipoles, changing to $\approx 0.9$ at the highest multipoles. This scaling indicates that the skewness can be reduced to $<0.1$ for maps larger than $400$ square degrees.

As we are only considering the clusters within the map size, we will not be including any effects from clusters that are positioned outside the map but extend into the map. This was true for the larger maps considered earlier, however this may become more important here as large clusters can extend up to a quarter of a degree in the maps. As the means from the three map sizes differ by less than a percent, however, and the standard deviations agree with predictions, this appears to be negligible.

\subsection{Effects of clustering} \label{sec:clustering}
\begin{figure}
\centering
\includegraphics[scale=0.68]{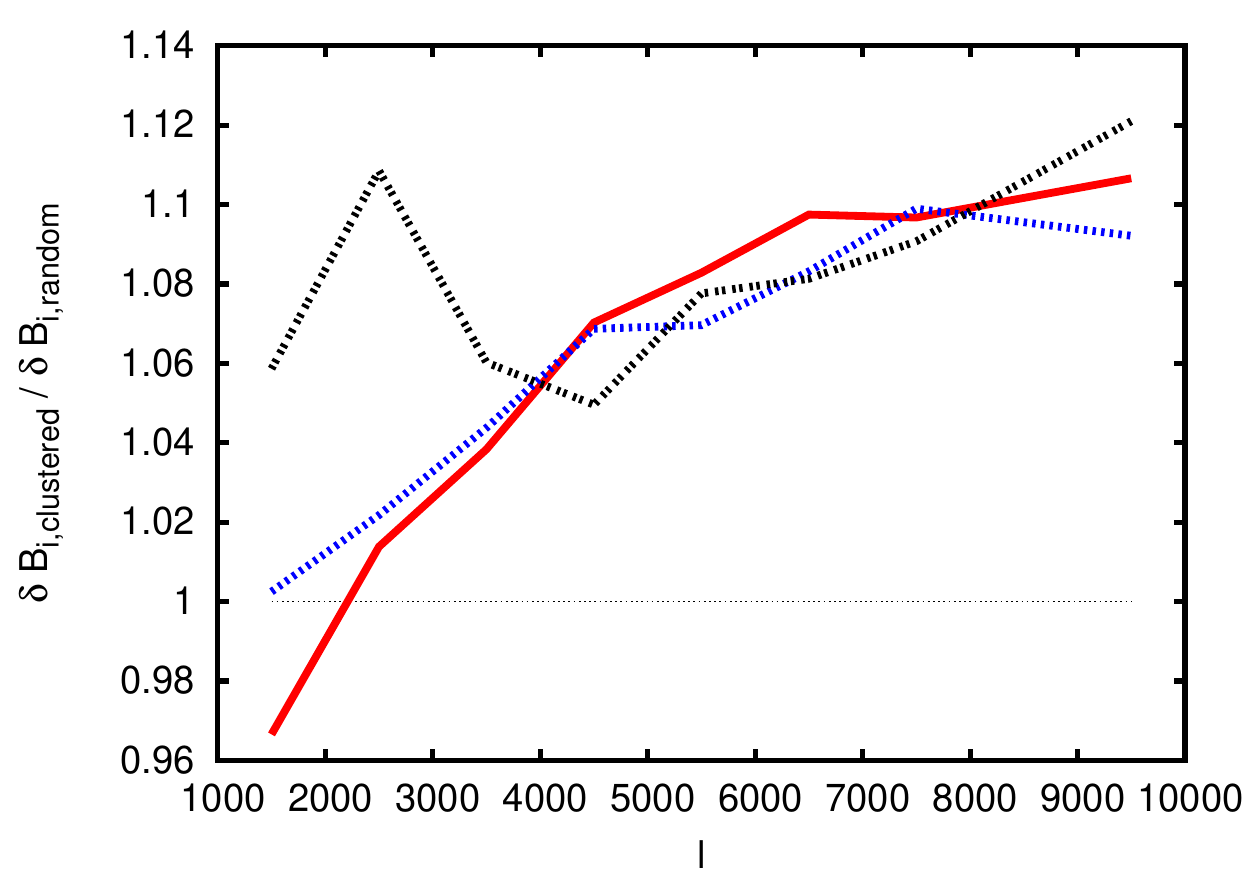}
\includegraphics[scale=0.68]{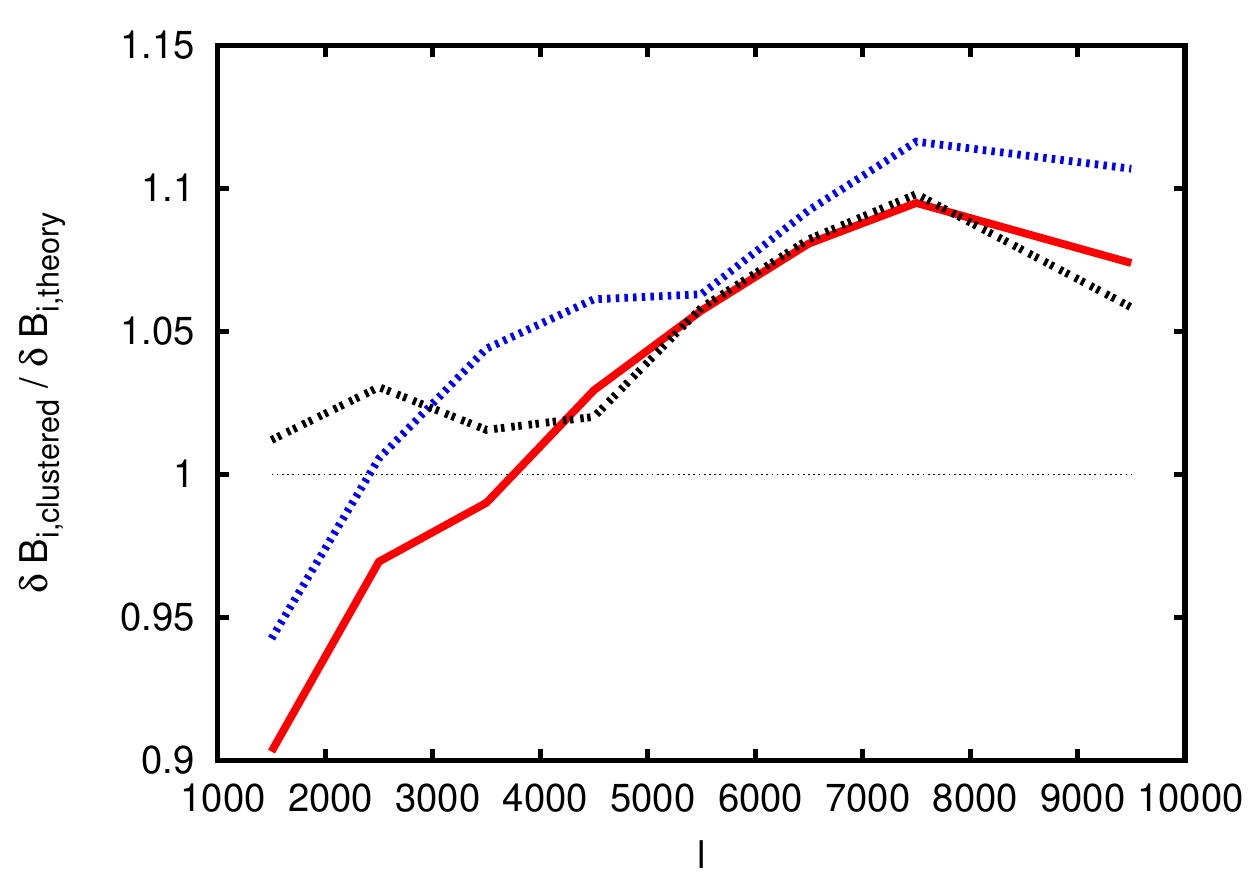}
\caption{Top: The ratio of $\delta B_i / \bar{B}_i$ from clustered and randomised $1 \times 1$ degree maps for the three different values of $\sigma_8$ (the red solid line is $\sigma_8 = 0.825$, the blue dotted line $0.9$ and the black double-dotted line $0.75$). Clustering increases the standard deviation by $\sim 10$ per cent. Bottom: The same as the top figure, but comparing the clustered realizations to the analytical expectation. The increase of $\sim 10$ per cent remains.}
\label{fig:compare_randomisation_methods}
\end{figure}

To check whether the clustering of the galaxy clusters can be responsible for the non-Gaussianity in the statistics, we randomize the positions of the clusters on the sky. This effectively modifies the angular correlation function to that expected of a Poissonian distribution, although this still includes the effects of clustering due to fluctuations that are larger than the map size. For the $3 \times 3$ degree maps, there is no significant change in any of the values of the mean, variance or normalized skew between the statistics of the randomized and the clustered maps. However, when the values of  $\delta B_i / \bar{B}_i$ are compared to those from the analytical calculations (as described in the Appendix), a $10-15$ per cent excess is present at high multipoles

Simply randomising the positions on the sky effectively destroys the clustering between galaxy clusters on scales smaller than the fiducial map size of $3 \times 3$ degrees. However, this will not remove the effects of clustering on the mean number density of clusters within the maps. To illustrate the effect of this, we create sets of $1 \times 1$ degree maps where we randomise the clusters by two methods: first, selecting the clusters within a $1 \times 1$ degree area then randomising their positions (analogous to the randomization performed on the $3 \times 3$ degree maps), and second, randomising the whole map and then selecting $1 \times 1$ degree patches.

Selecting the clusters that lie within the $1 \times 1$ degree area and then randomizing their positions provides largely the same statistical properties as the clustered realizations, as was the case for the $3 \times 3$ degree realizations. However, if the clusters are randomized prior to selection, then the ratio of $\delta B_i / \bar{B}_i$ is close to agreement with the analytical estimate, differing by a few percent. The effect of randomising and then selecting the clusters is shown in Fig. \ref{fig:compare_randomisation_methods}, which gives the ratio of $\delta B_i$ between the clustered and random realizations ($\bar{B}_i$ is the same for the two methods) and between the clustered realizations and the analytical values. Clustering increases $\delta B_i / \bar{B}_i$ by $\sim 10$ per cent at the highest multipoles, where the standard deviation is increased as the galaxy clusters group together in certain realizations and are absent in others, broadening the standard deviation. The lowest multipoles are not well sampled and are highly skewed, such that their standard deviation alone does not adequately represent the distribution.

\subsection{Upper mass limit} \label{sec:upper_mass}
\begin{figure}
\centering
\includegraphics[scale=0.68]{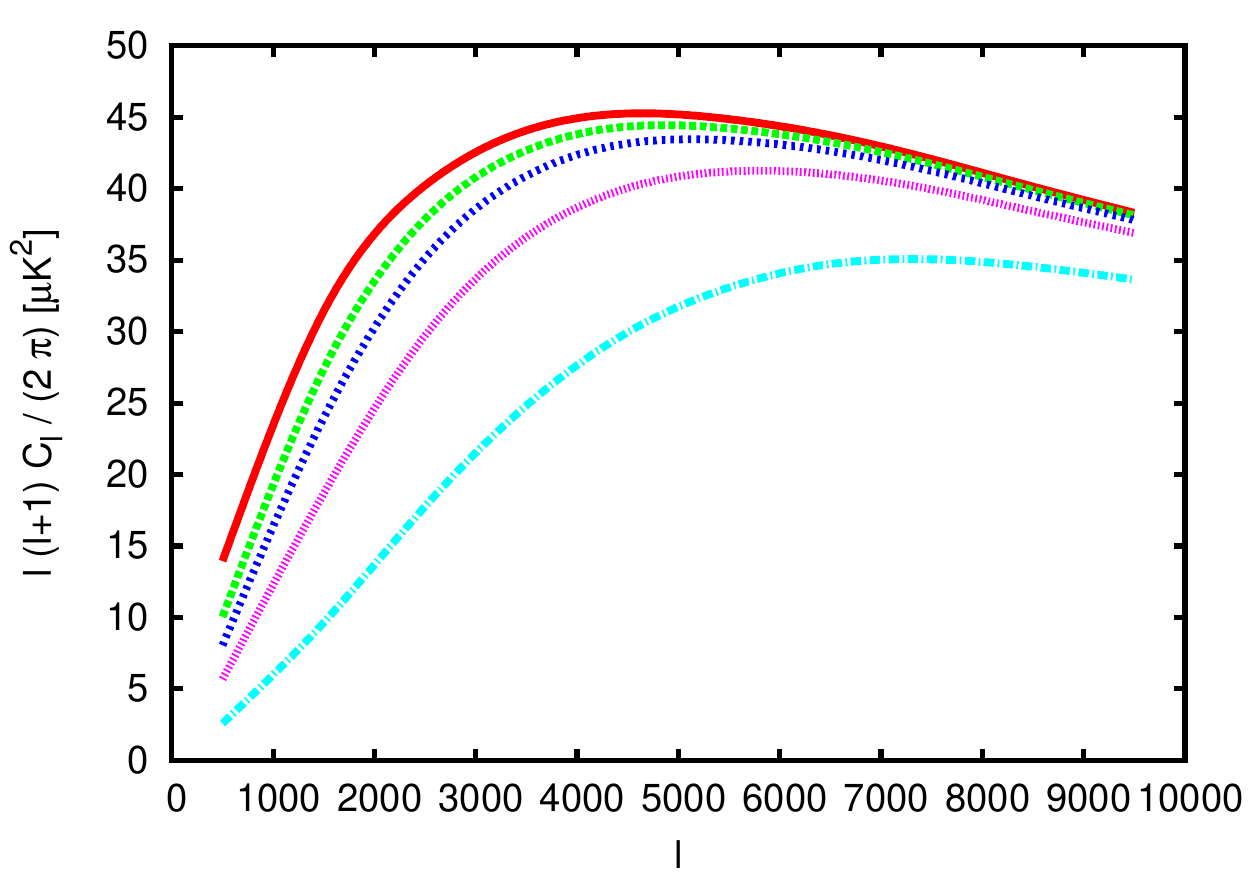}
\includegraphics[scale=0.68]{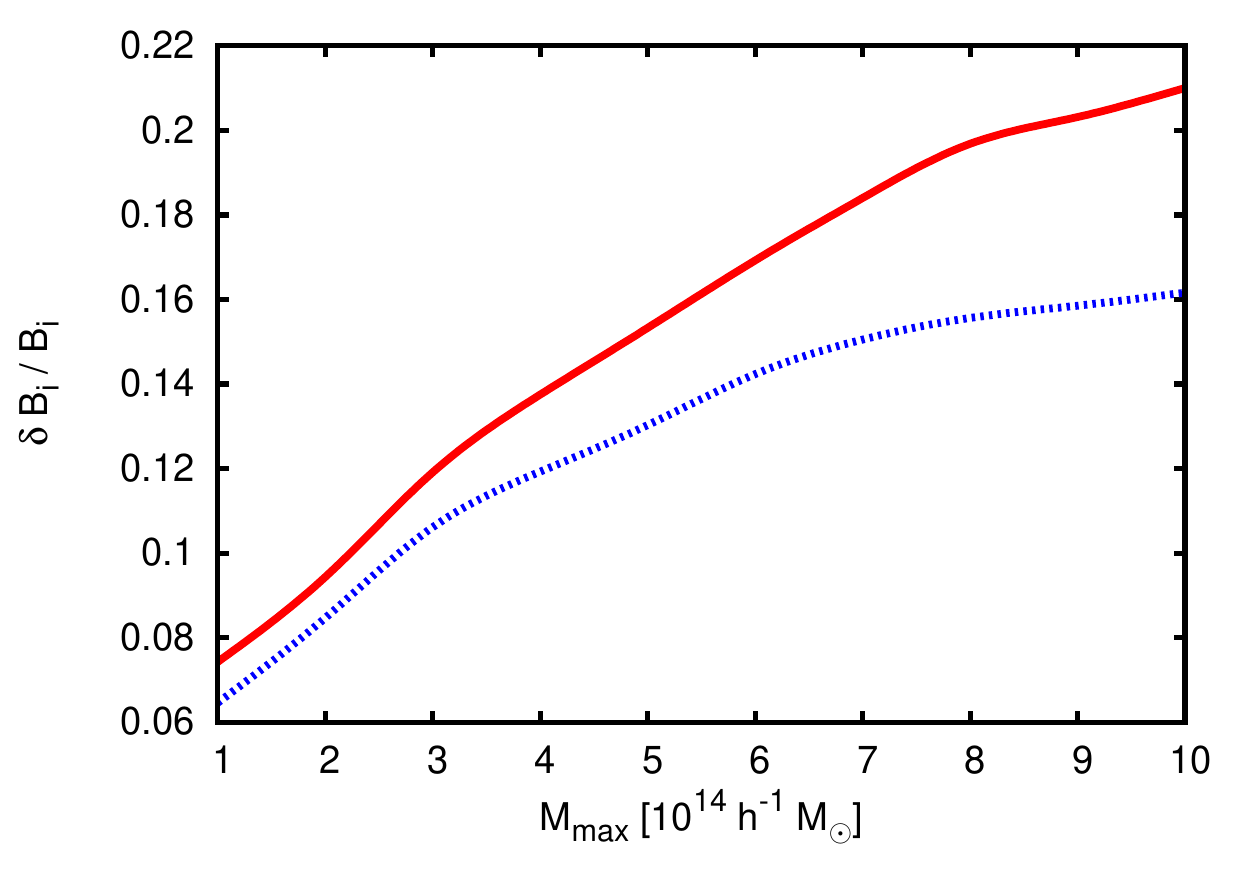}
\includegraphics[scale=0.68]{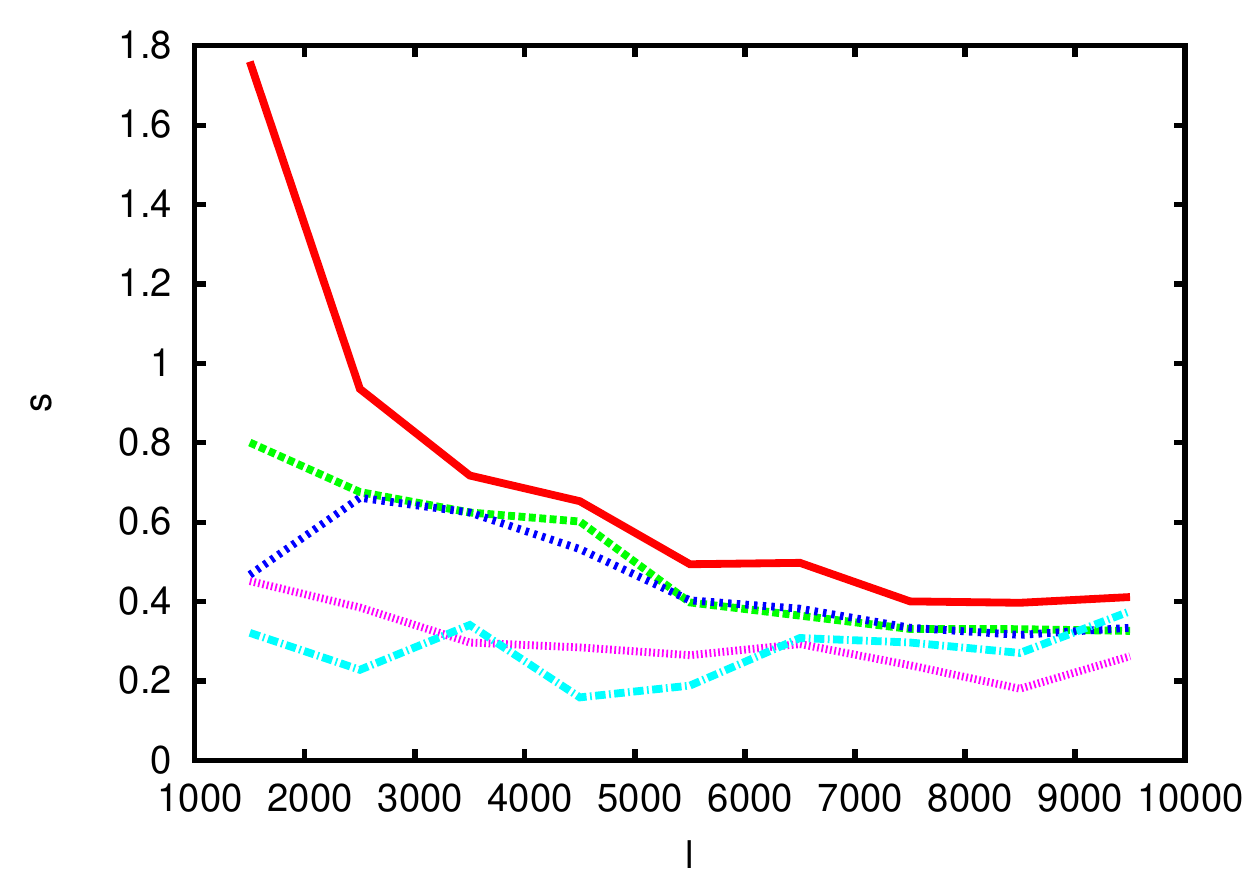}
\caption{Top: The mean power spectra from the SZ effect in the $\sigma_8=0.825$ cosmology with all clusters (red solid line) and after removing all clusters greater than $8 \times$ (green large-dashed), $6 \times$ (blue small-dashed), $4 \times$ (pink dotted) and $2 \times 10^{14} h^{-1} M_\odot$ (turquoise dot-dashed), from the top down. Middle: $\delta B_i / \bar{B}_i$ from the SZ effect as a function of the maximum mass in the realizations in two bins: $l=2000-3000$ (red solid line) and $l=3000-4000$ (blue dotted line). Bottom: the skewness $s$ of the distributions for the same mass cuts as the mean power spectra. All three of these quantities are significantly affected by the largest mass clusters in the realizations.}
\label{fig:sz_masscuts}
\end{figure}

The power spectrum of the SZ effect measured in different parts of the sky is very sensitive to the mass of the largest clusters within that field. Fields used for CMB observations might be preselected to avoid such objects, hence biasing our measurement of the power spectrum. In order to test the dependence on this, we apply a series of maximum mass cuts to the {\sc Pinocchio} catalogues; the mean spectra after these cuts are shown in Fig. \ref{fig:sz_masscuts}. Note that the effect of minimum mass cuts was discussed in Section \ref{sec:realization}.

The mean power spectrum is largely converged at maximum masses $\sim 6-8 \times 10^{14} h^{-1} M_\odot$, with the exact values depending on the multipole -- higher mass clusters contribute most to the lowest multipoles. The figure also shows the ratio of the standard deviation to the mean in the multipole bins $l=2000-3000$ and $3000-4000$ as a function of the maximum mass; for both multipole bins, the ratio increases as the maximum mass is increased. The skew also increases with the maximum mass of the cluster, approximately doubling between $M_\mathrm{max} = 10^{14} h^{-1} M_\odot$ and $10^{15} h^{-1} M_\odot$. Thus, it is the largest mass clusters that provide the largest amounts of non-Gaussianity within the power spectrum.

\subsection{Parameter dependence} \label{sec:parameter_dependence}
\begin{figure*}
\centering
\includegraphics[scale=0.58]{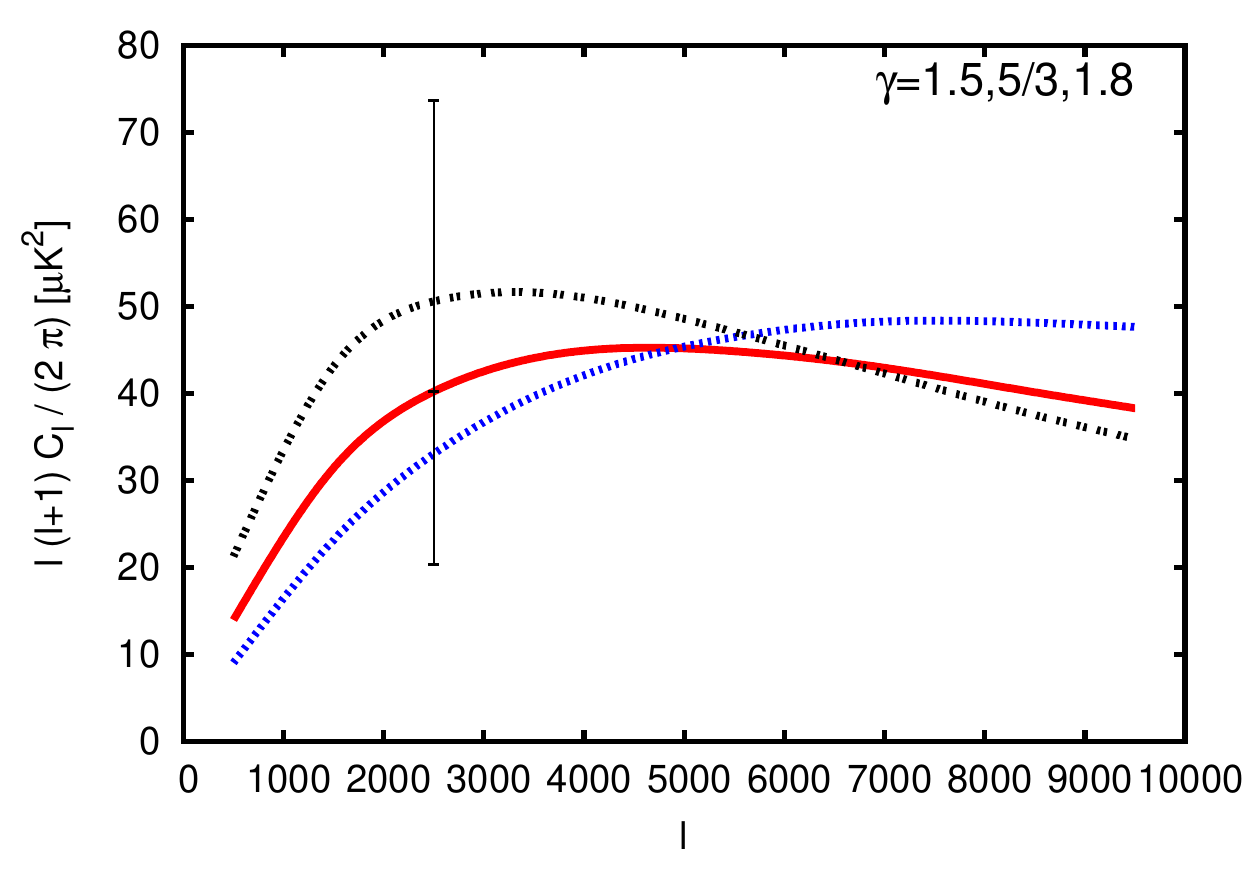}
\includegraphics[scale=0.58]{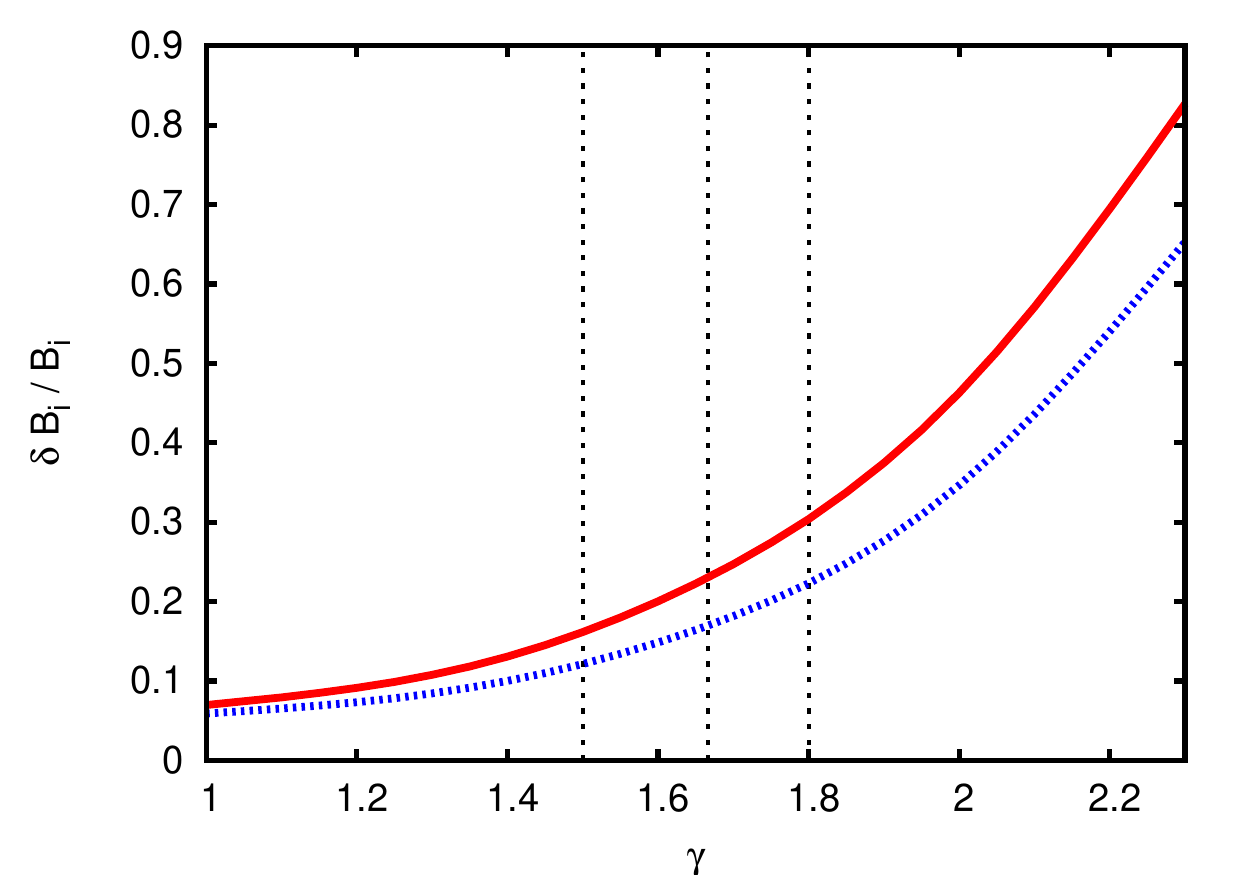}\\
\includegraphics[scale=0.58]{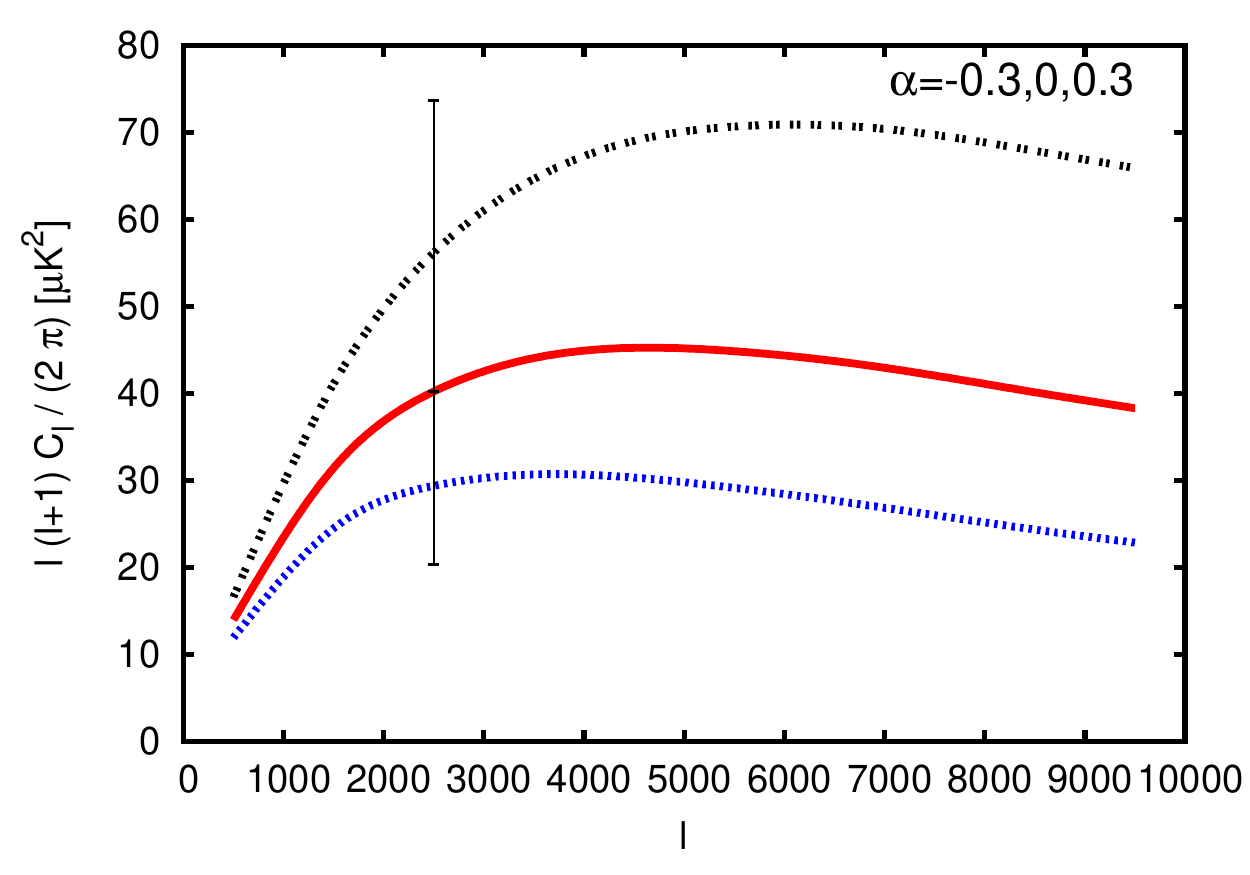}
\includegraphics[scale=0.58]{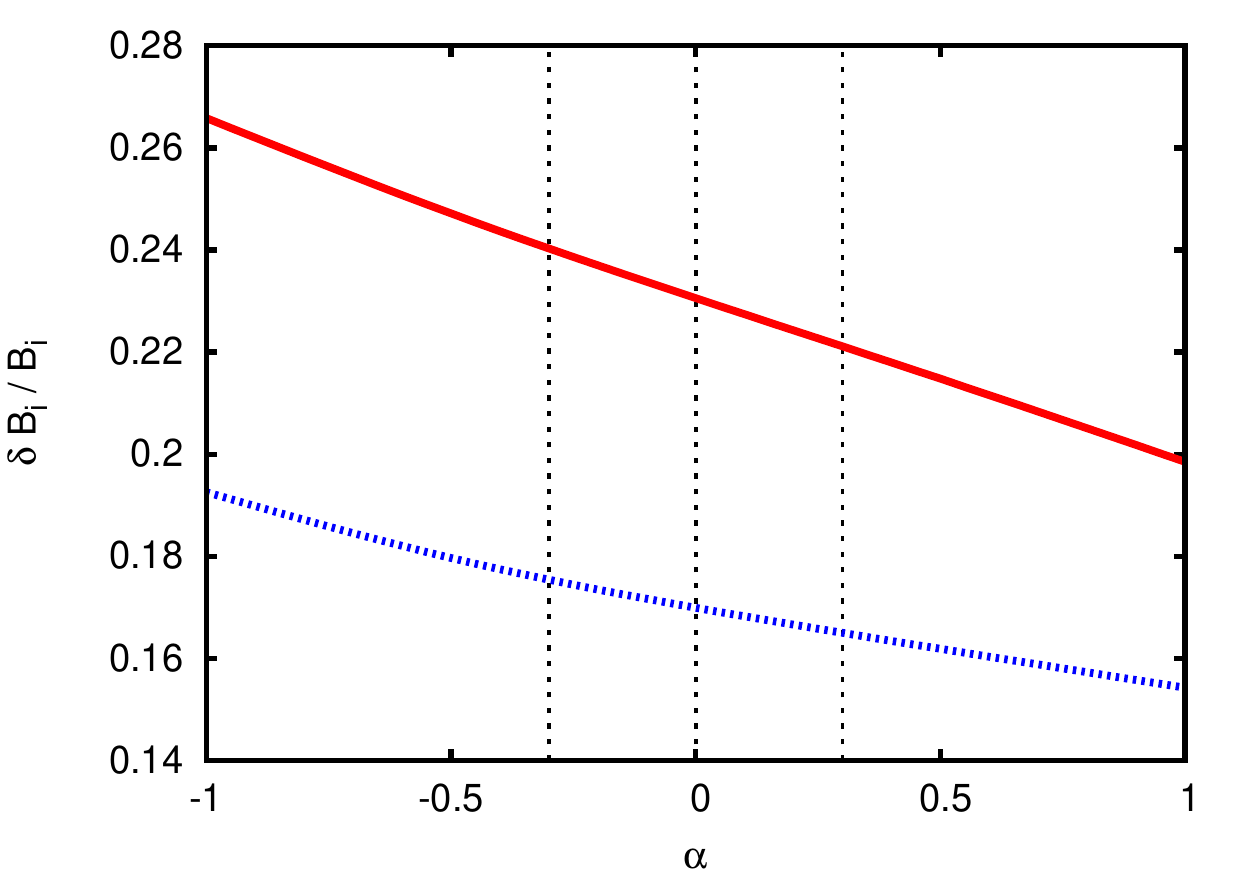}\\
\includegraphics[scale=0.58]{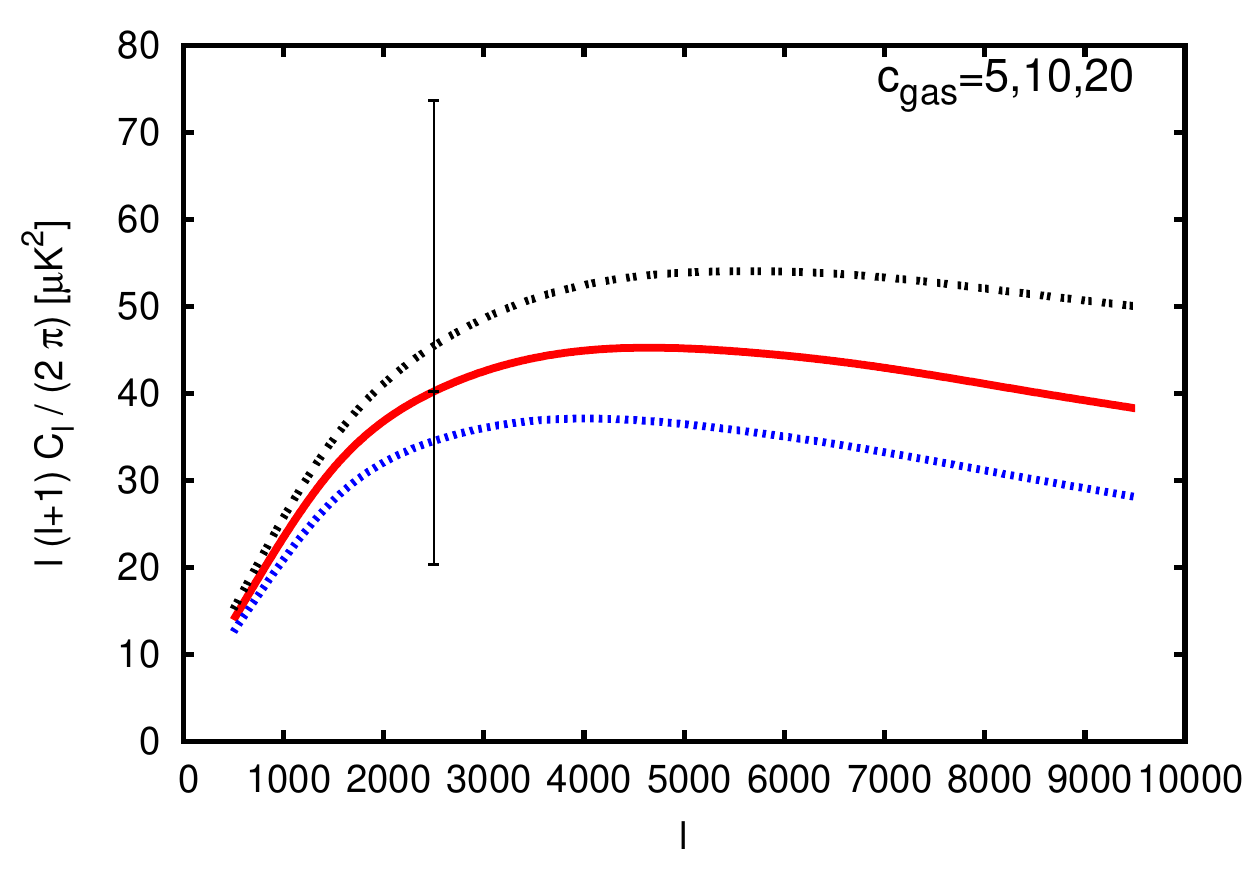}
\includegraphics[scale=0.58]{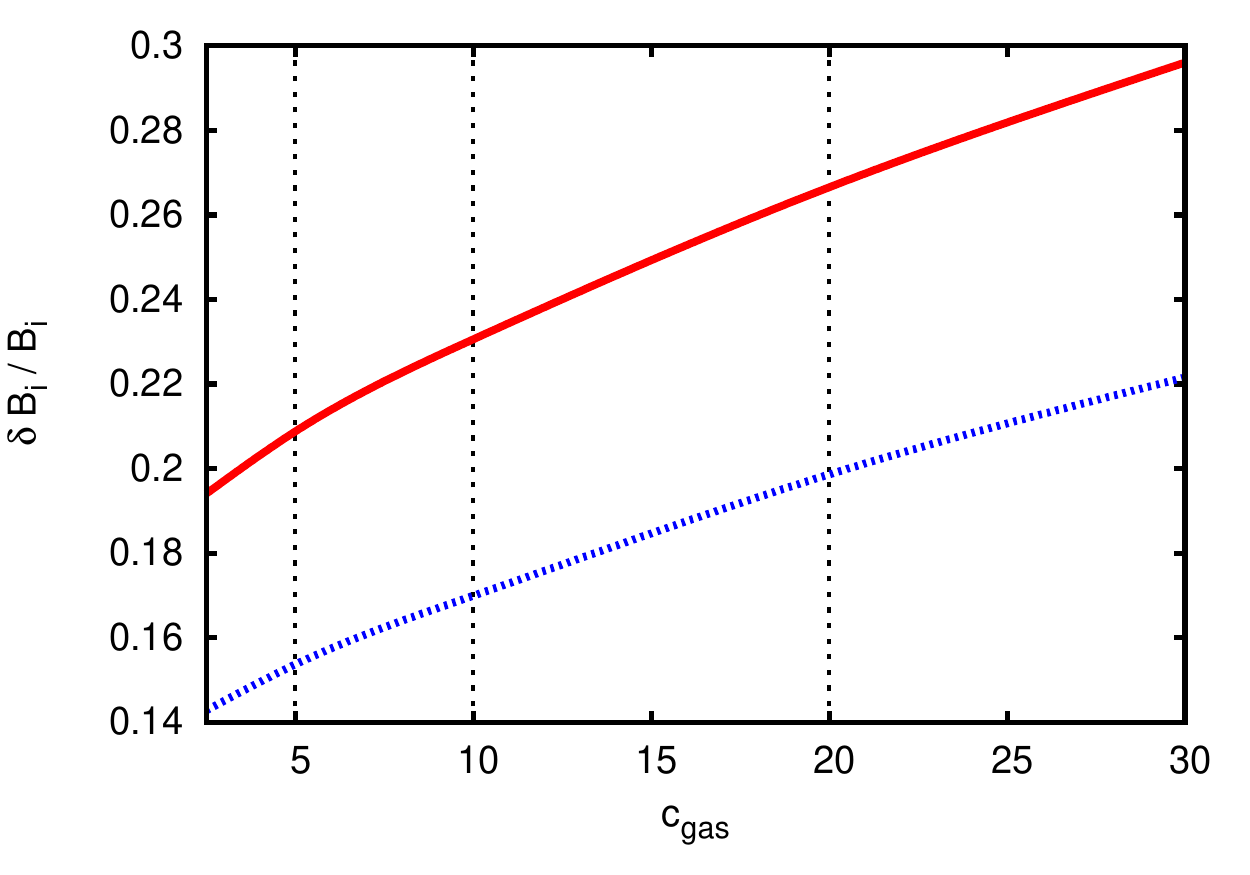}\\
\includegraphics[scale=0.58]{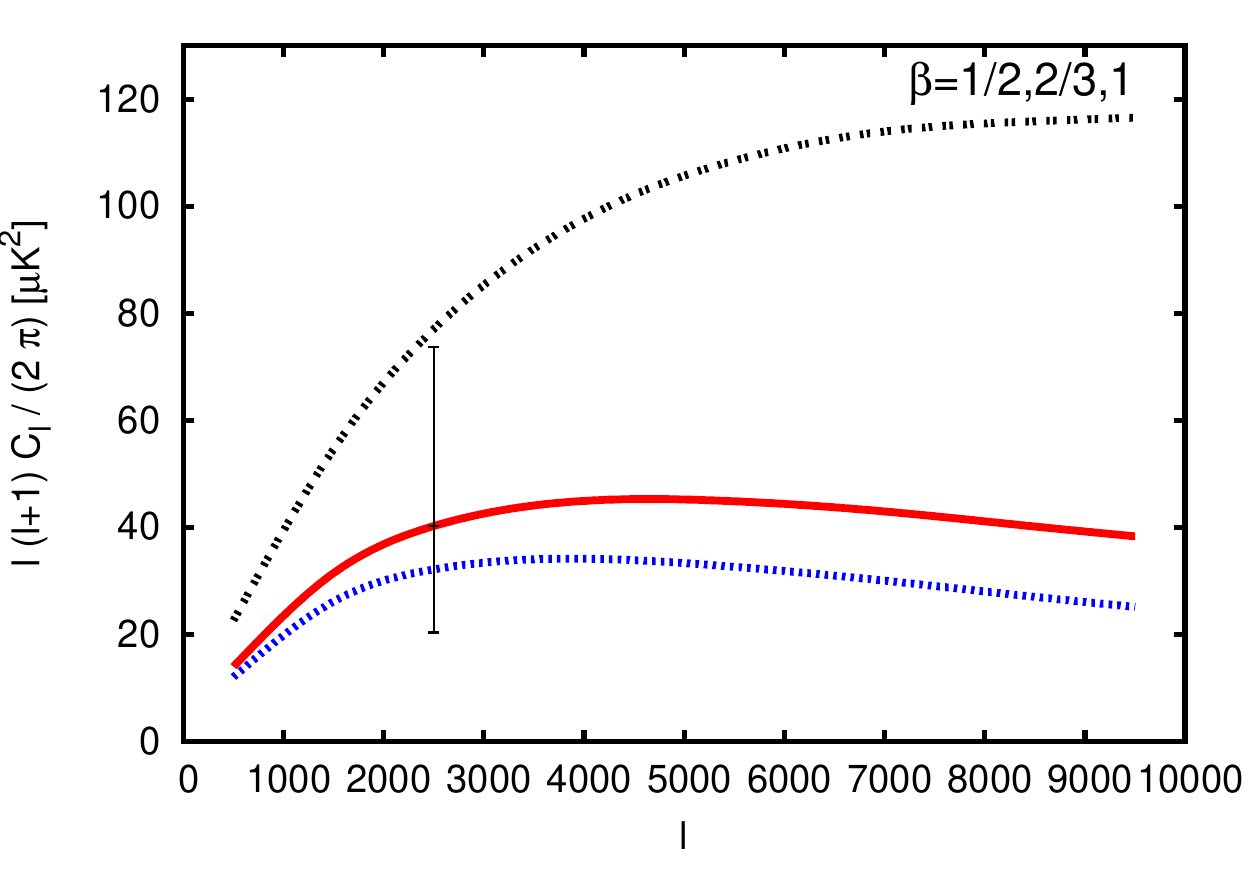}
\includegraphics[scale=0.58]{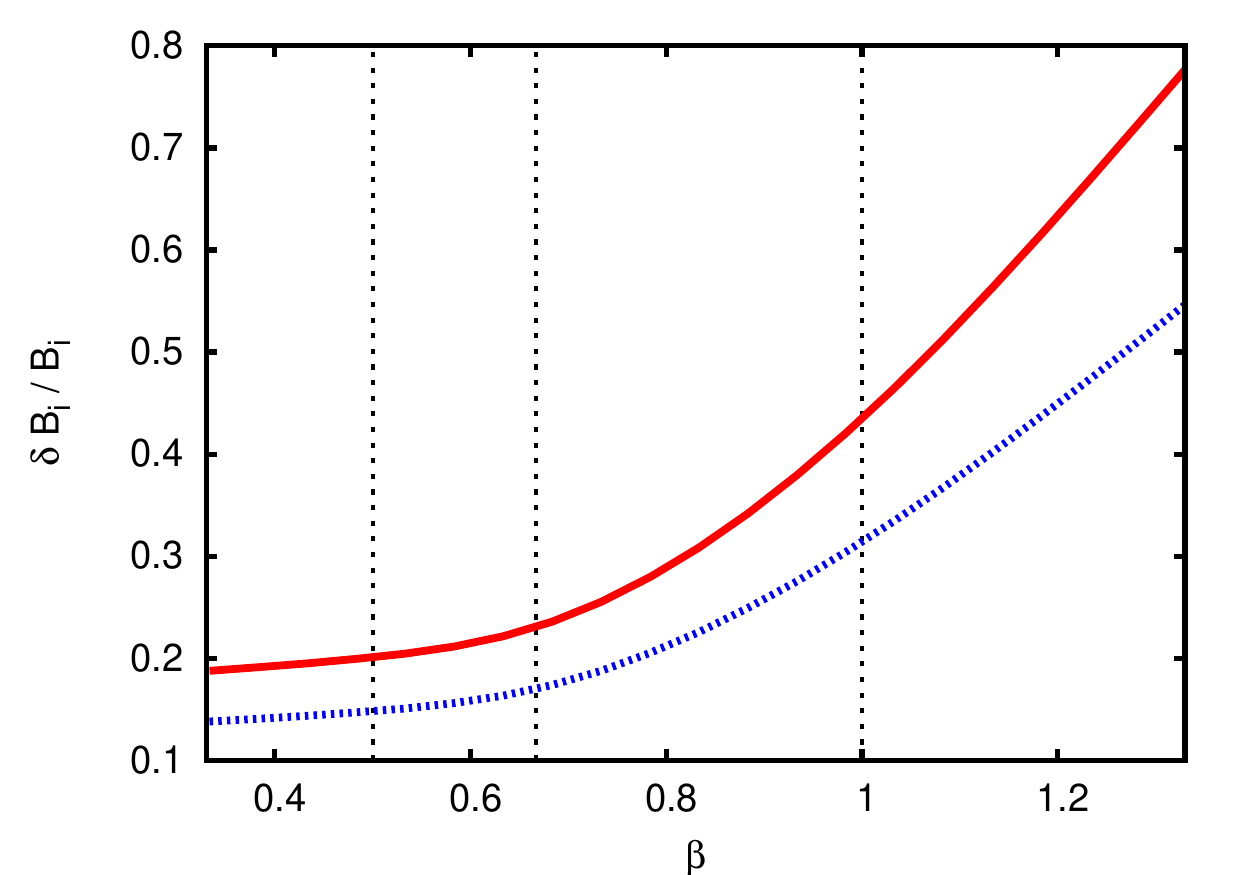}
\caption{The effect on the power spectrum from changing the models of the parameters. From top to bottom the parameters are $\gamma$ (the mass dependence), $\alpha$ (the redshift dependence), $c_\mathrm{gas}$ (the gas concentration) and $\beta$ (the outer gas profile). The left hand plots show the mean power spectrum as a function of multipole $l$. The fiducial value is shown by the red solid line, with the blue dotted line showing the lower value and the black double-dotted line showing the higher value (see text for the parameter values). The data point is the range of values from the three values of $\sigma_8$ in the 2000-3000 multipole bin. The right hand plots show $\delta B_i / \bar{B}_i$ vs. the parameter values. The solid red line is for the multipole bin $l=2000-3000$, and the blue dotted line is for $l=3000-4000$. The dashed vertical lines show the three values used in the left-hand plots. The central vertical line is the fiducial value; the $\delta B_i / \bar{B}_i$ for this value are between $0.23-0.27$ and $0.17-0.19$ for the $l=2000-3000$ and $3000-4000$ bins respectively for the three values of $\sigma_8$.}
\label{fig:param_dependence}
\end{figure*}

The five free parameters used in the cluster model (see equations \ref{eq:betamodel} and \ref{eq:clustermodel}) are not yet constrained to high accuracy, with different simulations of galaxy clusters finding different values. Observations to measure the SZ effect are still in their early stages, and no large scale SZ surveys of clusters yielding many clusters have been carried out yet. To see the effect of the uncertainty in these parameters on our results, we vary the parameters over a wide range of possible values.

\subsubsection{Parameters of the Y-M relation}
The overall amplitude of the SZ effect is determined by $Y_*$, which is fiducially set to $2 \times 10^{-6}$ Mpc$^2$. As discussed earlier, Kay et al. (in prep.) find that this can be $1.9$ or $2.3 \times 10^{-6}$ Mpc$^2$ for non-radiative or preheated gas. Changing this parameter simply scales the mean and standard deviation of the spectrum from the SZ effect, with the mean and standard deviation being multiplied by $\left( Y_{*,\mathrm{new}} / Y_{*,\mathrm{fiducial}} \right)^2$. This is because the increase does not depend on the structure or distribution of the clusters. There is no effect on normalized quantities such as $\delta B_i / \bar{B}_i$ and $s$ from the SZ effect only. Due to the increase in the power from the SZ effect, though, the cross-over point between the CMB and SZ effect will be shifted to lower multipoles as $Y_*$ is increased, which will increase $\delta B_i / \bar{B}_i$ and $s$ at those lower multipoles.

The evolution of the SZ effect as a function of mass is governed by $\gamma$. If the gas is assumed to only be gravitationally heated, then $\gamma = 5/3$. This slope could be steeper, reflecting a reduced amount of gas mass in smaller clusters, or it could be flatter due to an increase in temperature in smaller clusters from extra, non-gravitational energy. Via X-ray observations of the inner region of galaxy clusters, \citet{2007Arnaud} find $\gamma = 1.82^{+0.09}_{-0.08}$ for the $Y_X$-M relationship\footnote{$Y_X = M_\mathrm{gas} T_X$ is an X-ray proxy for the integrated SZ effect $Y$ that has recently been shown to correlate well with $M$ with little scatter \citep{2006Kravtsov}.}, in agreement with the preheating result from Kay et al. (in prep.). As this is measured for the inner region, whilst the SZ effect extends much further out in the cluster atmosphere, this is likely to be an upper limit. As such, we use $\gamma = 1.5$ and $1.8$ as the expected range for this parameter. The effect of these values on the power spectrum and $\delta B_i / \bar{B}_i$ is shown in the top row of Fig. \ref{fig:param_dependence}.

As $\gamma$ is increased, the peak in the SZ power spectrum shifts to lower multipoles, and $\delta B_i / \bar{B}_i$ increases. This is because the power from clusters with a mass over $10^{14} h^{-1} M_\odot$ will be increased as $\gamma$ increases, whilst that from smaller mass clusters will decrease. The larger clusters are dominant at lower multipoles due to their greater angular size, so the peak of the SZ effect will naturally shift to lower multipoles. The effect is similar to simply increasing the mass of all the clusters. Section \ref{sec:upper_mass} showed that $\delta B_i / \bar{B}_i$ increases as larger clusters are included in the maps; the same effect is present here, with high values of $\gamma$ providing the largest increase in the effective mass, hence the largest values of $\delta B_i / \bar{B}_i$.

The SZ effect is expected to be self-similar, that is, it will depend on the redshift of the cluster via the evolution of the mean matter density within the virial radius of the cluster as well as the Friedmann equation. The parameter $\alpha$ models evolution beyond that from self-similarity, and is expected to be close to zero. Using numerical simulations, \citet{2004Silva} find no evidence of evolution from self-similarity.

To illustrate the effect of changing this parameter, we choose a modest range of values: $\alpha = -0.3$ and $0.3$; the results are presented in the second row of Fig. \ref{fig:param_dependence}. Higher values of $\alpha$ will increase the SZ effect from clusters at the highest redshift. As the clusters at higher redshifts have a smaller angular size, this affects the power in the higher multipoles more than that in the lower ones. There are more clusters at higher redshifts, so when these provide increased power compared with the low redshift clusters the maps will become more similar to each other - it will become more difficult for a single cluster to stand out - hence the standard deviation of the maps will decrease.

\subsubsection{Parameters of the gas profile}
The final two parameters govern the profile of the SZ effect. The concentration of the gas within the cluster is determined by $c_\mathrm{gas}$, and  the slope of the cluster profile is governed by $\beta$ (see equation \ref{eq:betamodel}). For $\beta$ we use a fiducial value of $2/3$, which is in agreement with observations of X-ray emission from nearby clusters by \citet{2006Vikhlinin} that find $\beta \sim 0.6-0.9$ at $r_{500}$ (approximately half of the virial radius). Additionally, \citet{2008Croston} find a range of $\beta \sim 0.37-0.81$ for a range of radii between 0.3 and 0.8 $R_{500}$. However, steeper values than these are expected in the outer regions where the gas pressure profile is expected to trace the NFW profile; this is demonstrated by \citet{2007Hallman}, who use simulations of clusters to compare X-ray and SZ profiles, finding that $\beta=0.88$ for X-ray but $1.13$ for the SZ effect. As a result, we investigate the effects of $\beta = 1/2$ and $1$. For $c_\mathrm{gas}$, we use a fiducial value of $10$ and also try $5$ and $20$. Duffy et al. (in prep.) indicates that $c_\mathrm{gas}$ is similar to $c_\mathrm{DM}$ and depends on the cluster mass by around 10 per cent above $10^{14} M_\odot$, such that the values will lie around $5-10$.

The effects of $c_\mathrm{gas}$ and $\beta$ are shown in the bottom two rows of Fig. \ref{fig:param_dependence}. As these parameters only change the profile of the SZ effect, the average pixel value within the maps remains constant. The SZ effect is more concentrated within the centre of the cluster for higher values of $\beta$ and $c_\mathrm{gas}$, with the effect of increasing the mean power spectra across all multipoles, but particularly at the higher multipoles, in a similar way to the effect of point sources on the power spectrum (see section \ref{sec:ps}). As the parameters are increased, the ratio of the standard deviation to the mean also increases, as the power from the central part of the largest (and rarest) clusters becomes more important at these multipoles.

The assumption that all clusters are the same is unlikely to be true; in reality, there will be a certain amount of scatter within the cluster parameters, which may increase the standard deviation further when looking at smaller areas of the sky. Clusters are also likely to deviate from spherical symmetry. The extent of this scatter and its impact on the SZ effect is not yet known.

\subsubsection{Comparison to the effects of $\sigma_8$}
The changes in the gas physics have a comparable effect on the mean power spectrum as the range of $\sigma_8$ considered in this paper, especially for the outer gas profile $\beta$ and the evolution of the $Y-M$ relationship with redshift ($\alpha$). At the same time, the gas physics can substantially increase the standard deviation of the power spectrum to a much larger degree than is possible with $\sigma_8$, providing a much greater scatter in the power from different maps. These effects are obviously different to those from $\sigma_8$ when the whole of the power spectrum can be considered, but will be difficult to separate when only a few bins are measured.

\section{Point sources}\label{sec:ps}
Foreground emission from extragalactic point sources will contaminate the map. Synchrotron is the dominant process at frequencies below around 90~GHz, with emission from dust becoming important at higher frequencies.  Although these sources are physically extended, this is generally much smaller than the beam size of the telescopes used to measure the CMB and SZ effect, so that they are well approximated as point sources.

In our earlier results, we have not inserted point sources as CMB experiments generally attempt to subtract them. This procedure is, however, subject to uncertainties. Here, we discuss the effects of the inclusion of subdominant point source contributions for specific flux limits. We insert point sources statistically into these realizations, based on extrapolation of their observed number counts.

We calculate the total number of point sources with a flux density between $S_{\mathrm{min}}$ and $S_{\mathrm{max}}$ within an area $\Delta \Omega$ on the sky by
\begin{equation}
N_{\mathrm{tot}} = \Delta \Omega \int_{S_{\mathrm{min}}}^{S_{\mathrm{max}}} \frac{dN}{dS_{\nu}} dS_{\nu},
\end{equation}
where $dN/dS_\nu$ is the differential source counts as a function of flux density. To calculate the flux densities of individual sources, we randomly sample from a probability distribution defined by a normalized $dN/dS_\nu$ between the flux density limits.

The flux densities can be converted to the brightness temperatures at a given frequency $\nu$ using
\begin{equation}
T_\nu =S_{\nu} / \left( \theta_{\mathrm{pixel}}^{2} \frac{dB}{dT} \right),
\end{equation}
where $\theta_{\mathrm{pixel}}^{2}$ is the area of the pixel containing the point source and 
\begin{equation}
\frac{dB}{dT} = \frac{2k}{c^{2}} \left( \frac{kT_{\mathrm{CMB}}}{h} \right) ^{2} \frac{x^{4} e^{x}}{(e^{x} -1)^{2}},
\end{equation}
is the differential of the Planck function with respect to temperature, where $x$ is the dimensionless frequency, given by $x = h_\mathrm{P} \nu / k_{\mathrm{B}} T_{\mathrm{CMB}}$.

\subsection{Low frequency point sources}
\begin{figure}
\centering
\includegraphics[scale=0.68]{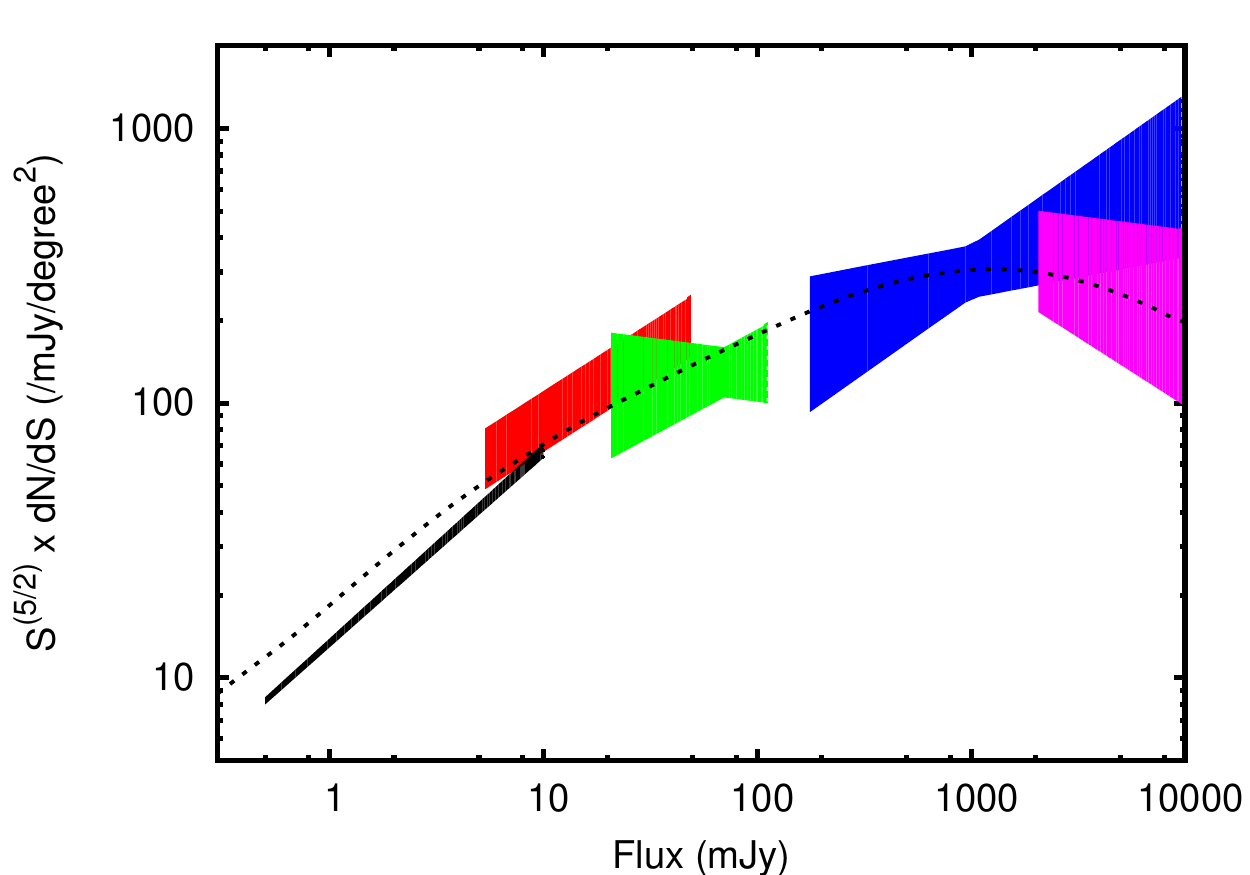}
\caption{Comparison of the 30GHz differential source counts from \citet{1998Toffolatti} rescaled by a factor of 0.7 (black dotted line) with the $1 \sigma$ ranges from measurements by (right-to-left)  {\it WMAP} Ka band \citep[][pink]{2003Bennett}, DASI \citep[][blue]{2002Kovac}, VSA \citep[][green]{2005Cleary}, CBI \citep[][red]{2003Mason} and GBT \citep[][black]{2009Mason}}
\label{fig:pointsource_comparison}
\end{figure}

\begin{table*}
\begin{center}\begin{tabular}{ccccccccccccc}
& \multicolumn{3}{c}{CMB+SZ}& & \multicolumn{3}{c}{CMB+SZ+PS} & & \multicolumn{3}{c}{CMB+SZ+CPS}\\
Multipole bin & $\bar{B}_i$ & $\delta B_i$ & $s$  & & $\bar{B}_i$ & $\delta B_i$ & $s$  & & $\bar{B}_i$ & $\delta B_i$ & $s$ \\
\hline 
1000-2000 & 720 & 56 & 0.40 & & 730 & 56 & 0.20 & & 720 & 57 & 0.25\\
2000-3000 & 150 & 13 & 0.59 & & 160 & 13 & 0.61 & & 150 & 13 & 0.65\\
3000-4000 & 58 & 7.7 & 0.70 & & 85 & 8.1 & 0.57 & & 75 & 7.7 & 0.73\\
4000-5000 & 47 & 6.1 & 0.68 & & 91 & 6.9 & 0.40 & & 82 & 6.3 & 0.56\\
5000-6000 & 45 & 5.0 & 0.50 & & 110 & 6.2 & 0.29 & & 100 & 6.0 & 0.30\\
6000-7000 & 44 & 4.3 & 0.49 & & 140 & 6.1 & 0.17 & & 130 & 6.0 & 0.23\\
7000-8000 & 42 & 3.7 & 0.40 & & 160 & 6.6 & 0.06 & & 160 & 6.3 & 0.23\\
8000-9000 & 40 & 3.2 & 0.39 & & 200 & 7.4 & 0.14 & & 190 & 7.0 & 0.09\\
9000-10000 & 38 & 2.8 & 0.41 & & 230 & 8.0 & 0.0 & & 230 & 8.2 & -0.03\\
\hline 
1000-2000 & 720 & 170 & 0.90 & & 730 & 160 & 0.72 & & 720 & 160 & 0.71\\
2000-3000 & 150 & 38 & 2.1 & & 160 & 40 & 2.0 & & 150 & 37 & 1.9\\
3000-4000 & 58 & 23 & 2.6 & & 84 & 24 & 2.2 & & 74 & 22 & 2.4\\
4000-5000 & 47 & 18 & 2.0 & & 91 & 20 & 1.5 & & 80 & 19 & 1.7\\
5000-6000 & 45 & 15 & 1.8 & & 110 & 19 & 1.2 & & 100 & 17 & 1.2\\
6000-7000 & 44 & 13 & 1.7 & & 140 & 19 & 0.84 & & 120 & 18 & 0.82\\
7000-8000 & 42 & 11 & 1.4 & & 170 & 20 & 0.49 & & 150 & 19 & 0.60\\
8000-9000 & 40 & 9.9 & 2.0 & & 200 & 22 & 0.30 & & 190 & 21 & 0.43\\
9000-10000 & 38 & 8.3 & 1.1 & & 230 & 24 & 0.34 & & 220 & 23 & 0.32\\
\hline 
1000-2000 & 700 & 52 & 0.21 & & 700 & 55 & 0.37 & & 700 & 55 & 0.11\\
2000-3000 & 120 & 7.7 & 0.26 & & 120 & 7.7 & 0.16 & & 120 & 7.8 & 0.18\\
3000-4000 & 24 & 2.0 & 0.51 & & 33 & 2.3 & 0.34 & & 33 & 2.3 & 0.47\\
4000-5000 & 13 & 1.5 & 0.66 & & 28 & 1.9 & 0.43 & & 27 & 1.8 & 0.39\\
5000-6000 & 11 & 1.2 & 0.47 & & 33 & 1.8 & 0.29 & & 33 & 1.8 & 0.18\\
6000-7000 & 11 & 1.0 & 0.49 & & 41 & 1.8 & 0.21 & & 41 & 2.0 & 0.02\\
7000-8000 & 10 & 0.89 & 0.39 & & 51 & 2.2 & -0.10 & & 51 & 2.3 & 0.03\\
8000-9000 & 10 & 0.76 & 0.39 & & 62 & 2.5 & 0.12 & & 62 & 2.6 & 0.13\\
9000-10000 & 9 & 0.67 & 0.42 & & 75 & 3.0 & 0.16 & & 75 & 3.0 & 0.14\\
\end{tabular}
\caption{The statistics of the $\sigma_8 = 0.825$ realizations containing CMB, SZ and point sources. The mean $\bar{B}_i$ and the standard deviation $\delta B_i$ within the bin $i$ are in $\mu$K$^2$; the skew $s$ is dimensionless. The top set are for $3 \times 3$ degree maps at 30~GHz using 0.2~mJy point sources; the middle for $1 \times 1$ degree maps also at 30~GHz and 0.2~mJy point sources, and the bottom set for $3 \times 3$ degree maps at 150~GHz with 2~mJy point sources. The three sets of values given are for no point sources (left), randomly positioned point sources (middle) and clustered point sources (right). Adding point sources increases the mean and standard deviation at high multipoles, as expected. Clustered point sources fill in some of the decrement from the SZ effect, reducing the mean and standard deviation.}
\label{tab:ps_stats_0.2mjy}
\end{center}
\end{table*}

For the low frequency point sources we use the 30~GHz model for $dN/dS_\nu$ from \citet[][henceforth T98]{1998Toffolatti}. This is an extrapolation of the observed number counts at 1.4~GHz, which have been measured to much lower flux densities than the number counts at the frequencies of interest here. We have normalized the model by a factor of 0.7 \citep[following][]{2005Cleary} to bring it into closer agreement with measurements of the number count at 30~GHz (see Fig. \ref{fig:pointsource_comparison}), although this still results in an over-prediction of the number of sources around 1~mJy compared with recent observations using the Green Bank Telescope \citep{2009Mason}.

To scale the flux densities to the desired frequency $\nu$, we use $S_{\nu} / S_{\mathrm{30GHz}} = (\nu / 30\mathrm{GHz})^{\alpha}$. $\alpha$ is determined for each source using a Gaussian distribution about a mean $\overline \alpha = -0.3$, with $\sigma_{\alpha} = 0.36$, for each source, which is the distribution for 15 to 30~GHz measured for the 9C sample (Cleary, private communication). We include point sources with fluxes between $10^{-4.5}$ and $10^{5}$~mJy at 30~GHz; sources above this flux range are very rare. Sources with weaker fluxes are not well characterized and should have a negligible contribution to the power spectra.

In addition to the total number of point sources and their flux densities, we need positions. The simplest method is to distribute them randomly over the map, so that the probability of finding a source at a particular position, $P(x,y)$, is constant across the map. However, galaxies emitting synchrotron radiation are thought to reside preferentially within galaxy clusters. Observations at 30~GHz by \citet{2007Coble} suggest that the number counts of these sources increase in the direction of known massive clusters; \citet{2007Lin} come to a similar conclusion at 1.4~GHz. To account for this, we correlate the positions of these sources with the galaxy clusters by using a power of the surface mass density $\Sigma (x,y)$ of the clusters. In particular we use the probability distribution
\begin{equation}
P(x,y) = \frac{\Sigma (x,y)^{1/3}}{\int \Sigma(x^\prime,y^\prime)^{1/3} \phantom{.} dx^\prime \phantom{.} dy^\prime}.
\end{equation}
The specific choice of $\Sigma (x,y)^{1/3}$ brings the number counts towards clusters into reasonable agreement with those estimated by \citet{2007Coble}.

\subsection{High frequency point sources}
Although negligible at the fiducial frequency of 30~GHz, dusty galaxies become important at higher frequencies. For these, we use the fit to SCUBA observations provided by \citet{2003Borys}. This is in the form of a double power law,
\begin{equation}
\frac{dN}{dS_{\nu}} = \frac{N_{0}}{S_{0}} \left[ \left( \frac{S}{S_{0}} \right) ^{\alpha} + \left( \frac{S}{S_{0}} \right) ^{\beta} \right]^{-1},
\end{equation}
where the parameters at 350GHz are $S_{0} = 1.8 \mathrm{mJy}$, $\alpha = 1.0$, $\beta = 3.3$ and $N_{0} = 1.5 \times 10^{4} \phantom{.} \mathrm{deg}^{-2}$. This is extrapolated to other frequencies using $S_{\nu} = \nu^{\gamma}$, where $\gamma = 2.5$. We use the flux range $10^{1}$ to $10^{5}$~mJy at 350~GHz. These sources are expected to be clustered, but they are at higher redshifts than the galaxy clusters we are interested in. Thus, they are randomly distributed in these realizations.

\subsection{Power spectrum statistics with point sources}
\begin{figure}	
\centering
\includegraphics[scale=0.80,viewport=80 30 300 215,clip]{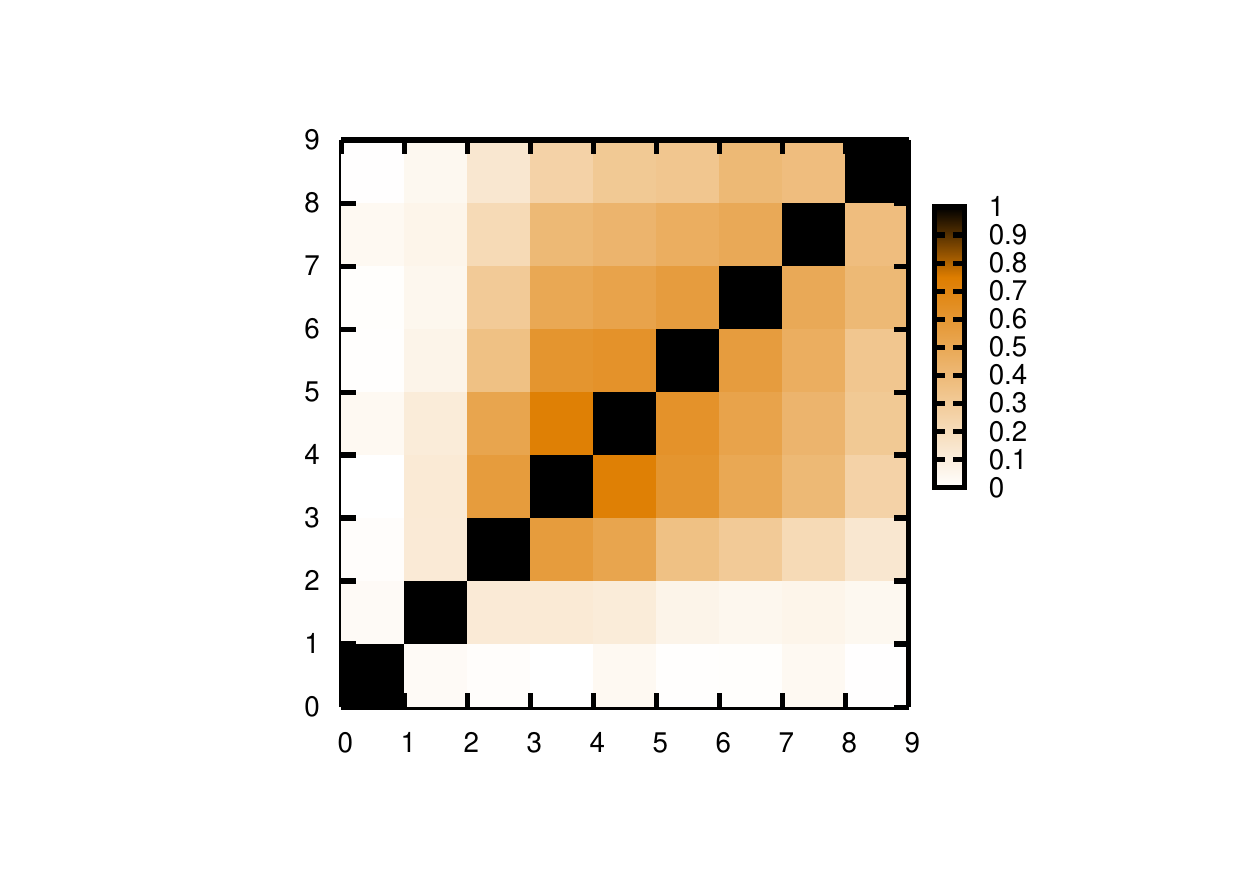}
\caption{The correlation matrix for 1000 realizations of the combined CMB, SZ effect and 0.2~mJy point sources at 30~GHz, using $3 \times 3$ degree maps. The addition of point sources decreases the off-diagonal terms (compare with Fig. \ref{fig:sz_covariance}).}
\label{fig:ps_covariance}
\end{figure}

Table \ref{tab:ps_stats_0.2mjy} shows the power spectrum statistics for maps containing point sources below 0.2~mJy in addition to the CMB and SZ effect. The addition of point sources naturally increases the mean and standard deviation of the spectra at higher multipoles. The point sources also decrease the normalised skewness and the off-diagonal terms of the correlation matrix at 30~GHz (see Fig. \ref{fig:ps_covariance}) as they can reduce the amount of power from the SZ effect by filling in the decrements \citep{2002Holder}. Clustering the point sources increases the chance of this happening, resulting in a slightly lower mean values for the power spectrum and a slightly reduced standard deviation. The effect at 150~GHz is similar to that at 30~GHz, except that the effect of clustering the low frequency point sources is reduced as the unclustered infrared sources are becoming more important.

We do not attempt to account for the effects of imperfectly removing strong point sources from the realizations, which may result in a distribution of effective point sources with a flux below the cut-off flux. This could be an important issue, especially if the point sources lie in galaxy clusters, in which case the residual point source flux could fill in the decrement from the SZ effect.

\section{Implications for the high multipole excess}

The original, and highest significance, measurement of a possible excess at large multipoles comes from the Cosmic Background Imager. An excess was first announced by \citet{2003Mason} and was refined by the addition of more observations by \citet{2003Pearson} and \citet{2004Readhead}. The final results were announced by \citet{2009Sievers} using the full 5 years of observations with the CBI. In total, the CBI has observed five $5 \times 5$ degree fields and one $5 \times 6$ degree field using mosaicked, shallow observations. Additionally, one $5 \times 0.75$ degree and three $0.75 \times 0.75$ degree ``deep'' fields were observed, which have reduced noise as they have been observed for longer. Following from \citet{2005Bond}, \citet{2009Sievers} find that if this excess is due to the SZ effect, then $\sigma_8$ must be between 0.9 and 1.0.

We do not attempt to carry out a direct comparison between our results and the CBI measurements as we omit a number of factors that will be important in this comparison. These include the difference in map sizes and the combination method - we investigate the statistics of $3 \times 3$ and $0.75 \times 0.75$ degree maps separately, whereas the CBI measurement results from the combination of a number of different areas of the sky and different map sizes. We do not take into account the effects of the UV coverage, beam shapes, nor the window function for the observations. We also exclude any effects of noise, point source subtraction and any other foregrounds such as galactic emission or atmospheric effects. Finally, we only briefly cover the effects of gas physics and limits on the maximum cluster masses on the statistics. As such, our results are only indicative.

\begin{figure}
\centering
\includegraphics[scale=0.68]{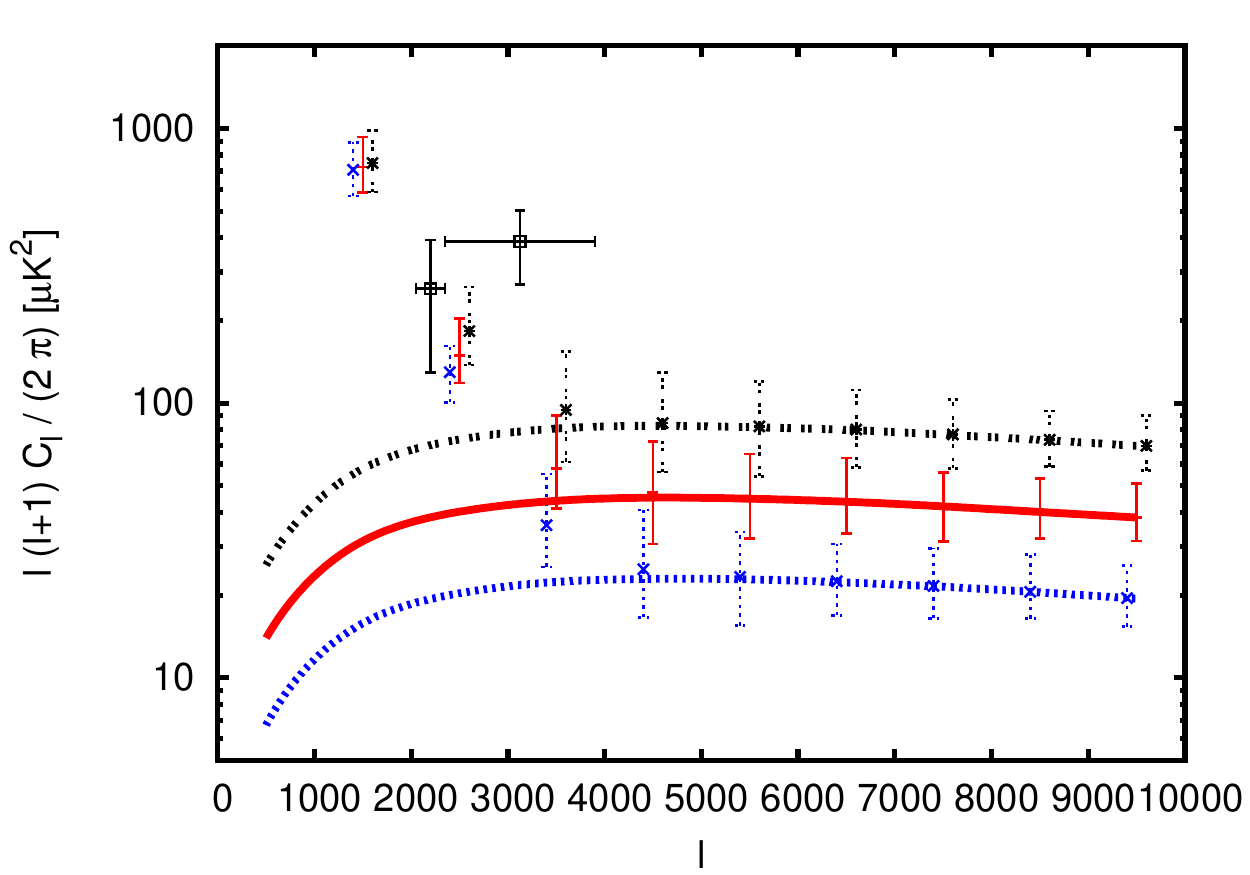}
\caption{The power spectra from the CMB+SZ effect for the three cosmologies, with the error bars showing the minimum and maximum values from all 1000 realizations with a map size of $3 \times 3$ degrees. The black double-dotted points (offset by 100 multipoles for clarity) are for $\sigma_8 = 0.9$, the red solid points for $\sigma_8 = 0.825$ and the blue dotted points (also offset by 100 multipoles) are for $\sigma_8 = 0.75$. The two black data points are the latest values for the highest multipole bins from CBI \citep{2009Sievers}. The three solid lines are (from top to bottom) the  mean power spectra from the SZ effect in with $\sigma_8=0.9$ (black double-dotted line), $0.825$ (red solid line) and $0.7$ (blue dotted line).}
\label{fig:cmb_sz_powerspectra}
\end{figure}

We concentrate on two of the bins measured by the CBI, namely $l=2050-2350$ and $l=2350-3900$; these are the bins where the SZ effect will be most significant. For these bins, the CBI has measured $261 \pm 132 \mu$K$^2$ and $387 \pm 117 \mu$K$^2$ respectively. They find similar values, but with larger error bars, when the deep fields are included or considered on their own, and find low significance excesses in each of the four fields seperately. The CBI data points are shown in Fig. \ref{fig:cmb_sz_powerspectra}, along with the power spectra from the SZ effect on its own for each three values of $\sigma_8$, as well as the mean values from the combined CMB and SZ realizations for $3 \times 3$ degree maps with the spectrum binned with bin widths of 1000; the minimum and maximum values from the realizations are shown by the error bars. The highest CBI data point is clearly in excess over what would be expected from the mean and scatter of these realizations.

\begin{figure}
\centering
\includegraphics[scale=0.68]{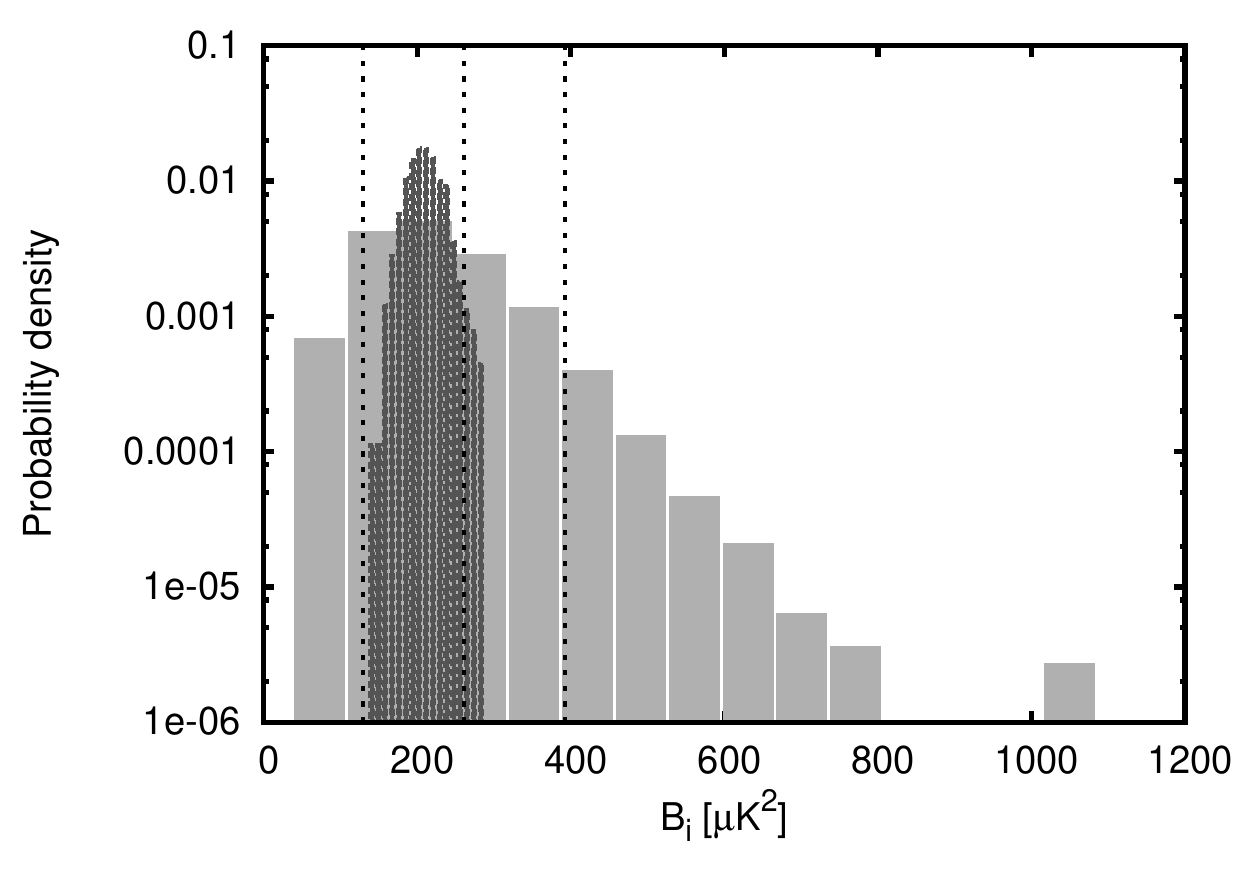}
\caption{The statistics of the combined CMB and SZ effect realizations with $\sigma_8 = 0.825$ for the multipole bin 2050-2350. The large grey histogram is for the $0.75 \times 0.75$ degree realizations (bin width of 75 $\mu$K$^2$); the small, darker histogram is for the $3 \times 3$ degree realizations (bin width of 9 $\mu$K$^2$). The CBI value for this bin is $261 \pm 132 \mu$K$^2$ (mean and $\pm 1 \sigma$ represented by the three dashed vertical lines); 26.5 per cent of the $0.75\times0.75$ degree realizations have this central power or higher, and 5 per cent have $376.5 \mu$K$^2$ or higher.}
\label{fig:cbi_histogram_lowerbin}
\end{figure}

\begin{figure}
\centering
\includegraphics[scale=0.68]{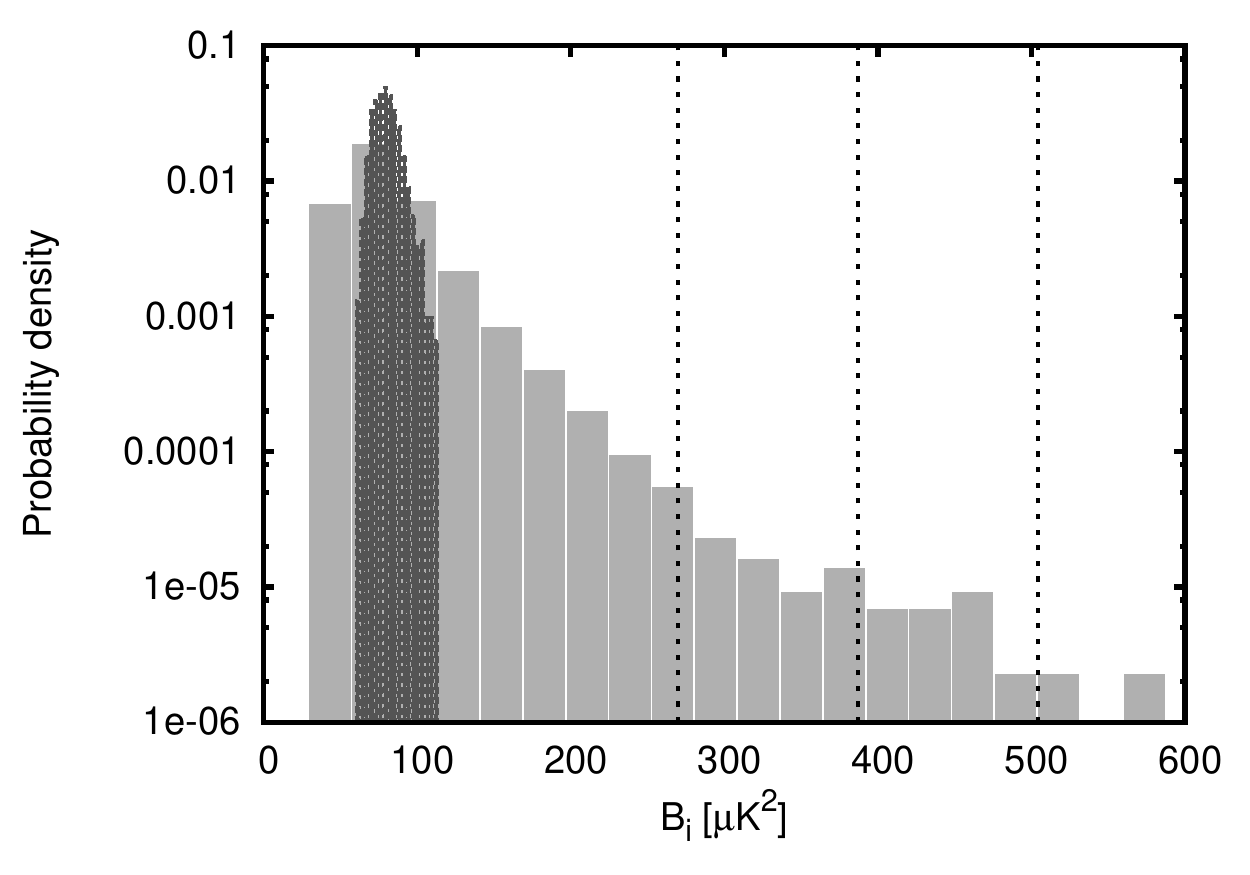}
\caption{As Fig. \ref{fig:cbi_histogram_lowerbin}, but for the multipole bin 2350-3900 with bin widths of 28 and 3 $\mu$K$^2$ for the $1 \times 1$ and $3 \times 3$ degree realizations respectively. The CBI value for this bin is $387 \pm 117$ $\mu$K$^2$ (mean and $\pm 1 \sigma$ represented by the three dashed vertical lines); 0.36 per cent of the $0.75 \times 0.75$ degree maps have this central power or higher, with 5 per cent having $137.7 \mu$K$^2$ or higher.}
\label{fig:cbi_histogram_upperbin}
\end{figure}

\begin{table*}
\begin{center}
\begin{tabular}{lcccccccccccc}
& & & & \multicolumn{4}{c}{$3 \times 3 \deg$} & & \multicolumn{4}{c}{$0.75 \times 0.75 \deg$}\\
\hline
Components & $\sigma_8$ & $S_\mathrm{max}$ & & $B_i$ & $\delta B_i$ & $s_i$ & \% & & $B_i$ & $\delta B_i$ & $s_i$ & \%\\
\hline
CMB only & -- & -- & & 170 & 17 & 0.19 & 0 & & 180 & 65 & 0.77 & 12.1\\
CMB, SZ & 0.75 & -- & & 190 & 20 & 0.27 & 0 &  & 200 & 74 & 1.1 & 18.7\\
CMB, SZ & 0.825 & -- & & 210 & 23 & 0.34 & 1.4 & & 220 & 86 & 1.5 & 26.5\\
CMB, SZ & 0.9 & -- & & 240 & 30 & 0.40 & 24.6 &  & 250 & 110 & 3.0 & 38.1\\
CMB, ST99 SZ & 0.825 & -- & & 210 & 23 & 0.46 & 2.7 & & 220 & 88 & 1.5 & 26\\
CMB, SZ, PS & 0.825 & 0.2 & & 220 & 24 & 0.41 & 5.6 &  & 230 & 89 & 1.3 & 30.6\\
CMB, SZ, PS & 0.825 & 1.0 & & 270 & 28 & 0.24 & 60.8 &  & 280 & 110 & 1.1 & 51.0\\
CMB, SZ, CPS & 0.825 & 0.2 & & 210 & 22 & 0.29 & 2.7 &  & 220 & 86 & 1.4 & 27.2\\
CMB, SZ, CPS & 0.825 & 1.0 & & 260 & 28 & 0.39 & 43.5 &  & 270 & 100 & 1.2 & 47.2\\
CMB, SZ, $Y_*=2.3 \times 10^{-6}$ & 0.825 & -- & & 220 & 25 & 0.5 & 7.6 & & 200 & 120 & 0.74 & 27.2\\
CMB, SZ, $\beta=1$ & 0.825 & -- & & 230 & 42 & 2.1 & 14.7 & & 250 & 170 & 10 & 34.9\\
\hline

CMB only & -- & -- & & 37 & 2.0 & 0.17 & 0 & & 37 & 6.6 & 0.41 & 0\\
CMB, SZ & 0.75 & -- & & 59 & 5.2 & 0.71 & 0 & & 59 & 20 & 4.7 & 0.07\\
CMB, SZ & 0.825 & -- & & 80 & 8.6 & 0.70 & 0 & & 80 & 33 & 3.5 & 0.36\\
CMB, SZ & 0.9 & -- & & 120 & 15 & 0.73 & 0 & & 120 & 59 & 5.3 & 2.4\\
CMB, ST99 SZ & 0.825 & -- & & 80 & 8.5 & 1.3 & 0 & & 80 & 35 & 5.4 & 0.14\\
CMB, SZ, PS & 0.825 & 0.2 & & 100 & 8.8 & 0.66 & 0 & & 100 & 34 & 3.1 & 0.5\\
CMB, SZ, PS & 0.825 & 1.0 & & 200 & 12 & 0.24 & 0 & & 200 & 49 & 1.5 & 10.3\\
CMB, SZ, CPS & 0.825 & 0.2 & & 93 & 8.2 & 0.83 & 0 & & 91 & 32 & 3.4 & 0.36\\
CMB, SZ, CPS & 0.825 & 1.0 & & 190 & 12 & 0.26 & 0 & & 190 & 46 &1.5 & 6.0\\
CMB, SZ, $Y_*=2.3 \times 10^{-6}$ & 0.825 & -- & & 94 & 11 & 0.66 & 0 & & 94 & 43 & 3.6 & 0.18\\
CMB, SZ, $\beta=1$ & 0.825 & -- & & 120 & 31 & 2.0 & 0 & & 120 & 120 & 9.1 & 2.4\\
\end{tabular}
\end{center}
\caption{Values for the mean and standard deviation at 30~GHz for the $l=2050-2350$ (top) and $l=2350-3900$ (bottom) CBI bins for $3 \times 3$ degree (1000 realizations) and $0.75 \times 0.75$ degree (16000 realizations) for various cosmologies and components, as well as the minimum and maximum values found and the normalized skew. The mean $\bar{B}_i$ and the standard deviation $\delta B_i$ within the bin $i$ are in $\mu$K$^2$; the skew $s$ is dimensionless and and the values of $S_\mathrm{max}$ given are in mJy. The components are the Cosmic Microwave Background (CMB), the SZ effect (SZ), point sources (PS) and clustered point sources (CPS). The CBI data points are $261 \pm 132$ and $387 \pm 117$ $\mu$K$^2$ for the lower and higher bin respectively; the final column gives the percentage of realizations at or above the central values of these measurements.}
\label{tab:excess_bin_values}
\end{table*}

To investigate the excess more systematically, we sample the power spectra for the same bins as used by the CBI. The histograms of the power spectrum from the CMB and SZ effect within the multipole bin $l=2050-2350$ for $\sigma_8 = 0.825$ and map sizes of $0.75 \times 0.75$ degree (representative of the deep fields) and $3 \times 3$ degree are shown in Fig. \ref{fig:cbi_histogram_lowerbin}. The measured value for the CBI power spectrum is compatible with the simulations for this multipole range; 26.5 per cent of the realizations have the same or higher value as the central value from the CBI measurement. The same histogram for the $l=2350-3900$ bin is shown in Fig. \ref{fig:cbi_histogram_upperbin}. In this histogram, the distribution is significantly skewed towards higher values, however the probability of getting the CBI excess from the realizations is low, with only 0.36 per cent of the realizations having the same or more power than the central CBI value. This would be reduced further if high-mass clusters are removed from the maps.

Table \ref{tab:excess_bin_values} gives the values for the various components and maps sizes for the two CBI high-multipole bins. Although the mosaicked fields extend to 5 degrees across, we use $3 \times 3$ fields due to the limits of the simulation box size at the highest redshifts. For the lower multipole bin, we find that the central value can be easily obtained using the CMB and SZ effect with $\sigma_8 = 0.9$, and realizations remain with the same amount of power for $\sigma_8 = 0.825$. Due to the large error on the measurement, however, it is in agreement with a large range of values of $\sigma_8$. For the higher multipole bin, there are no realizations with the central power for any of our three values of $\sigma_8$ using $3 \times 3$ degree realizations. There are $0.75 \times 0.75$ degree realizations that match the central value, even down to $\sigma_8 = 0.75$, but these are very unlikely, making up less than a percentage of the realizations for the lower values of $\sigma_8$.

Adding point sources to the realizations at the level of 0.2 or 1 mJy significantly increases the mean of the realizations in both bins, greatly increasing the number of realizations with the central power or higher as measured by the CBI in both bins. Clustering the point sources, however, slightly decreases this probability.

Due to the uncertainty in the parameters controlling the power and profile of the SZ effect (see Section \ref{sec:parameter_dependence}), the mean power spectra given here are uncertain to around 50 per cent. If this effect is in the positive direction, then this would significantly increase the number of realizations which agree with the CBI excess measurements. We change two of the parameter values, setting $Y_*=2.3 \times 10^{-6}$ Mpc$^{2}$ and $\beta = 1$ in two different sets of realizations. We find that the change in $Y_*$ has little effect, however the change in $\beta$ has a dramatic effect on the standard deviation and the skew of the distributions, as well as on the mean.

As shown earlier in Fig. \ref{fig:sz_masscuts}, a large part of the skew and increased standard deviation compared to the mean comes from the largest clusters in the maps. Although there is no formal mass limit for the CBI fields, they were selected to avoid known large clusters, making it unlikely that any local, high mass clusters are present in the fields. This will have the effect of reducing the number of realizations matching the power measured by the CBI in accordance with the results of section \label{sec:upper_mass}.

In addition to the CBI, the Berkeley-Illinois-Maryland Association (BIMA) interferometer operating at 28.5~GHz has also measured an excess. \citet{2006Dawson} surveyed eighteen 6.6 arcminute fields with a total area of $\sim 0.2$ square degrees, and found $220^{+140}_{-120} \mu$K$^2$ with an average multipole of $l=5237$ and a FWHM of 2870. We model this as a constant bin between $l=3800$ and $6670$. For a thousand realizations with $\sigma_8 = 0.825$ and the same map size as BIMA's $6.6$ arcminute field of view, we find a mean within this bin of 41 $\mu$K$^2$, a standard deviation of 82 $\mu$K$^2$ and a normalized skew of 5.5. Using just the mean and standard deviation, this makes the central value of the measurement a $\sim$ 2.2 $\sigma$ excess. We find that 3.3 per cent of the realizations have greater than 220 $\mu$K$^2$.

The Sunyaev-Zel'dovich Array (SZA) has also measured the power spectrum at higher multipoles, putting an upper limit of $149 \mu$K$^2$ (95\% confidence) on the power spectrum from secondary anisotropies between $l=2000-6000$ \citep{2009Sharp} using fourty-three 12 arcminute fields. As 68 per cent of their multipole coverage is between $l=2929-5883$, we use a constant bin with a width of $l=2900-5900$. We find a mean of $50 \mu$K$^2$, with a standard deviation of $64$ and a skew of 5.8. 5.6 per cent of the realizations have a power greater than $149 \mu$K$^2$. This is compatible with the measurement from SZA.

At higher frequencies, the Arcminute Cosmology Bolometer Array Receiver (ACBAR) has also measured an excess of $34 \pm 20 \mu$K$^2$ at 150GHz in the range $l = 1950-3000$; this 1.7 $\sigma$ excess is compatible with the CBI and BIMA measurements if the signal has the spectral distribution of the SZ effect \citep{2008Reichardt}. These measurements cover nearly 700 square degrees over 10 fields; we are limited by the map size of the Pinocchio simulations, which are 9 square degrees, so cannot make a prediction for these fields. For comparison, however, we find that the SZ effect alone contributes $9.6 \mu$K$^2$ with a standard deviation of $2.2 \mu$K$^2$ and a skew of 0.94 within the bin $l=1950-3000$ for $3 \times 3$ degree maps. We note that QUaD \citep{2009Friedman} also measure the 150~GHz power spectrum at high multipoles; they find values that are compatible with $\Lambda$CDM on its own with no SZ effect required.

\section{Conclusions}
We have simulated the microwave sky, including the CMB, SZ effect and point sources, and have used {\sc Pinocchio} to generate large numbers of cluster catalogues with realistic distributions for three different values of $\sigma_8$. Using these maps, we have investigated the statistics of the power spectrum between multipoles of 1000 and 10000.

We find that the inclusion of the SZ effect increases the standard deviation of the power spectrum by a factor of 3 over that expected from cosmic variance, in agreement with the predictions from an analytical calculation based on the halo formalism. The mean and standard deviation vary as $1/f_\mathrm{sky}^{1/2}$ as expected, and scale approximately as $\sigma_8^7$ over the range of values sampled here. We also find that the distributions are non-Gaussian, and are skewed by large mass clusters, with the degree of this skewness increasing as the map size is decreased.  Additionally, we find that correlations between galaxy clusters play a small role in the statistics of the power spectrum at the level of $\sim$10 per cent.

Several instruments have measured an excess at high multipoles, which may be due to the SZ effect with a large value of $\sigma_8$. We cannot explain the central values of these measurements with the range of $\sigma_8$ investigated here, however the increased standard deviation and the presence of skewness in the distribution means that these measurements could be explained by a lower value of $\sigma_8$ than has been suggested so far. There is also a large uncertainty in the parameters describing the cluster gas physics, which can have a large effect on the mean of the distributions, comparable to that from the different values of $\sigma_8$, and can also significantly effect the standard deviation and the skew of the distributions.

The next generation of CMB instruments are currently being commissioned, and are expected to provide more data at multipoles comparable to those probed by the CBI. The {\it Planck} spacecraft will measure the power spectrum from the whole sky out to multipoles of 2500 within the next few years, and instruments such as the South Pole Telescope \citep[SPT;][]{2004Ruhl}, the Arcminute Microkelvin Imager \citep[AMI;][]{2008Zwart} and the Atacama Cosmology Telescope \citep[ACT;][]{2004Fowler} will observe large numbers of galaxy clusters using the SZ effect. These measurements will provide much more information on the SZ effect and may provide a resolution to the discrepancy in $\sigma_8$ from the measurements to date. Our results should also be relevant to these observations.

\section*{Acknowledgments}

The authors thank the creators of {\sc Pinocchio}, {\sc CAMB} and {\sc FFTW} for making their software freely available. M Peel acknowledges the support of an STFC studentship. During the final stages of this work we became aware of work by \citet{2009Shaw} which addresses similar issues to those studied in this paper using a complementary approach.

\appendix
\section{Analytical calculation of the SZ power spectrum}

Rather than calculating the mean power spectrum from maps of the SZ effect, it can be evaluated directly from the number density and cluster profile using the halo formalism \citep[see for example][]{2001Cooray,2002Komatsu} under the assumption that the clusters are Poisson distributed. The mean binned spectra can be calculated by
\begin{equation}
B_i = \frac{1}{N}\sum_{l \in \mathrm{bin}} \frac{l(l+1)}{2 \pi} \int_{\mathrm{z_{min}}}^{\mathrm{z_{max}}} \int_\mathrm{M_{min}}^\mathrm{M_{max}} \frac{dV}{dz} \frac{dn}{dM}  y_{l}^2 dM dz,
\end{equation}
where $i$ is the bin number and the sum is over the multipoles $l$ in the bin.
\begin{equation}
\frac{dV(z)}{dz} = \frac{c^3 \left( \int_0^z E(z^\prime) dz^\prime \right)^2}{E(z) H_0^3}
\end{equation}
is the comoving volume of the Universe at a given redshift, where $E(z)$ is as in equation \ref{eq:hubble}  and $dn / dM$ is the comoving cluster number density. The parameter $y_{l}$ is the value of the Fourier transform of the cluster model at the multipole $l$, given by
\begin{equation}
y_{l}(M,z) = 2 \pi g(x) \int_0^{\varphi_\mathrm{vir}} \varphi_c y(\varphi_c) J_0 \left( (l+0.5) \varphi_c \right) d \varphi_c,
\end{equation}
where $g(x)$ determines the frequency dependence of the SZ effect, as per equation \ref{eq:conv_y_t}, $y(\varphi_c)$ is from equation \ref{eq:y_theta} and $J_0$ is zeroth-order cylindrical Bessel function.

Using the covariance matrix, the standard deviation can also be calculated in a similar way. Here, the angular trispectrum $T_{ll^\prime}$ is used, defined by \citep{2001Cooray,2002Komatsu}
\begin{equation}
T_{ll^\prime} = \int_{\mathrm{z_{min}}}^{\mathrm{z_{max}}} \int_\mathrm{M_{min}}^\mathrm{M_{max}} \frac{dV}{dz} \phantom{.}  \frac{dn}{dM} \phantom{.}  y_{l}^2 \phantom{.} y_{l^\prime}^2 \phantom{.}  dM dz.
\end{equation}
This is then combined with the expected Gaussian cosmic variance from the mean spectrum to calculate the covariance matrix $M_{ll^\prime}$ via
\begin{equation}
M_{ll^\prime} = \frac{1}{f_\mathrm{sky}} \left( \frac{2 C_l C_{l\prime}}{(2l+1) \Delta l}\delta_{ll^\prime} + \frac{T_{ll^\prime}}{4 \pi} \right),
\end{equation}
where $\Delta l$ is the width of the multipole bin and $f_\mathrm{sky}$ is the fraction of the sky being considered. The standard deviation can then be calculated from the diagonal of the covariance matrix, that is, $\sigma_l = l(l+1) \sqrt{M_{ll}} / 2 \pi$. This effectively includes both Gaussian and Poissonian terms. Note that \citet{2007Zhang} add an additional component to the calculation of the variance to model the clustering of galaxy clusters but find that the Poissonian term dominates over the clustering term; for this reason we do not consider it here.

\bsp

\label{lastpage}

\end{document}